碩士學位論文

# Localization using Optical Camera Communication and Photogrammetry for Wireless Networking Applications

# 무선 네트워크 응용을 위한 광학카메라 통신과 사진측량을 이용한 위치 추적

국민대학교 일반대학원

전자공학과 전자공학전공

**Md. Tanvir Hossan**

**2017**

# Localization using Optical Camera Communication and Photogrammetry for Wireless Networking Applications

*by*

*Md. Tanvir Hossan*

A thesis Submitted to the Department of Electronics Engineering, Graduate School, Kookmin University in partial fulfillment of the requirements for the degree of Master of Science

*Supervised by*

Professor Yeong Min Jang

2018 년 6 월

**국민대학교 일반대학원**

**전자공학과 전자공학전공**

**2017**

# Localization using Optical Camera Communication and Photogrammetry for Wireless Networking Applications

A thesis Submitted in partial fulfillment of the requirements for the degree of Master of Science

*by*

*Md. Tanvir Hossan*

June 2018

This is certified that it is fully adequate in scope and quality as a thesis for the degree of Master of Science

Approved by

---

Professor Sang-Chul Kim (Chair, Thesis committee)

---

Professor Sunwoong Choi (Member, Thesis committee)

---

Professor Yeong Min Jang (Thesis supervisor & member, Thesis committee)

국민대학교 일반대학원

**Md. Tanvir Hossan 의 碩士學位 請求論文을**

**認准함**

**2018 年 6 月**

審查委員長　　金 尙 澈 ㊞

審查委員　　崔 善 雄 ㊞

審查委員　　張 暎 民 ㊞

國民大學校 一般大學院

*Dedicated*

*to*

*My Mother, Sister and Grandmother*

# Acknowledgment

Foremost, I would like to express my earnest gratefulness to my advisor Prof. Yeong Min Jang for the enormous support of my master degree study and, for his endurance, inspiration, enthusiasm, and immense knowledge of my research in the field of wireless communication. I could have imagined that how lucky a student could be to have a such advisor and mentor. Furthermore, I would like to thank the rest of my thesis board members: Professor Sang-Chul Kim, and Professor Sunwoong Choi, who deploy their valuable time to weigh my critique. Their reviews, insightful comments, encouragement, and legitimate guidelines aid me to develop the contents of my research work to a great level.

I would like to acknowledge Kookmin University along with the Department of Electronics Engineering for providing me a chance to study under student friendly environment. Alongside, Government of the Republic of Korea opens the door of opportunities for the student like me to do research work under this research harbor.

Additionally, this is a great privilege for me to be a graduate scholar in Wireless Networks & Communication Laboratory (WNCL) where I have a chance to enrich my research expertise profoundly. I would want to express my appreciation from the deep of my heart to lab members, especially Amirul Islam, Partha Pratim Banik, and Eunbi Shin for their unconditional support. Concurrently, I would like to express my most sincere gratitude to Dr. Mostafa Zaman Chowdhury and Nam Tuan Le for their supportive attitude and research guidelines. I want to provide my acknowledgement to all of my Bangladeshi groups living in Korea, for their endless care and support during my study in Korea.

Finally, I must express my very insightful gratitude to my beloved mother, my elder sister and grandmother for providing me with trustworthy support and continuous reassurance throughout my years of study and through the process of researching and writing thesis. This achievement would not have been conceivable without them.

*Md. Tanvir Hossan*
*Kookmin University, Seoul, South Korea*
*June 2018*

# Contents

# List of Figures

# List of Tables

# Acronyms

| **Abbreviation** | **Full Form** |
|---|---|
| 2D | Two Dimension |
| 3D | Three Dimension |
| AI | Artificial Intelligence |
| AOA | Angle of Arrival |
| AP | Access Point |
| AWGN | Additive White Gaussian Noise |
| BER | Bit Error Rate |
| CMOS | Complementary Metal Oxide Semiconductor |
| CV | Computer Vision |
| FOV | Field of View |
| fps | Frame Per Second |
| FSK | Frequency Shift Keying |
| FV | Forwarding Vehicle |
| FV-ID | Forwarding Vehicle-Identity |
| GPS | Global Positioning System |
| HV | Host Vehicle |
| HV-ID | Host Vehicle-Identity |
| IC | Integrated Circuit |
| IEEE TG7m | IEEE 802.18.7m Task Group |
| IM/DD | Intensity-Modulation or Direct-Detection |
| IoT | Internet of Things |
| IS | Image Sensor |
| ITS | Intelligent Traffic System |
| KF | Kalman Filter |
| LBS | Location Based Services |
| LED | Light Emitting Diode |
| LED-ID | LED-Identity |
| LiDAR | Light Detection and Ranging |
| LOS | Line of Sight |

| **Abbreviation** | **Full Form** |
|---|---|
| MIMO | Multiple-Input and Multiple-Output |
| ML | Maximum Likelihood |
| NIR | Near Infrared |
| LOS | Line-of-Sight |
| OBD | On-Board Diagnostic |
| OCC | Optical Camera Communication |
| OFDM | Orthogonal Frequency Division Multiplexing |
| OOK | On-Off Keying |
| OWC | Optical Wireless Communication |
| PPM | Pulse Position Modulation |
| PSK | Phase Shift Keying |
| PWM | Pulse Width Modulation |
| RF | Radio Frequency |
| ROI | Region of Interest |
| RSSI | Received Signal Strength Indication |
| S2-PSK | Spatial-2-Phase-Shift Keying |
| SL | Street Light |
| SL-ID | Street Light-Identity |
| SNIR | Signal to Noise plus Interference Ratio |
| SNR | Signal to Noise Ratio |
| TDOA | Time Difference of Arrival |
| TOA | Time of Arrival |
| TOF | Time of Flight |
| V2I | Vehicle-to-Infrastructure |
| V2V | Vehicle-to-Vehicle |
| VANET | Vehicular ad-hoc Network |
| VLC | Visible Light Communication |
| WAVE | Wireless Access in Vehicular Environments |
| Wi-Fi | Wireless Fidelity |

# Notation

| | |
|---|---|
| $\varphi$ | Luminous flux |
| $I(0)$ | Luminous intensity at the center of LED lighting fixtures |
| $I(\varphi)$ | Luminous intensity |
| $\lambda_1$, $\lambda_2$ | Minimum and maximum wavelength of used optical spectrum |
| $K_m$ | Maximum spectral efficacy of vision |
| $V(\lambda)$ | Standard curve of luminosity |
| $P_t$ | Transmitted optical power |
| $\varphi_e$ | Energy flux |
| $\phi_e(\lambda)$ | Energy flux with the function of optical wavelength |
| $E_b$ | Energy-per-bit |
| $N_0$ | Spectral-noise-density |
| $\rho$ | Unit pixel value of the image sensor |
| $n$ | Noise term |
| $S$ | Signal amplitude |
| $\Delta$ | Camera exposure time |
| $\alpha$, $\beta$ | Model fitting parameters for system noise |
| $\kappa$ | Conversion efficiency from optical-to-electric at the receiver |
| $P_{opt}$ | Average optical power |
| $H$ | Optical channel DC gain |
| $\iota$ | Conversion between average optical power and average electrical power |
| $B$ | Modulation bandwidth |
| $P_{elec}$ | Average electrical power |
| $H_{else}$ | Channel gain for ambient light sources |
| $C$ | Channel capacity |
| $F_{fps}$ | Frame per second of camera |
| $W_s$ | Spatial bandwidth |
| $m$ | Lambertian index |
| $A_{IS}$ | Physical area of image sensor |
| $d$ | Distance from transmitter to receiver |
| $\theta$ | Incidence angle |
| $\theta_c$ | FOV of camera semi-angle |
| $\phi$ | Irradiation angle |
| $v$ | Transmitter |
| $u$ | Receiver |
| $\mathbb{N}$ | Normal number set |
| $N$ | Signal characteristic |
| $P_r$ | Average received optical power |
| $X, Y, Z$ | Euclidean three-space coordinates |
| $\mathbb{R}$ | Euclidean coordinate system |
| $f$ | Focal length of the camera |
| $p_x, p_y$ | Camera principal point coordinates |
| $R$ | Camera orientation of real-world coordinates |
| $P_{cam}(x, y, z)$ | Coordinates of camera |
| $f_x, f_y$ | Camera focal length in terms of pixel area at $x$ and $y$ direction |
| $m_x, m_y$ | Pixels number per unit distance at $x$ and $y$ direction |
| $s$ | Skew parameter |
| $K$ | Camera calibration matrix |
| $I$ | Identity matrix |
| $X$ | Coordinate matrix in the coordinate frame of world |
| $e$ | Distance from the focal length to the projected image on the image sensor |
| $a_i, b_i$ | Height and width of the projected image of LED light |
| $a, b$ | Height and width of the target LED light |
| $M$ | Lens magnification |
| $A_{LED}$ | Area of the target LED light source |
| $\eta_i$ | Number of pixels occupied by the projected image on the image sensor |
| $A_{circle}, A_{rec\tan gle}, A_{square}$ | Projected image area of circle, rectangle, and square shape light, respectively |
| $\tau$ | Constant for certain camera and LED light fixtures |
| $d_{r1}, d_{g1}, d_{b1}$ | Direct distance from camera lens to |
| $\eta_{ir1},\ \eta_{ig1},\ \eta_{ib1}$ | Number of pixel areas on IS of three different LEDs i.e., red, green and blue LED lights for position 1 |
| $d_{r2},\ d_{g2},\ d_{b2}$ | Direct distance from camera lens to three different LEDs i.e., red, green and blue LED lights for position 2 |
| $\eta_{ir2},\ \eta_{ig2},\ \eta_{ib2}$ | Number of pixel areas on IS of three different LEDs i.e., red, green and blue LED lights for position 2 |
| $P_i(x_{Lj},\ y_{Lj},\ z_L^{'})$ | Coordinates of any LED |
| $j$ | Set of $\mathbb{N}$ |
| $x_p$ | Particular solution of homogeneous system (trilateration) |
| $\varepsilon$ | Real parameter of homogeneous system (trilateration and multilateration) |
| $x_h$ | Solution of homogeneous system |
| $\gamma$ | Real parameter of homogeneous system (multilateration) |
| $P_1,\ P_2,\ P_3$ | Three positions of three LED lights within a 2D space |
| $X_{k_p}$ | Preliminary projected location |
| $J$ | Adoption (or state) matrix |
| $X_{k-1}$ | Initial location of detected object |
| $w_k$ | Additive noise from the initial location |
| $P_{k_p}$ | Error in position approximation (or process covariance matrix) |
| $P_{k-1}$ | Initial process covariance matrix |
| $Q_k$ | Additive noise |
| $K_g$ | Kalman filter gain |
| $T$ | Transformation matrix to convert a covariance matrix into Kalman filter gain matrix |
| $O_{error}$ | Error in measurement or observation |
| $X_k$ | Current estimation |
| $X_{k_p}$ | Previous approximation |
| $Y_k$ | Measured coordinates |
| $P_k$ | Present error in the approximation |
| $V$ | Transformation matrix |
| $s_1(t)$ | Bit interval for one of the pairs of LED |
| $k$ | Unsigned integer |
| $T$ | Signal cyclical interlude |
| $T_{bit}$ | Bit interval |
| $s_2(t)$ | Bit interval for one of other pairs of LED |
| $t_s$ | Sampling time |
| $s_1(t_s),\ s_2(t_s)$ | States of the pairs LED |
| $P_{e,S2-PSK}$ | Nonlinear XOR classifier to remove residual BER from modulation scheme |
| $\delta$ | Enhancement of error rate |
| $p_e$ | Probability of bit error of the LED state |
| $a_j$ | Flat distance between street lights |
| $a_1,\ a_1$ | Flat distance for street light 1 and street light 2 |
| $c$ | Space between the shortest distance from the SL to the cross section and a cross section of the flat line |
| $h$ | Flat distance between the pavement and camera |
| $d_{SL-SL}$ | Constant distance between street lights |
| $t$ | Position shifting time |
| $\Delta t$ | Fraction of position shifting time |
| $\theta_{SL_j-HV}$ | Angular position camera of host vehicle to street light |
| $V_{HV}$ | Host vehicle’s velocity |
| $n_{IS_SL}$ | Area of the street light’s LED on host vehicle’s camera |
| $D_{SL_j-HV}$ | Direct distance from LED of the street lights to the camera in the host vehicle |
| $\theta_{FV_k}$ | Angular displacement of forwarding vehicle with respect to the host vehicle |
| $Hd_{IS_FV_k}$ | Flat displacement of the projected image of forwarding vehicle on the image sensor |
| $n_{IS_FV}$ | Image area of the forwarding vehicle’s taillight on host vehicle’s camera |
| $P_{HV}$ | Virtual position of the host vehicle |
| $P_{FV}$ | Calculate position of forwarding vehicle |

# Abstract

## Localization using Optical Camera Communication and Photogrammetry for Wireless Networking Applications

Md. Tanvir Hossan
Department of Electronics Engineering
Graduate School, Kookmin University
Seoul, Korea

Localization defines a term to describe the identifying process of a location within the space of two-dimensional (2D) space or three-dimensional (3D). A localization scheme is an important concern for connecting sensor nodes in remote locations. The demand of localization in wireless networking is increased due to the availability of mobile devices as well as scope for billion-dollar market in e-commence sector. Moreover, a new era has written with internet-of-things, which boost this demand 100 times than ever before. Importance of localization applications is considered in both indoor and outdoor environments. Due to several advantages, LED and camera based positioning is more demanding over radio frequency based localization. Using the existing light-emitting diodes (LEDs) based illumination infrastructure it is possible to compute the coordinates of the camera, whereas cameras are embedded/installed in mobile objects, such as smartphone, vehicle. Optical camera communication (OCC) and photogrammetry are two important technologies to measure the position of these mobile objects. These technologies based localization scheme for both smartphone and vehicle should be cost effective and can be deployed with little modification of the existing infrastructures. I proposed two different schemes to localize these objects with OCC and photogrammetry techniques.

The important concern during localize very dynamic device like smartphone should extra attend and the proposed scheme should adapt with its dynamic scenarios. Though the proposed scheme use LEDs as the reference nodes to compare the position of the smartphone. At OCC, the receiver is camera to receive signal from LED lighting fixtures, which acts as the transmitter. At least three LEDs should locate within the field of view of smartphone camera to localize it. The location is calculated from the respective location from nearest LEDs where each and every LED is identified with their LED-ID. After receiving IDs, the change direct distance between the LEDs and camera calculate with the help of photogrammetry on the image sensor (IS). The algorithm uses a Kalman filter to track the next possible position to eliminate the error from the estimation.

To describe the proposed localization scheme, the vehicles on the road are classified into two sections: host vehicle (HV), which approximation the position of the other vehicles, such as forwarding vehicles (FVs), which travels in front of the HV. The taillight of the FV uses to transmit modulated signal to the HV and the camera of the HV receives the signal using OCC technology. OCC technology ensures the region of interest as well as some basic information from the FV. In the same scenario, both HV and FV are moving; therefore, HV identifies its position with respect to the street lights (SLs). These distances from the camera of the HV to the SLs or FVs is measured by applying photogrammetry. It helps to calculate the variation of the projected image area on the IS with the variation of distance.

Before moving at the conclusion of this research works, the simulation result for the proposed schemes (i.e., indoor and outdoor environments) are analyzed and improved performance has achieved.

# Chapter 1
# Introduction

## 1.1 Introduction

The term localization holds a meaning of identifying a location of an object within the space of two-dimensional (2D) space or three-dimensional (3D). There are other coordinate systems, e.g., homogeneous, curvilinear, cylindrical and spherical coordinate system. Above all, Cartesian coordinate system and polar coordinate system are most considerable coordinate system. Currently, the necessity of a localization scheme is increased due to the growing demand of mobile devices or internet-of-things (IoT) especially smart devices e.g., smartphone, autonomous vehicles are expanding dramatically. Massive number of nodes are interconnected with each other through wireless or wire mediums through the IoT [1]. Localization is an important feature where it enables one node to communicate with other nodes accurately and remotely. The reasons, methodology or even the characteristics of localization schemes vary with the requirements of indoor and outdoor environments [2].

In this chapter, importance of localization scheme for both indoor and outdoor (i.e., vehicular environment) is defined. The novelty of Optical camera communication (OCC) and photogrammetry based localization schemes is presented. Finally, at the termination of this chapter, the organization of this research works is discussed.

## 1.2 Importance of Localization

### 1.2.1 Indoor localization

Several sectors such as e-commerce, and e-banking are targeting these mobile devices as a medium to make communication with the consumers. The approaches of consumer-facing commercial applications will boost with location based services (LBS). A dynamic, reliable, adaptive, secure, and accurate localization technique is essential for LBS [3]. Moreover, this technique should provide less interruption for other available techniques. The scope of LBS is found in those places such as shopping malls, supermarkets, and transit stations (i.e., railway, bus, subway) where the access density of smartphones is the highest. Localization in indoor environments acts as a hub for almost all web based business cases.

### 1.2.2 Vehicle localization

It is no doubt that localize wireless sensors in indoor is a promising and crucial obligation for modern commerce and businesses. Moreover, localizing wireless sensors in outdoor environments, especially vehicle localization is more prioritized over indoor localization. Recently, the concern of road traffic safety [4] is coming once again under the light after the increasing number of fatal road accidents. From one statistic of the World Health Organization [5] shows that between the age of 15 and 29 years the number of deaths lies in 1.3 million due to the traffic-related accidents worldwide and this number is 15–40 times greater (between 20 and 50 million) for the non-lethal wounds. Top 10 causes of death, such as comparable to suicide, HIV/AIDS, homicide, and other diseases; traffic fatalities take a rank in the chart. Higher vehicle speeds (above 80 km/h) on the road [6] is the most common reason for traffic fatalities, which is around 60%. Autonomous vehicle has seen as important solution and its dramatic growth in the aspects of traffic accident reduction [7].

## 1.3 Novelty of OCC and Photogrammetry in Localization

Instead of radio frequency (RF) to avoid the typical problems, the optical frequency band in the electromagnetic spectrum is a novel approach to deal with the prevailing challenges. Optical wireless communication (OWC) [8] is an emerging and promising communication technology uses optical frequency bands (Figure 1) that can be considered a possible aspirant for tackling worst scenarios where RF faces challenges. OWC is not intended to replace RF; however, the coexistence of both technologies can offer a better performance in wireless application [9].

OCC [10] is the promising sub-system of OWC, which boost the localization technique one step ahead. It is a reliable, safe, fast, and secure essential for commun-ication as well as localization [12]. Therefore, deploying a single technology can use for localization and communication makes it unique than other technology. It uses a transmitter, such as light-emitting-diodes (LEDs) to transmit modulated binary data beam by varying its states to the camera or image sensor (IS), which is the receiver of the same technology [13]. Generally, LED uses for lighting purpose to illuminate indoor as well as outdoor environment. With little modification in the existing infrastructure, can make ready for communicating between devices, e.g., communication between camera of the smartphone and LEDs, or communication between two vehicles using the taillight of one vehicle and camera of the other vehicle [14], [15], [16], [17], [18], [19]. The biggest advantages of OCC are high signal-to-noise (SNR), secure line-of-sight (LOS) communication, un-interrupted communication channel except the ambient lights, multiple-input and multiple-output (MIMO) support, and simple signal processing technique [12] - [20]. Most prominent features of OCC are its MIMO functionality [21], which permits to communicate with multiple device at the same time.

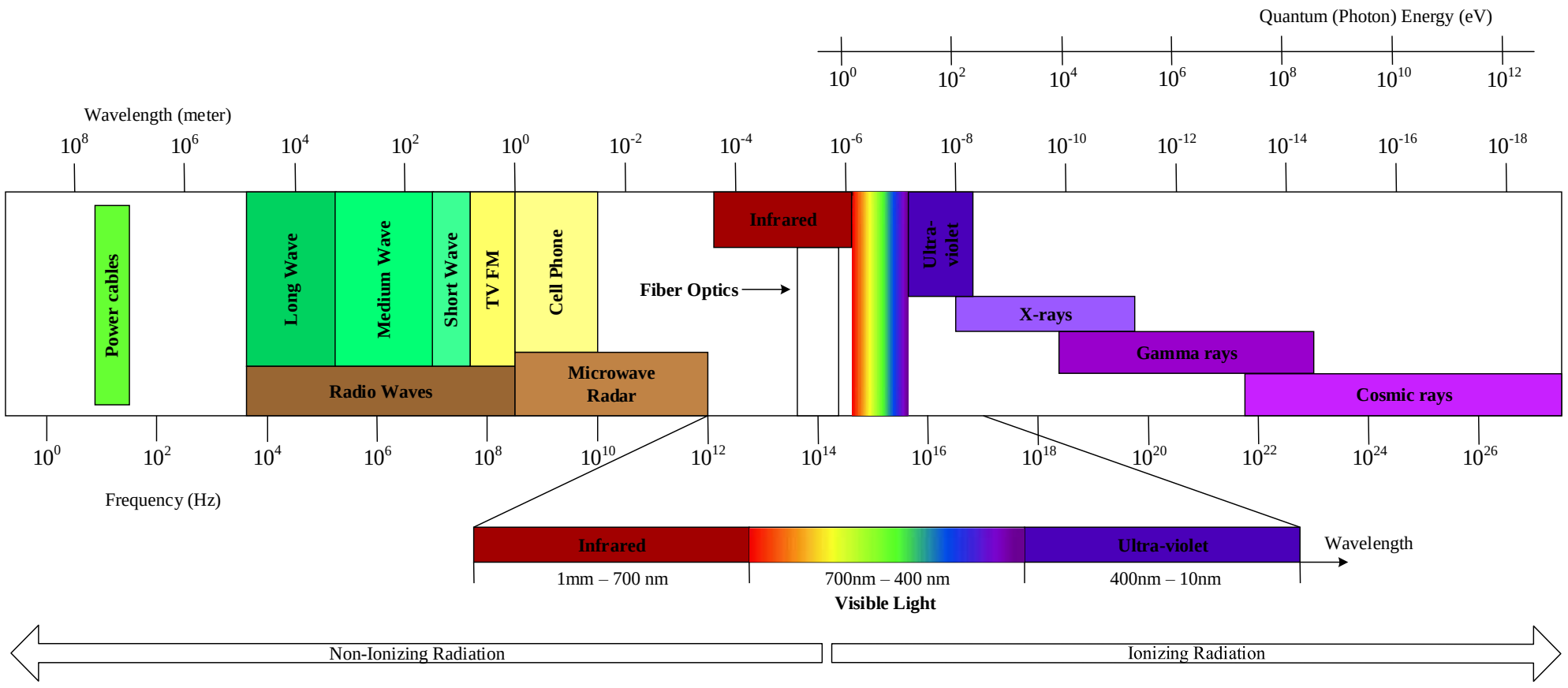

**Figure 1.** Electromagnetic spectrum [11].

To finalize the standard specification for OCC, IEEE has formed a new working group named IEEE 802.15.7 since 2011 [22]. This standard will be finalized by the middle of 2018. On the contrary, the camera output can be utilized by photogrammetry method to measure the space between objects and the camera.

Therefore, it is conceivable to uphold communication as well as localization at the same time with the wireless sensor using both OCC and photogrammetry techniques. Photogrammetry [23], [24] holds a division of geometry, which helps to measure the space from objects to IS of the camera and the way of measurement is done by quantifying the photon intensities of different wavelengths of light on an area, i.e., unit pixel of camera. It helps to collect information on the geometric properties, semantic, and variety of comparative distances of an object, i.e., LEDs, taillight of vehicles.

## 1.4 Contribution

Main contributions of this work are enlisted as follows:

1) Developing localization techniques for both smartphones in indoor environments and vehicular in outdoor environments
2) Optical camera communication and photogrammetry
3) Calculation of localization error and accumulating statistical data to mitigate localization
4) Accumulating sensor localization information for wireless networking

## 1.5 Thesis Organization

Rest of the thesis is structured as follows:

Chapter 2: briefly discusses about the related work on the issue of localization.

Chapter 3: describe about optical camera communication and photogrammetry.

Chapter 4: states indoor localization technique.

Chapter 5: represents vehicle localization technique.

Chapter 6: discuss about the results.

Chapter 7: importance of localization for wireless networking.

Chapter 8: conclude with future research approaches.

# Chapter 2
# Related Work for Localization

## 2.1 Introduction

There is a rising attention on both indoor and outdoor localization for smart sensors from both academic and industry areas. They provide several solutions for localization by using the existing infrastructures [25]- [26]. Global positioning system (GPS) is utmost common and reliable positioning system [27]. The application of GPS for vehicle localization cannot ignore. Moreover, due to several limitations in GPS i.e., poor GPS signal reception, loss of GPS signal, and limited localization accuracy [28]; makes the overall system unfit for the indoor localization and in some case for vehicle localization. Time of flight (TOF) cameras are thinking as an alternative solution for wireless sensors localization cases [29]. Beside the advantages TOF cameras; an expensive, complex system requires to localize wireless sensors. Sometime, dictation and ranging is not enough for localization sensors which position changes dynamically whereas communication with the sensor node ensures localization accuracy. TOF cameras use only for detection and ranging purposes and unable to keep communication with sensors [30]. There are several physical parameters of radio signal use for localization with special distributed monitors such as angle of arrival (AOA), time of arrival (TOA), time difference of arrival (TDOA), received signal strength indication (RSSI) [31] - [32]. There is also a vision based, i.e., computer vision (CV) or artificial intelligence (AI) based [33] localization approach exists. They have their own advantages along with their several limitations. For instant, RSSI has to consider several environmental effects such as interference from neighboring cells, signal fading, path loss, shadowing [34] and these will include an error in the overall localization measurement. Concurrently, reduced data transmission rate, lower location precision, and narrowband signals can be inhibited in TDOA parameter [35]. There is another approach like combining RSSI and AI [36] separate or together approaches to improvise the localization challenges, but it fails since subjective and objective data from AI have an impact on new input data makes the scheme more critical [37]. It feedbacks to the input of the system for performance improvement. At the same time, with the fast variations of the input variables, these feedbacks cause a huge impact on the new data. Photogrammetry is another essential approach for distance calculation by generating a map from sequentially taken photographs [38]. In Table 1, visible light communication (VLC) based positioning approaches are enlisted.

In this chapter, available localization schemes for indoor and vehicular environments are discussed. Though the condition of the scenarios, requirements, scopes, and localizing devices are differing with the environments, therefore, application of similar technique vary with the environments.

## 2.2 Available Localization Techniques for Indoor and Vehicular Cases

### 2.2.1 Indoor localization

Today, everyone brings and uses their phone either home or outside of the home. The camera is embedded with almost all mobile phones. These mobile devices become the hub for e-commerce. Therefore, LBS has become an important issue. To perform better in LBS, it is mandatory to measure the position of these mobile devices accurately especially in indoor.

GPS is a pseudo lite system, which faces challenges in the point of indoor localization. This system is LOS based localization solution accumulating sensor information from satellites far from the ground 20,000 km. Its signal interrupted by the obstacles on the ground such as trees, building; due to the LOS channel characteristic and longer distance between satellites and sensors. Therefore, a huge modification has done within this system to make it suitable for indoor localization. For example, integrating GPS signal with the indoor transmitting antennas to localize sensor nodes [39], which is not cost effective solution.

Recently, OWC based localization techniques are considered as a unique solution for indoor localization and navigation. In [40], [41], [42], [43], [44], [45], [46]; the authors are explained several approaches of visible light based localization and navigation schemes for mobile, photo-diodes or wearable computers. Still, there is plenty of debate on the value of localization resolution. The lower value of localization resolution has found with simulation results for various conditional approaches. Important sub-part of the OWC is OCC, where camera receives signal from the modulated light beam of LED. In [48], IS detects intensity of the metameric LEDs within a fixed indoor environment. The localization performance shows 1-degree orientation measurement and calculated resolution has found 1.5 cm and 3.58 cm in 2D and 3D space, respectively. Without measuring the angular displacement, the author in [49] claims 0.001 m localization resolution. Similar approach using LED (as transmitter) and camera (as receiver) for localization have discussed in [50], [51]. Additionally, the author in [52] added information from accelerometer sensor as well as demodulated data from IS to improve the overall performance in 3D indoor environment.

**Table 1.** Cataloguing visible light communication based localization methodologies. [47]

| Geometry (indoor) | Research objective | Parameter | Methodology |
|---|---|---|---|
| 2D space | Calculation of the impulse response of the optical channel and localization | RSSI | Triangulation |
| 3D space | Localization | RSSI | Triangulation |
| 3D space | Localization | RSSI | Triangulation |
| 3D space | High data rate communication and localization | RSSI | Triangulation |
| 3D space | Calculating RSS by the image sensor for localization | RSSI | Image sensor based on collinearity condition |
| 3D space | Localization | RSSI | Accelerometer and image sensors |
| 3D space | Localization | RSSI | Image sensor based on only off-the-shelf devices |
| 3D space | Localization | RSSI | Triangulation |
| 3D space | Localization | AOA and RSSI | Triangulation |
| 3D space | Localization | Light signals | Fingerprinting |
| 3D space | Localization | AOA and RSSI | Visible light communication and ZigBee network |
| 3D space | Amorphous Cell construction and localization | RSSI | Fingerprinting and Triangulation |
| 3D space | Localization based on intensity modulated or direct detection (IM/DM) and carrier allocation methods | RSSI | Triangulation |
| 3D space | Tracking | RSSI | Particle and Kalman filters |
| 3D space | Localization | TDOA | Sinusoidal signal properties |
| 3D space | Localization | TDOA | Based on phase difference |
| 3D space | Optimizing localization and optical channel characteristics | Light signals | Lambertian equation group |
| 3D space | Localization | RSSI | Bayesian approach |

Popular methods for localization to measure the receive signal strength at the receiver, i.e., photodiode, camera. This signal can be either visible light or RF. Author [53] proposed an RSSI based localization scheme for aperture based receiver, which has better angular diversity and a wider field-of-view (FOV). They derive Cramer-Rao lower bound on the position approximation for improving overall performance. Meanwhile, in [54], generic framework of RSSI has introduced for enriching the positioning performance. Author in [55] combining spatio-temporal constraints in RSSI fingerprint-based localization for the same purposes. Compressive sensing is applied in [56] to improve this performance in noisy scenarios. The implication of RSSI approach also found in visible light based positioning schemes [57], [58], [59]. Here, the localization resolution found 4 m [60]. Importantly, artificial intelligence algorithm has applied to collect location labeled fingerprints for optimizing the overall performance as in [61], [62]. Though the positioning accuracy for RSSI fingerprint has obtained 80% in [63]; moreover, 3D space fingerprint causes more overhead for positioning a high speed object compare with 2D space.

TOA and TDOA, these two approaches are several times for bringing a proper solution of localization in indoor environments. In [64], [65], [66], the TDOA based localization scheme has proposed where the light beam with unique frequency has transmitted in different phases. The TOA based localization algorithm has approached in [67]. The most important issue is that TOA and TDOA based localization schemes are considered as cost-effective and accurate scheme for localization. Moreover, this scheme depends on the location information of central node as well as other reference nodes in the same indoor space. Author in [68] proposed extended Kalman filter based TDOA approach to ignore the impact of this dependency. Deployment of such tracking algorithm is not always advantageous. For a first order approximation, the extended Kalman filter failed to approximation the location accurately.

AOA is another technique to apply for indoor localization. Transmitting gain difference from LED has considered by AOA in [60], [69], [70] for indoor localization. Simulation result shows the average value of localization resolution is 3.5 cm. Furthermore, it has some disadvantages e.g., accuracy degrades with increment of space between transmitter and receiver, reflection of signal from multipath will add error in the location measurement. Therefore, the AOA is recommended to avoid from implementation for any purpose especially in indoor scenarios.

### 2.2.2 Vehicle localization

For over a decade, researchers have been searching for a complete outdoor localization scheme. So far, GPS has been considered as most prominent solution in this regard. This provides a LOS vehicle localization solution by incorporating sensor information from the host vehicle (HV) [71], [72], [73] with data from a satellite orbiting at an altitude of almost 20,000 km. GPS uses RF band for positioning the HV on the road. Nevertheless, GPS does not enable the HV to measure its distance from other vehicles (e.g., forwarding vehicles) except to provide the current location information of HV. It has been found that, within a 10 sec periods, localization using GPS can generate a localization error of up to 1 m [74]. There is a wireless network standard for vehicle states as IEEE 802.11p, stated as wireless access in vehicular environments (WAVE) [75]. This standard is meant to maintain a communication network among vehicles within vehicular ad hoc networks (VANETs) and support intelligent transport systems (ITS) applications. RF signals in VANET systems are used not only for communication but also for vehicle localization [76]. Due to the multipath nature of the network and various environmental effects, the non-Gaussian noise is included with the transmitted signal, whose strength shows nonlinear characteristics over distance. The WAVE standard uses a license-free RF band (i.e., 2.4 GHz) [77], which is open to interference from other signal sources, making the whole network vulnerable. There are some other existing technologies for localizing vehicles, e.g., light detection and ranging (LiDAR) [78], [79], [80], and ToF camera technique [81], [82], [83], [84]. LED and cameras or photodiodes are embedded in same infrastructure of LiDAR and ToF system, but are used only for detection and ranging purposes. However, they are not useful for vehicle to infrastructure (V2I) or vehicle to vehicle (V2V) communications [30], [85], [86], [87], [88], these technologies are not economically convenient in a vehicular environment.

# Chapter 3
# Optical Camera Communication and Photogrammetry for Localization

## 3.1 Introduction

OCC is the subsection of OWC, whereas applying advance image processing technique on camera image to receive information as modulated light beam from the lights sources, i.e., LED. The features of OCC open enormous opportunities for the camera to use not only as a tool for capturing, monitoring but also as a receiver for receiving information from light sources for wireless communication purposes. On the other hand, using the technique of photogrammetry, occupied areas of the light source on the IS from captured images is measured. The size of occupying image area on the IS changes with the relative distance between camera and the light sources. Therefore, with the help of this technology it is possible to measure the direct distance, which is one of the coordinate from the Cartesian coordinate system to localize any mobile object.

In this chapter, mathematical model for OCC is described as well as the key equation for distance calculation is derived using photogrammetry.

## 3.2 Optical Channel Modeling and Communication

### 3.2.1 Light propagation model from LEDs

The geometry of the encapsulating lens and roughness of the chip faces effect on the radiation pattern of LED lights. This radiation pattern is explained by several models. Among all Monte Carlo ray tracing is one of the popular models [89]. The light ray is diffusely refracted or reflected in a cosine (or Gaussian) power distribution **[90]**. In Figure 2, the final light radiation pattern should appear linear super due to diffusely refraction or reflection characteristic of light. This radiation pattern is angularly shifted in function of the angle of incidence of every traced ray.

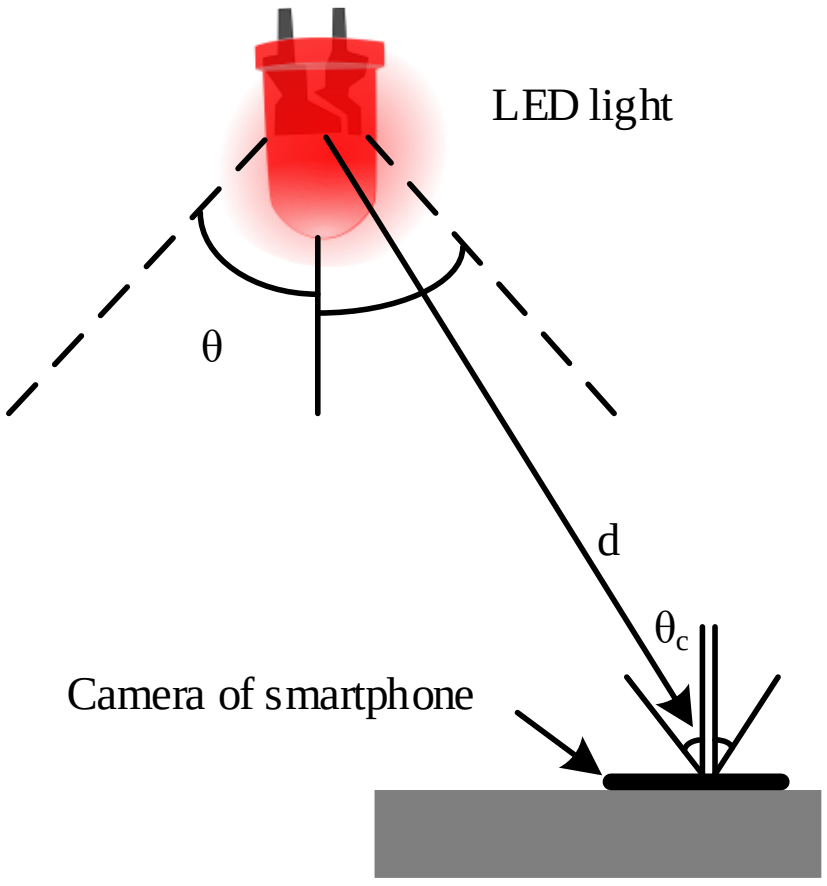

**Figure 2.** Geometric model of transmission and reception.

Light source, i.e., LED light fixture has two basic properties (a) transmitted optical power and (b) the luminous intensity also known as energy flux per solid angle. The luminous intensity is stated as follows:

$$I(\varphi) = I(0)\cos^{m}(\varphi) \quad (1)$$

where $I(0)$ is LED lighting fixtures center luminous intensity and $\varphi$ is the luminous flux. The value of luminous flux will be obtained by integrating the working optical spectrum between the limit of minimum wavelength $\lambda_1$ and maximum wavelength $\lambda_2$.

$$\varphi = K_m \int_{\lambda_1}^{\lambda_2} V(\lambda)\varphi_e(\lambda)d\lambda \quad (2)$$

where $K_m$ is concentrated spectral efficacy of vision and $V(\lambda)$ is standard curve of luminosity. $P_t$ is the optical power during data transmission, which is integrating the energy flux $\varphi_e$ in all directions stated as follows:

$$P_t = \int_0^{2\pi} \varphi_e d\vartheta d\lambda \quad (3)$$

The propagation direction of the radiated single light is shown in Figure 3(a) whereas its strength shown in Figure 3(b). At the center of the single light source, yellow color shade represents a maximum radiated power level (i.e., 40 dBm) and at the edge of the sphere minimum radiated power level (i.e., -80 dBm) represents with violet color shade. Joining average power strengths of light at certain x- and y-coordinate is forming an ellipse shape in Figure 3(b). Therefore, signal interference may happen from weak signal of neighboring light sources. Controlling the camera exposure time, the background of the capture area converts into dark space to mitigate the influence of ambient light.

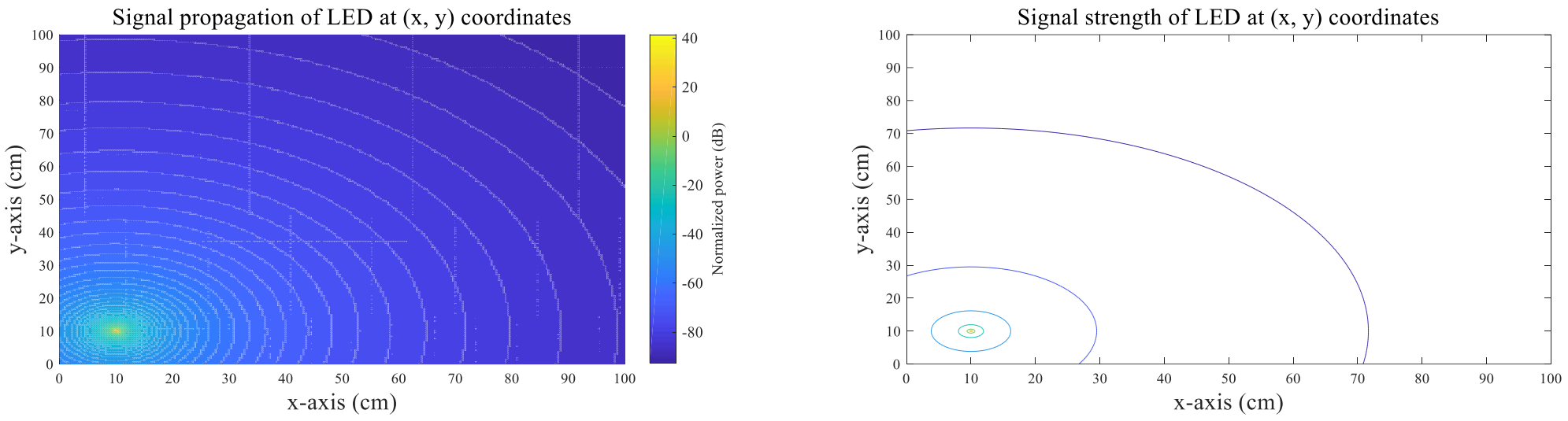

**Figure 3.** (a) Propagation and (b) strength of transmitting signal at (x, y) direction.

### 3.2.2 LED-ID for optical camera communication

A single LED light has a fixed coordinate numeric value, which is different from the other LED light fixtures in the same room. Each and every LED in that room transmits its own coordinates (i.e., x- and y-coordinates) as LED-identity (ID) to the camera on a smartphone. Integrated circuit (IC) of the LED driver sends data as modulated LED-ID by varying the light intensity. The driver controls dimming of LEDs in various ways to control the intensity of light. The encoded data signal phase has changed by regulating the LED lights on or off. However, it is not possible to tune the LED light when it will completely turn off. Therefore, dimming the intensity of the LED is a solution to get rid of this problem. This modulation technique is known as IM/DD modulation [91]. Following technique of IM/DD modulation, it can be classified as follows:

- On-off keying (OOK) [92]: In digital data transmission at the transmitter, high and zero voltage represents two logic data signal i.e., '1' and '0' respectively. The flickering illumination of the LED light helps to achieve these switching status.

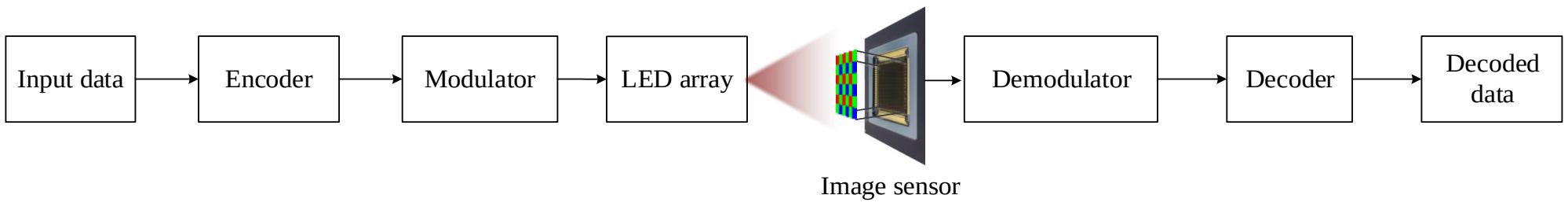

**Figure 4.** Architecture of transmitter and receiver in the OCC system.

- Pulse width modulation (PWM) [93]: In this modulation scheme, the light signal is transmitting as a square wave where the desire level of gained by the dimming control of LED light sources.
- Pulse-position modulation (PPM) [94]: With a single pulse in one time shifts, modulated light beam transmits to the receiver.
- Orthogonal frequency division multiplexing (OFDM) [95]: Data send as parallel sub-streams of modulated data using multiple orthogonal subcarriers in a channel.
- Frequency-shift keying (FSK): Instantaneous frequency shifting of the baseband signal with a constant amplitude carries the modulated digital signal.
- Phase-shift keying (PSK) [96]: Instantaneous phase shifting of the baseband signal carries the digital signal.

OFDM is successful for high speed data transmission, where there is a enormous chance of inter-symbol interference and multipath fading probability [97]. Moreover, this modulation scheme is not appropriate for indoor localization because of its high speed supportability. By considering all sorts of advantages and drawbacks,

OOK is the best choice for sending data to the camera for localizing smartphone.

In Figure 4, several blocks before receiving data at the camera of the smartphone shows the step-by-step processes of data encoding, modulation, and transmission by the LED light fixtures. The transmitted data have extracted after demodulation and decoding within several steps.

### 3.1.3 Channel modeling for OCC

The pixel of the model of IS [21] found as below:

$$Pixel\,\frac{E_b}{N_0}=\frac{E[\rho^2]}{E[n^2]}\approx\frac{S^2\Delta}{\alpha S\Delta+\beta} \tag{4}$$

where $E_b$ states energy-per-bit, $N_0$ implies the spectral-noise-density, $\rho$ is unit pixel value of the IS, $n$ is noise term, $S$ represents as the signal amplitude, $\Delta$ initialize the camera exposure time, and $\alpha$, $\beta$ are the model fitting parameters for system noise.

Considering environmental possessions, the signal-to-noise-plus-interference ratio (SNIR) [98] within a distort channel state as follows:

$$SNIR=\frac{(\kappa P_{opt}H)^2}{\iota^2 N_0 B+\sum(\kappa P_{opt}H_{else})^2} \tag{5}$$

where $\kappa$ states as the conversion efficiency from optical-to-electric at the receiver i.e., camera, $P_{opt}$ average optical power, $H$ is optical channel DC gain, $\iota$ states the conversion between the average optical power and average electrical power (i.e., $P_{elec}$ ), $B$ is modulation bandwidth, and $H_{else}$ channel gain for ambient light sources.

Since the additive white Gaussian noise (AWGN) characteristic of camera channel [99], the Shannon capacity formula has explained the channel capacity of the space time modulation as follows:

$$C=F_{fps}W_s\log_2(1+SNIR) \tag{6}$$

where $F_{fps}$ indicates the frame per second of camera, $W_s$ denotes as the spatial bandwidth. The spatial bandwidth ensures the amount of data flow by the pixels of each image frame. In a MIMO system, the number of parallel or orthogonal channels considers as the analogous term of this spatial band.

Both SNIR and the modulation scheme have an impact on bit error rate (BER) value. Which helps to measure the channel impact. The list of noise sources such as thermal noise, background and transmitter LED shot noise, and inter-symbol interference have direct and/or indirect impact on the transmission of the light signal.

Available smartphone's camera is rolling shutter based camera, which IS makes with the complementary metal oxide semiconductor (CMOS). Light intensities capture row by row on the IS and the whole image is composed of different pixel array with this shutter technique. Therefore, the exposed time delay between pixel array lines records the changing state of illumination of the LED light as a group of pixels in one image.

In the interim, the channel characteristic between LED lights on the camera is modeled by the optical channel DC gain $H$ [100], which states as follows:

$$H = \begin{cases} \dfrac{(m+1)A_{IS}}{2\pi d^2} g(\theta)T_s(\theta)\cos^m(\phi)\cos(\theta), \ 0 \le \theta \le \theta_c \\ 0, \quad \theta \ge \theta_c \end{cases} \tag{7}$$

where $m$ implies the Lambertian index, $A_{IS}$ is physical area of IS, $d$ is distance from transmitter to receiver, $\theta$ is incidence angle, $\phi$ is irradiation angle, and $\theta_c$ is the FOV of camera semi-angle.

Herein, $m$ is expressed as follows:

$$m = -\frac{\ln 2}{\ln(\cos_{\theta_c/2})} \tag{8}$$

For $\mathbb{N}$ number of LED-ID signal from $u$ number of transmitters, the channel output of each transmitter $v$ depends on the independent of signal characteristic $N$ as follows:

$$v = R_{cam}\sum_{i=1}^{\mathbb{N}} h_i u_i + N \tag{9}$$

where $u = [u_1 \ u_2 \ ... \ u_n]^T$ and $R_{cam}$ is camera responsivity.

Average received optical power on IS of the camera can be stated as follows:

$$P_r = \frac{I(0)\cos^m(\phi)\cos(\psi)}{d^2} \tag{10}$$

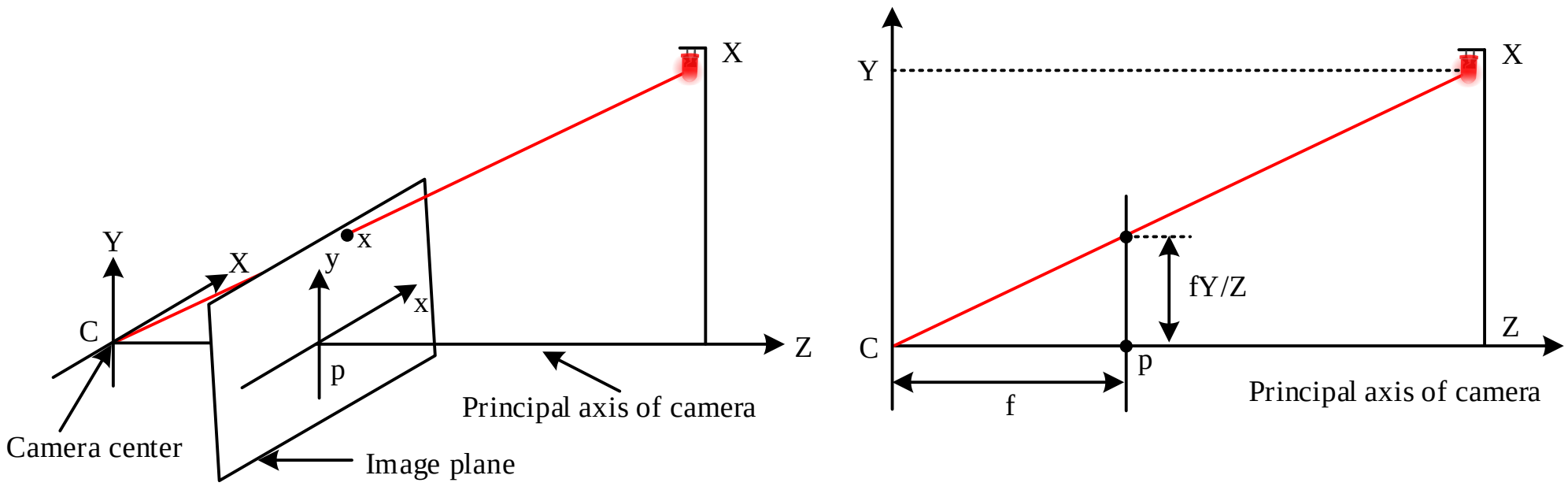

**Figure 5.** Camera calibration for vehicle localization.

### 3.2.4 Camera calibration and photogrammetry

Camera calibration is an essential tool in computer vision applications where it converts the 3D image from the real-world coordinates into 2D image. A pinhole camera model is simply discussing a perspective transformation for camera calibration purposes [101]. As in Figure 5, the starting coordinates of the projected object shift from the principal point of the camera's image plane in a Euclidean coordinate system. Mapping an object in Euclidean three-space $\mathbb{R}^3$ coordinates $(X,Y,Z)^T$ to Euclidean two-space $\mathbb{R}^2$ coordinates help to calibrate the camera as follows:

$$(X,Y,Z)^T \rightarrow \left(\frac{fX}{Z}+p_x,\ \frac{fY}{Z}+p_y\right)^T \tag{11}$$

where $f$ is camera's focal length and $(p_x, p_y)^T$ are camera principal point coordinates.

Homogeneous vectors allow to map the coordinates from 3D world into the 2D image in terms of matrix multiplication as follows:

$$\begin{pmatrix} fX + Zp_x \\ fY + Zp_y \\ Z \end{pmatrix} = \begin{bmatrix} R & -RP_C \\ 0 & 1 \end{bmatrix} \begin{bmatrix} f_x & s & p_x & 0 \\ & f_y & p_y & 0 \\ & & 1 & 0 \end{bmatrix} \begin{pmatrix} X \\ Y \\ Z \\ 1 \end{pmatrix} \tag{12}$$

where $R$ is the camera orientation of real-world coordinates, $P_{cam}$ states as the coordinates of camera; at the $x$ and $y$ direction $f_x (= fm_x)$ and $f_y (= fm_y)$ are camera focal length in terms of pixel area, respectively; $s$ implies as the skew parameter, whereas $m_x$ and $m_y$ state as number of pixels per unit space at $x$ and $y$ direction, respectively. Equation 12 be written simply as follows:

$$x = RK[I \mid -P_{cam}]X \tag{13}$$

where $K$ camera calibration matrix, $I$ is identity matrix, $X$ is coordinate matrix in the coordinate frame of world.

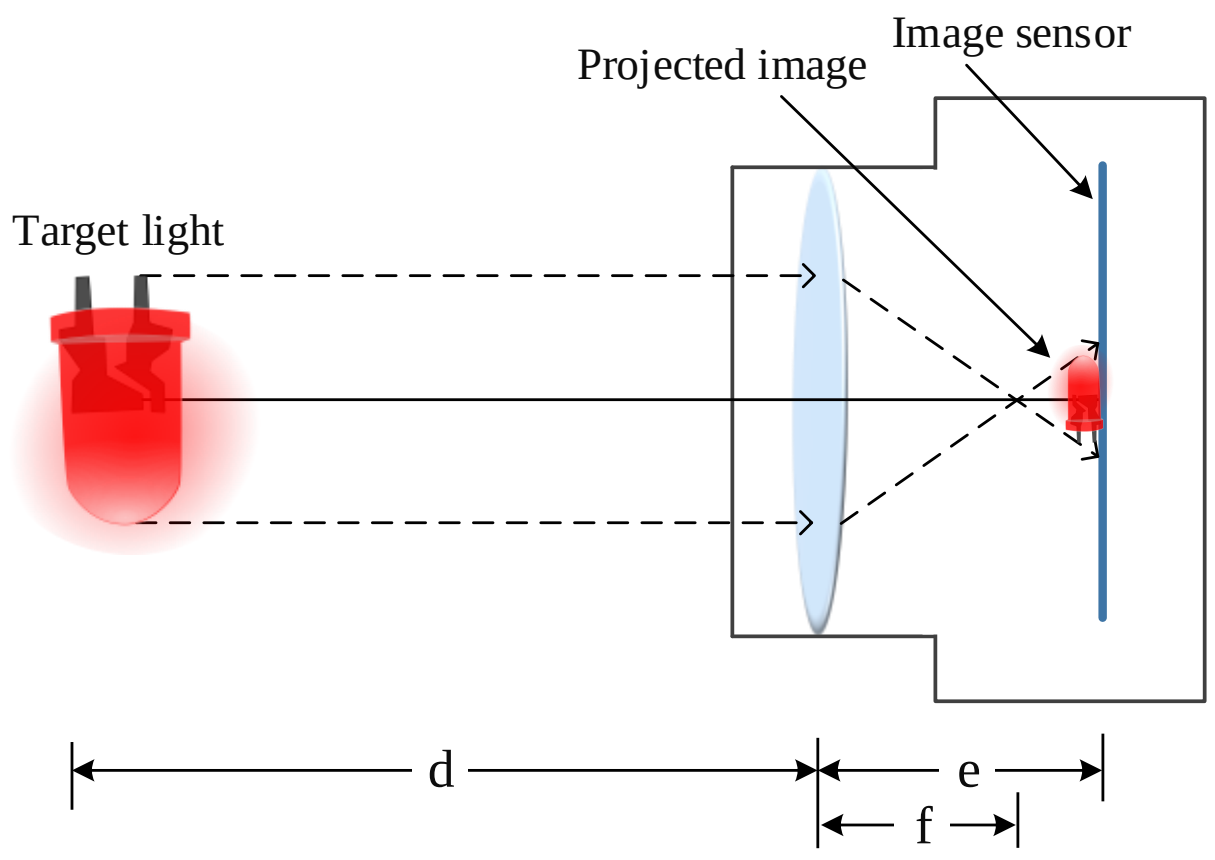

**Figure 6.** The basic architecture of the image projected on the IS of the camera.

## 3.3 Distance Calculation and Localization

### 3.3.1 Distance calculation between LED light and camera

Figure 6 shows the fundamental operational function of the camera where after passing through lens, the light from a target LED projects on the IS. The projected image is projected as up-side-down from the target on the image plate. If $e$ is the distance from the focal length to the projected image on the IS, $d$ is space from camera lens to target LED light, and $f$ is focal length of camera; it can state as:

$$\frac{e}{d} = \frac{f}{d-f} \tag{14}$$

The lens magnification is defined as the ratio of geometric span of projected image to span of the target LED light. For a square projected image on the IS, height and width of the projected image and target light are $(a_i, \mathrm{b}_i)$ and $(a, \mathrm{b})$, respectively. This lens magnification can be expressed mathematically as follows:

$$M = \frac{a_i}{a} = \frac{b_i}{b} = \frac{e}{d} = \frac{f}{d - f} \tag{15}$$

In a lossless optical channel, the value of distance $d$ is much bigger than the value camera's focal length $f$, i.e., $d >> f$. Therefore, denominator of Equation 15 is written as $d$ (i.e., $d - f \approx d$). Merging the Equations (14) and (15), following mathematical expression has accumulated:

$$a_i b_i = M^2 ab \tag{16}$$

The projected image size of the IS equals to the unit pixel area times to a pixel number of IS. Therefore, the pixel number for projected image is stated as follows:

$$\eta_i = \frac{\textit{projected area} \text{ on the image sensor}}{\textit{unit} \text{ pixel area}} = \frac{a_i b}{\rho^2} \tag{17}$$

If $A_{LED}$ is an area of the target LED light source, $\rho^2$ is the unit pixel area, and $\eta_i$ is the number of pixels; then the space from LED to camera is resolute by combining (15),(16), and (17) as follows:

$$d = \frac{f}{\rho}\sqrt{\frac{A_{LED}}{\eta_i}} \tag{18}$$

The typical available shapes of LED light fixture are rectangular, square, and circle in the market. For circular shape LED light, its area is $A_{circle} = \pi a^2$ where $a$ is the radius of that circle. Concurrently, the area of the rectangular and square LED light panel is $A_{rectangle} = ab$ and $A_{square} = a^2$ where $a$ and $b$ are the width and length of a rectangle, and $a$ is the length of the square, respectively.

The distances from LED light fixtures to the camera of the smartphone are different due to a fixed distance between every LED light fixture where all those LED light fixtures stay within the FOV of the camera. The distance will be minimum when the LED face straight to the camera than the LEDs which are situated at an angle to the same camera. Equation 18 gives a simple mathematical expression to calculate this distance. With the relative movement between the camera and the LED light fixtures, distance calculation should be updated accordingly. The focal length of the camera $f$ and the unit pixel area $\rho^2$ of IS are fixed for certain camera specification. . If the camera focal length $f$ and unit pixel length $\rho$ are maintained constant for a certain camera, the distance of the LED is kept proportional with respect to the square root of LED's area and disproportional with respect to square root of pixel area of that LED on the IS Therefore, Equation 18 can be written as follows:

$$d = \tau \sqrt{\frac{1}{\eta_i}} \tag{19}$$

where $\tau \left( = f\sqrt{A_{LED}} \big/ \rho \right)$ is constant for certain camera and LED light fixtures in a certain environment [19].

# Chapter 4
# Indoor Localization

## 4.1 Introduction

The smartphone is localized by using LED light fixtures as reference node as well as taking technical support from both OCC and photogrammetry. Each and every LED light fixture transmits modulated data signal as LED-ID to the camera using OCC technology. The coordinates information about these lights sends via these LED-IDs. This coordinates contains the information about the x- and y-coordinates only and never provides any idea about the direct distance about the direct distance information between smartphone and LED light fixtures. Therefore, the information about the z-coordinate does not appear in this context. There is a unique approach to measure these missing coordinate information. The size of the projected image on the IS of the camera change with the relative position displacement between camera and light sources. Though the position of the LED light fixtures is almost constant with the respect to the position of the smartphone. Photogrammetry helps to measure the change in the projected image area by calculating its contour on the IS. Furthermore, the position of the smartphone changes frequently over a fraction of time. Therefore, it requires an algorithm, e.g., Kalman filter to estimate its next position.

In this chapter, overall architecture of the proposed scheme is presented. It describes every steps from getting coordinate information from the light fixtures to finalizing the location of the smartphone progressively. When there is no smartphone within the indoor environment, the localization process terminates, which is described at the end of this chapter.

## 4.2 Receiving Coordinates of LEDs and Calculating Distances

### 4.2.1 Localization infrastructure

The proposed localization scheme [102] enable to find smartphones inside the indoor environment with the help of LED light fixtures and the camera of the smartphone. As in Figure 7, several factors have to consider such as all LED light fixtures are attached to the ceiling, constant inter distance between celling to floor for avoiding calculation complexities, more than two LEDs under the camera FOV, continuous communication link with the lighting server; to localize smartphone in indoor site. The LED light fixture coordinates and the world coordinates are parallel with each other. Concurrently, the distance between ceiling to the floor represents the z-coordinates which always equal in our case, which is inverse to the world z-coordinates. To a certain spot in a room, a single LED coordinates are also fixed. The only varying coordinates are smartphone coordinates with respect to either LED light fixture coordinates or world coordinates. The smart- phone has the API-based access of reproducibility, web- based visualizations; which is provided by a lighting server. FOV is an important parameter for smartphone camera through its solid angle, it can sense the electromagnetic radiation (e.g., visible light). Increase the number of LED lights within the FOV of the camera ensure a solid improvement of the system performances.

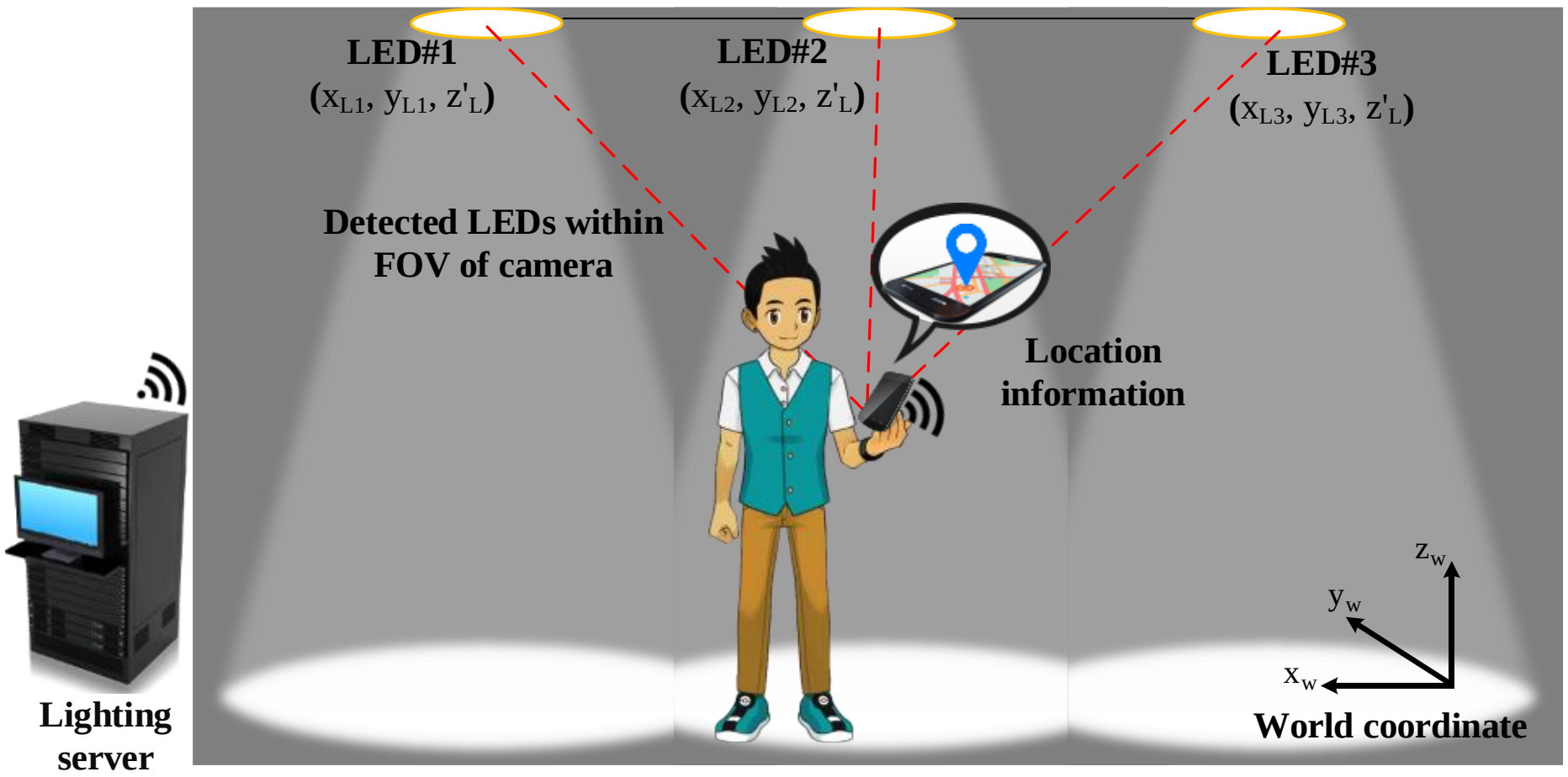

**Figure 7.** Overall scenarios of indoor localization scheme.

In Figure 8, each LED light fixture broadcast its own coordinates information as a modulated light beam. Though the z-coordinate is similar for every space inside a room, the coordinate information contains only the numerical value of x- and y-coordinate. It will make the transmitting data packet simple and small. Camera of the smartphone is the receiver of the OCC system. After demodulation and decoding, there are two different steps that smartphone process the data: (a) retrieve the LED-IDs from the received signal, (b) using photogrammetry, measures the distance of LED light fixture of the corresponding LED-ID. In photogrammetry, pixels number of an LED on the IS changes with variation of direct distance between smartphone's camera and light fixtures. The number of pixels analogous to the area of the projected image on the IS. This area has disproportion relation to the distance of the light source from the camera. The area is smaller for far distant light sources and alternatively larger for near distance light sources.

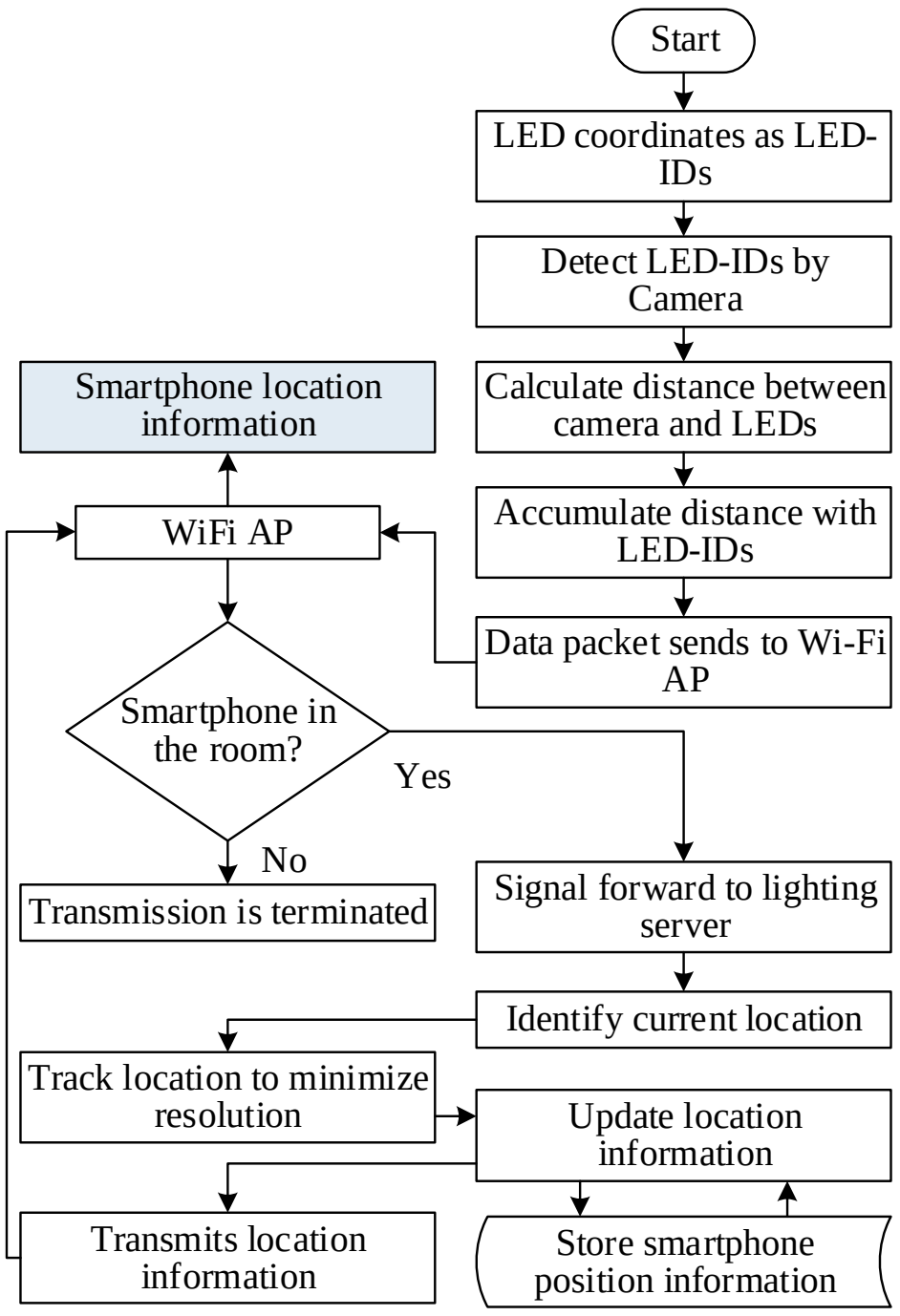

**Figure 8.** Flow diagram visualizes the indoor localization progressive development.

A wireless fidelity (Wi-Fi) access point (AP) keeps a wireless communication link between smartphone and lighting server. Final measurement of distance with the corresponding LED-ID send to the lighting server as data packet by the smartphone. It helps to estimate the position of the smartphone around some specific LED light fixtures. All LED light fixture coordinates stored in lighting server. The main function of the lighting server is to compare these position information of the smartphone with store coordinate information of the LEDs.

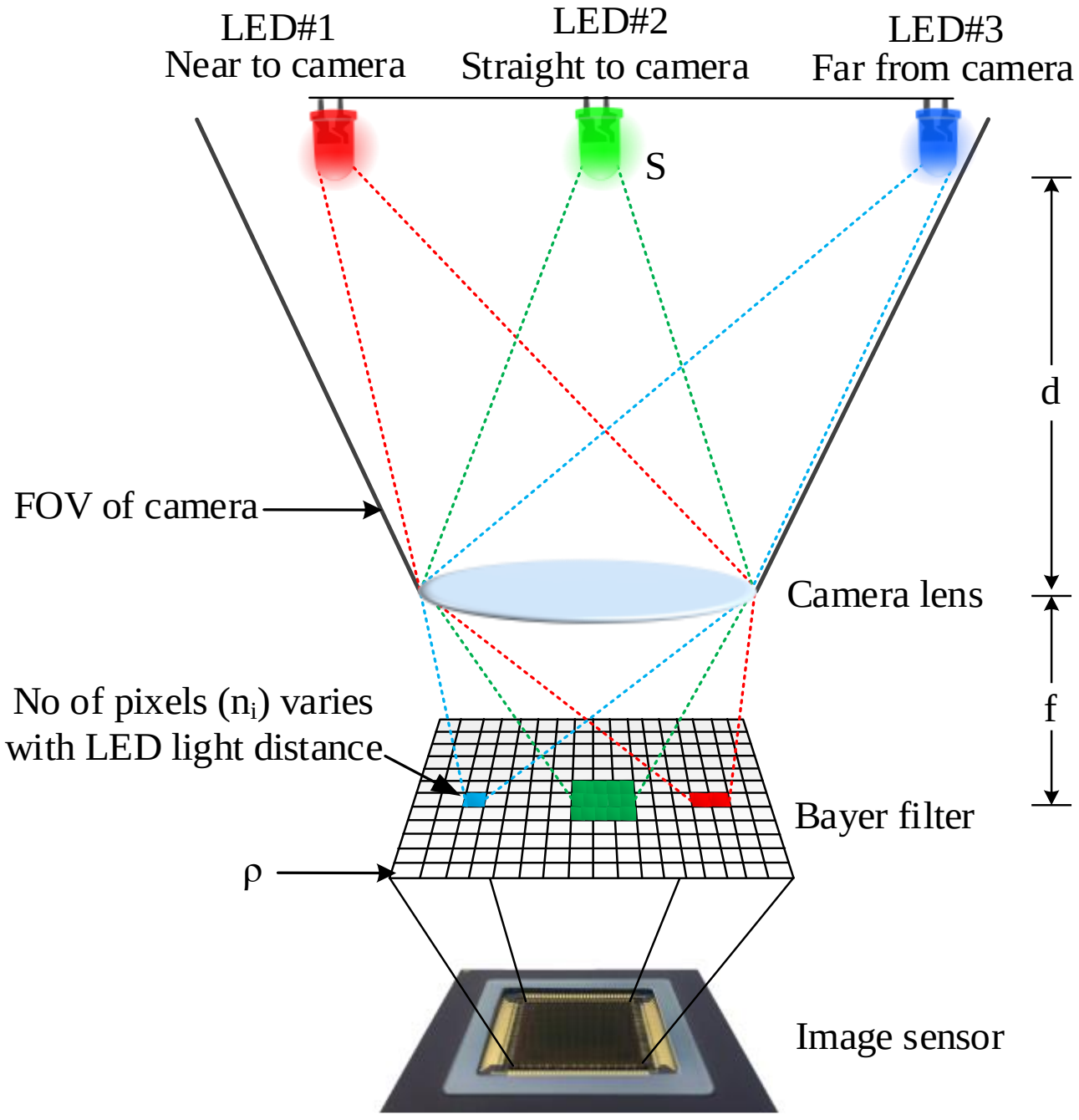

**Figure 9.** Difference of projected image area with distance.

Localization of smartphone is comparatively critical then other static sensors due to its dynamic nature. During data processing for localization, the smartphone may not find at the estimate position. For a long time, the use of Kalman filter for tracking dynamic devices or sensors is well known, which helps to predict the next possible position precisely with currently measured position information and from other parameters e.g., velocity, acceleration. The calculated location information in the lighting server may store or share with the smartphone to further approaches, e.g., navigation instruction, location based services.

For various positions of the LED light fixtures with the FOV of the smartphone's camera; the distance will vary accordingly, which is shown in Figure 9. It can conclude from Equation 19 that the distance between LED light fixture to the camera is contrariwise proportionate to square root of the projected image area. Therefore, when the distance increases, the size of the projected image area of the IS will increase accordingly.

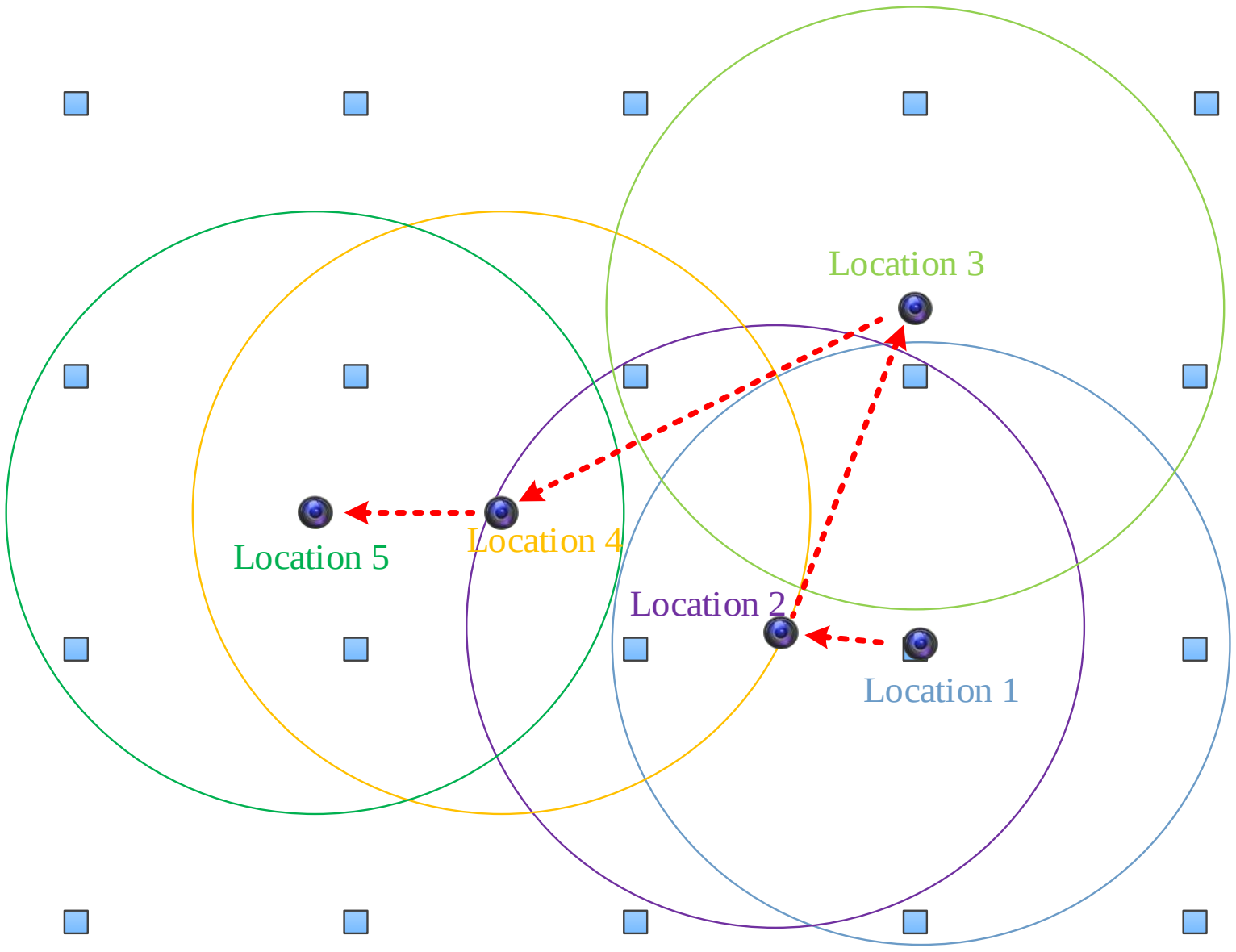

**Figure 10.** Change of projected LEDs with camera's FOV.

### 4.2.2 Changing detection area on the IS for camera movement

In indoor environment, smartphone changes its position with respect to the LED light fixtures. With the change of smartphone position, the FOV of the camera also changes. Therefore, projected image of the light sources also changes accordingly. On the other way, the position among the LED light fixtures should be kept in a way to maintain at least three LEDs within the camera FOV. The reason of getting data from three LED simultaneously explains at the end of this section. A scenario in Figure 10 describes the changes of detecting LED light fixtures due to the movement of the camera from location 1 to location 5. Here, black dots, large circles, and square blocks represent camera, FOV of the same camera, and LED light block, respectively. In all cases of image projection, the number of LED is more than three. More than three LED will ensure localization accuracy more precise.

Considering the movement between position 1 to position 2 in Figure 11; to analysis the projected image area with the change of camera position. In Figure 11(b), three LEDs (i.e., red, blue, green) locate constantly either FOV of camera shifts with the smartphone. Initially, the camera was at location 1 (Figure 11(c)) and projecting three

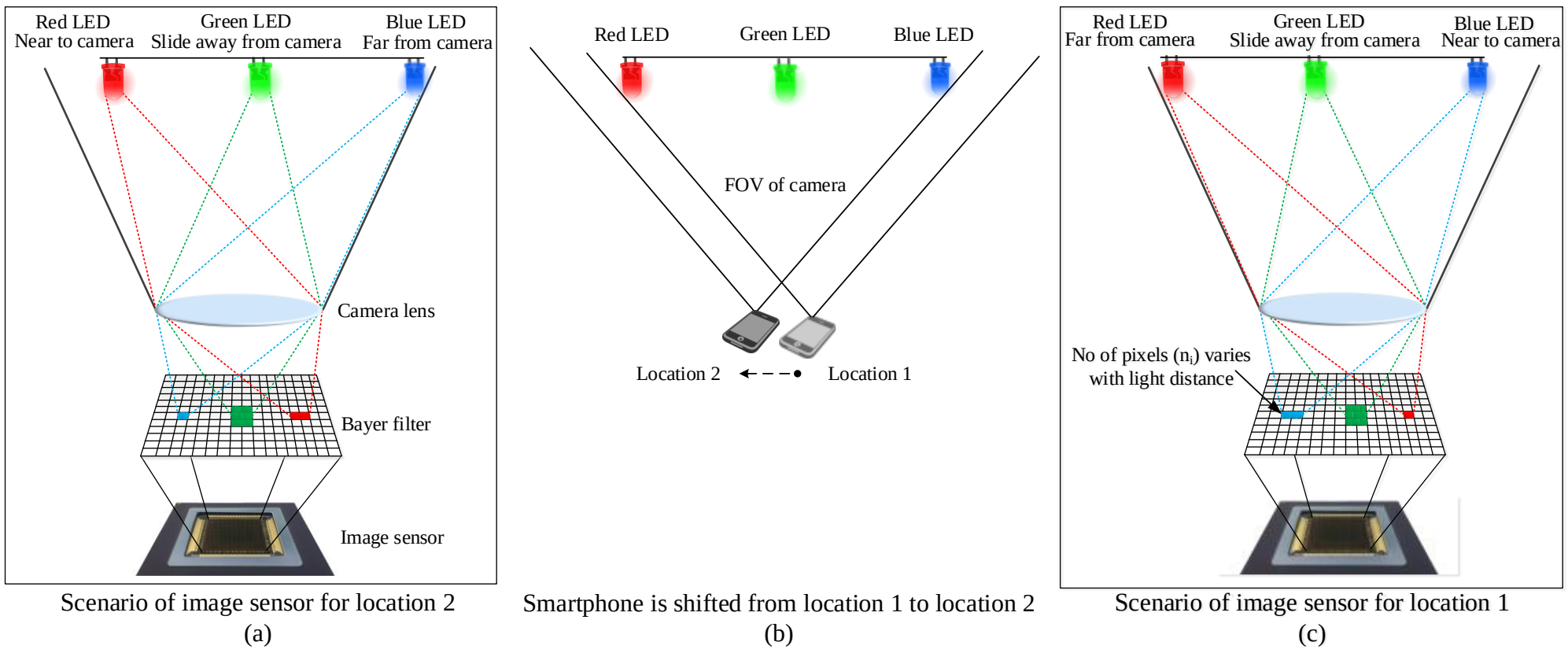

**Figure 11.** Change of projection with smartphone's position.

LEDs at various angles, e.g., green LED stays slide away from the straight line to the camera whereas blue LED stays near to the camera compare with the red LED. Considering the angular displacement between camera and LEDs; this value is higher in red LED compare with the blue LED. Concurrently, this value is near to ignorable for green LED compared with the red and blue LEDs.

The direct distance from camera lens to three different LEDs i.e., red, green and blue LED lights represents as $d_{r1}, d_{g1}, d_{b1}$ whereas the corresponding number of pixel areas of those LED lights on IS represent as $\eta_{ir1}, \eta_{ig1}, \eta_{ib1}$.

For location 1, Equation 19 can be written as follows:

$$d_{r1} \propto \frac{1}{\sqrt{\eta_{ir1}}} \tag{20}$$

$$d_{g1} \propto \frac{1}{\sqrt{\eta_{ig1}}} \tag{21}$$

$$d_{b1} \propto \frac{1}{\sqrt{\eta_{ib1}}} \tag{22}$$

In Figure 11(c), the project image area for the green LED light comparatively larger than blue to red LED lights. In another point of view, this area is smaller for the red LED light due to the highest direct distance. In the point of size of the projected image area on IS, it can easily conclude with the mathematical relation $\eta_{ig1} > \eta_{ib1} > \eta_{ir1}$. Though these image areas completely depend on the distance; therefore, the mathematical relation to the point of distance states as $d_{r1} > d_{b1} > d_{g1}$.

In Figure 11(a), perspective position between camera and LED light fixtures has changed due to the movement of smartphone from position 1 to position 2. In this case, blue LED remains far from the camera compared with the red LED. Additionally, distance from green LED to camera lens remains same as position 1.

For location 2, the direct distance from camera lens to three different LEDs i.e., red, green and blue LED lights represents as $d_{r2}$, $d_{g2}$, $d_{b2}$ whereas the corresponding number of pixel areas of those LED lights on IS represent as $\eta_{ir2}$, $\eta_{ig2}$, $\eta_{ib2}$. Equation 19 can be written as follows:

$$d_{r2} \propto \frac{1}{\sqrt{\eta_{ir2}}} \tag{23}$$

$$d_{g2} \propto \frac{1}{\sqrt{\eta_{ig2}}} \tag{24}$$

$$d_{b2} \propto \frac{1}{\sqrt{\eta_{ib2}}} \tag{25}$$

In Figure 11(a), the project image area for the green LED light comparatively larger than blue to red LED lights whereas this area is smaller for the blue LED light due to the highest direct distance. In the point of size of the projected image area on IS, it can easily conclude with the mathematical relation $\eta_{ig2} > \eta_{ir2} > \eta_{ib2}$. Though these image areas completely depend on the distance; therefore, the mathematical relation to the point of distance states as $d_{b1} > d_{r1} > d_{g1}$.

### 4.3 Uploading Information to the Lighting Server

As explained earlier, the direct space from the camera to LED light fixtures differs. Therefore, after measuring the distance from the projected image area, smartphone combine the measured distance value with corresponding LED-ID as data packet and sends to the lighting server via Wi-Fi AP. The lighting server stores the coordinates information on all LED light fixtures. It starts to compare the receive information from the smartphone with the store coordinate information. It calculates the location of the smartphone with the mathematical model of trilateration (compare with three known points) or multilateration (compare with more than three known points).

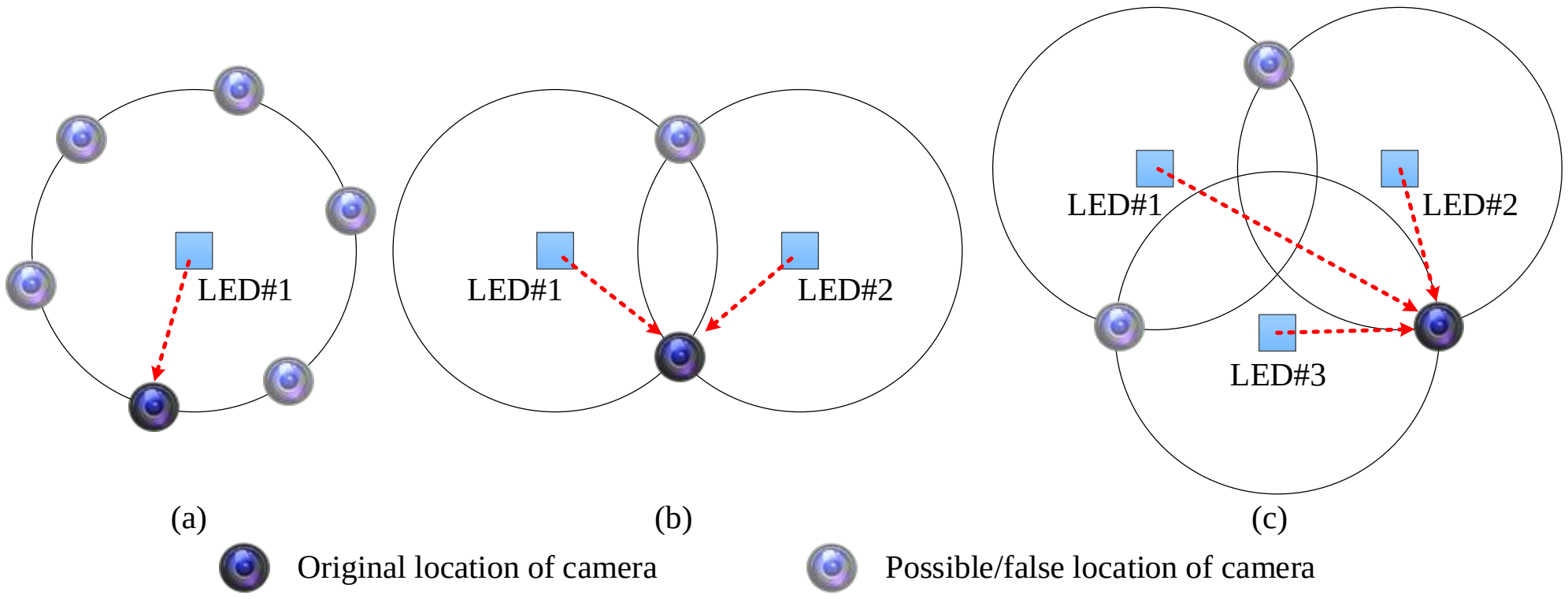

**Figure 12.** Narrowing down localization estimation error with 3 LEDs.

### 4.3.1 Computing smartphone location

The light from LEDs propagates and illuminates 360° areas from the center of a light source. Within these areas, the intensity of illuminated light is equal. If a receiver, e.g., camera located anywhere within these areas; it observes and measures distance will carry error. The possibility of false location measurement will be higher. In Figure 12(a) the possible false position around a single LED light is shown as faded camera images whereas the original camera also locates with them. When the lighting server compares the location information with its stored data, it gets confusing to finalize the coordinates of the smartphone.

If there are two LED lights within the FOV of the camera, it may narrow down the position estimation error. As shown in Figure 12(b), the comparative position of each LED to the camera helps the lighting server to get understand the actual position of the smartphone. Moreover, there are few places within the territory of two LED lights, which will arise confusion during position estimation. The faded camera indicates the false position around those LEDs.

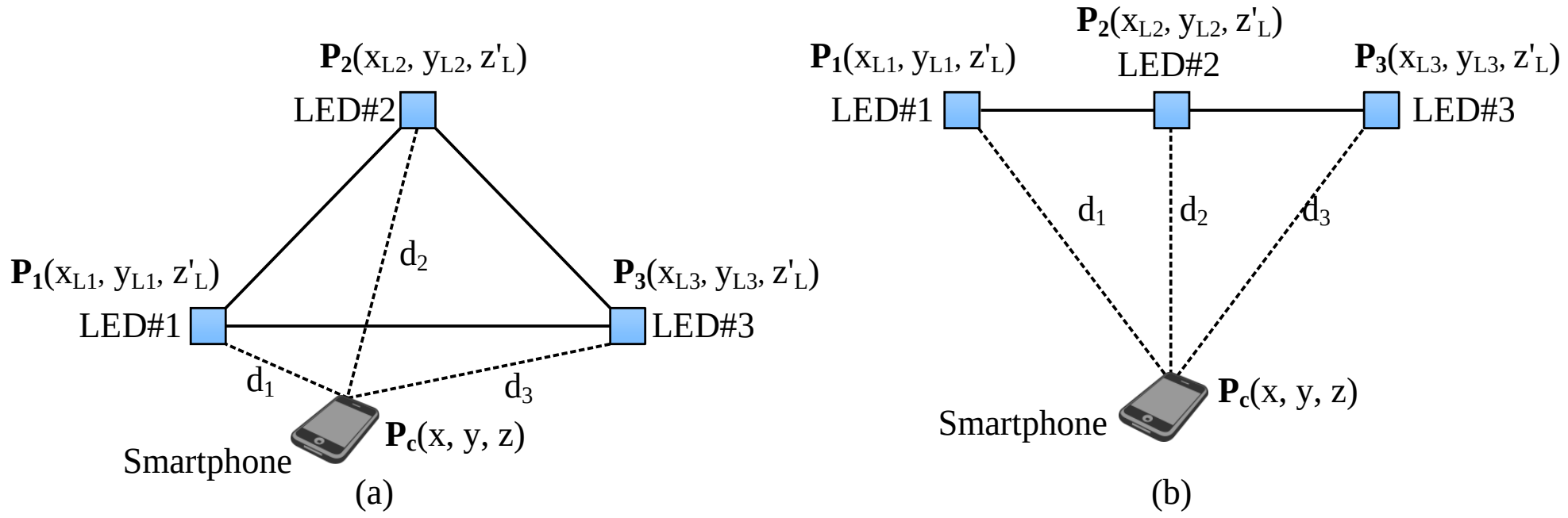

**Figure 13.** Trilateration for smartphone's coordinate calculation.

With a third LED light we can narrow the location of the smartphone to one possible location. In Figure 12(c), three circles centered on each of the landmarks overlap at three different locations where the radius of each circle is equidistant from each landmark. Therefore, two other locations of the smartphone along with the original are still possible even with three landmarks. Moreover, other location information may not arise any confusion for smartphone position estimation. It is possible to estimate the smartphone's location accurately by comparing information from two LED lights with information from a third light.

Trilateration or multilateration is a technique to define the position of a sensor, e.g., smartphone by means of three or more than three known positions, e.g., position of LED lights [103]. Three relevant nonlinear equations, measure the platform simultaneously in trilateration. As in Figure 13, the light fixtures align in a triangle or straight line with each other.

Under the ceiling, if $P_i = \left(x_{Lj},\, y_{Lj},\, z_L^{'}\right)$ is the coordinates of any LED light where $P_c = \left(x,\, y,\, z\right)$ is the coordinate of smartphone camera. Here, $j$ is set of $\mathbb{N}$ and its value is greater or equal to three $(j:\mathbb{N};\, j \geq 3)$. Therefore, the direct distance from camera to LED light states with considering as follows:

$$d_j^2 = (x - x_{Lj})^2 + (y - y_{Lj})^2 + (z - z_L^{'})^2 \tag{26}$$

For the value of $j$ (i.e., $j = 1,\ 2,\ 3$), Equation 26 can define as follows

$$\begin{bmatrix} 1 & -2x_{L1} & -2y_{L1} & -2z_L^{'} \\ 1 & -2x_{L2} & -2y_2 & -2z_L^{'} \\ 1 & -2x_{L3} & -2y_{L3} & -2z_L^{'} \end{bmatrix} \begin{bmatrix} x^2 + y^2 + z^2 \\ x \\ y \\ z \end{bmatrix} = \begin{bmatrix} d_1^2 - x_{L1}^2 - y_{L1}^2 - z_L^{'2} \\ d_2^2 - x_{L2}^2 - y_{L2}^2 - z_L^{'2} \\ d_3^2 - x_{L3}^2 - y_{L3}^2 - z_L^{'2} \end{bmatrix} \tag{27}$$

The simplified form of Equation 27 can be written as follows:

$$Z_0 x = q_0 \tag{28}$$

Trilateration problem solve on focusing on two different scenarios: distributed position of the LED lights as in Figure 13(a) and aligned LED lights in a row as revealed in Figure 13(b). The general solution of Equation 28 helps to identify the location of the smartphone where LED lights are located at the edge of a triangle. The solution states as follows

$$x = x_p + \varepsilon x_h \tag{29}$$

where $x_p$ is the particular solution and $\varepsilon$ is the real parameter. For homogeneous system, $Z_0 x = 0$, then $x_h$ is its solution.

The pseudoinverse matrix $Z_0$ helps to determine a solution for $x_p$. The value of $\varepsilon$ can be determined using the expression of $x_p = \left[x_{p0},\ x_{p1},\ x_{p2},\ x_{p3}\right]^T$, $x_h = \left[x_{h0},\ x_{h1},\ x_{h2},\ x_{h3}\right]^T$, and $x = \left[x_0,\ x_1,\ x_2,\ x_3\right]^T$.

There are two solutions of Equation 29,

$$x_1 = x_p + \varepsilon_1 x_h \tag{30}$$

$$x_2 = x_p + \varepsilon_2 x_h \tag{31}$$

The general solution for localizing smartphone as follows:

$$x = x_p + \varepsilon x_{h1} + \gamma x_{h2} \tag{32}$$

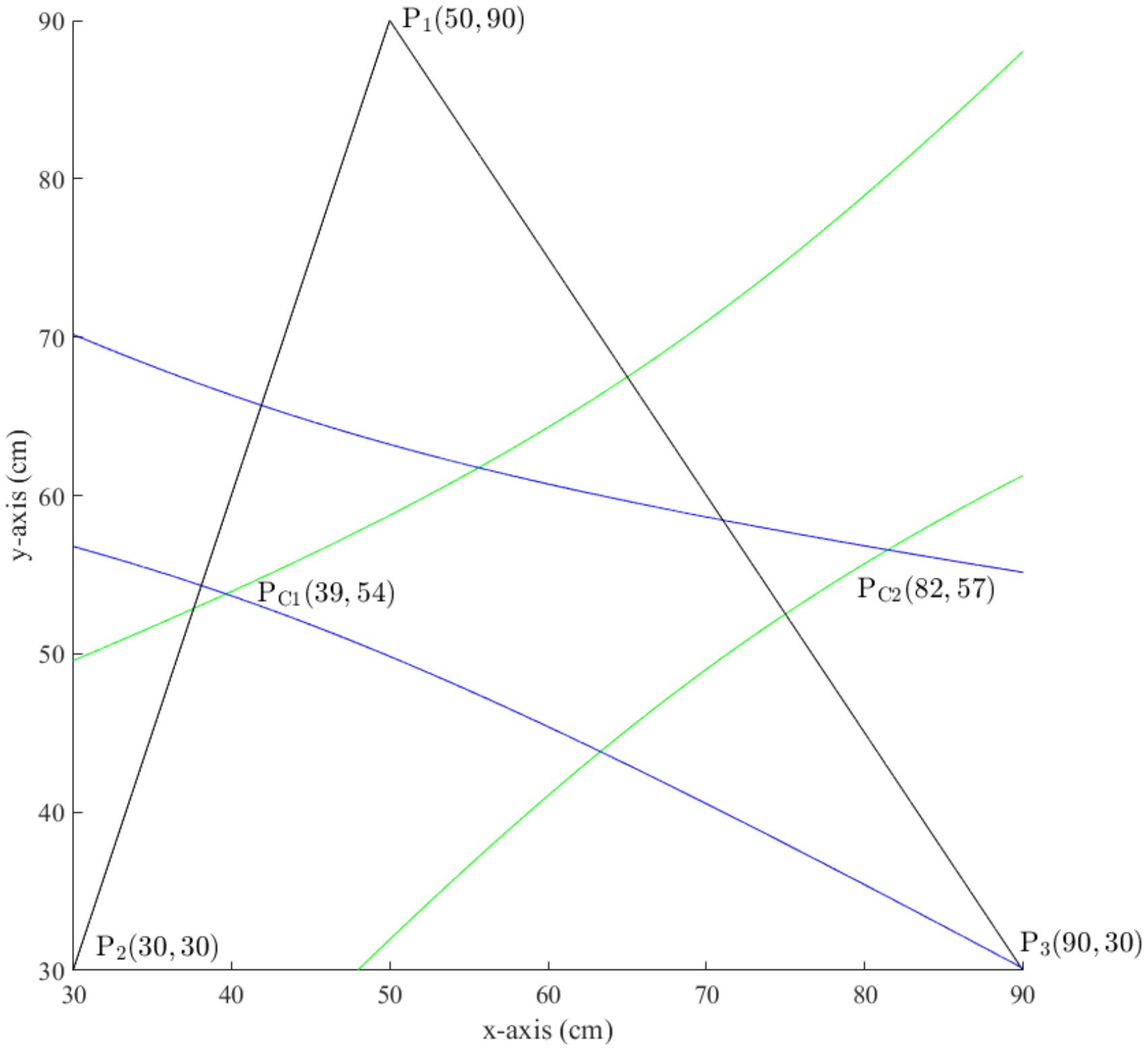

**Figure 14.** Position of camera within the three reference LEDs in 2D space.

where $x_{h1}$ and $x_{h2}$ are two solutions for the homogeneous system $Z_0 x = 0$. Here $\gamma$ the real parameter.

If there is more than three reference points, the relevant equation instead of Equation 28 can be expressed as follows:

$$Zx = q \tag{33}$$

Solution of Equation 33 can be found on the base of the least squares methods as follows:

$$\hat{x} = (Z^T Z)^{-1} Z^T q \tag{34}$$

As in Figure 14, $P_1$, $P_2$, and $P_3$ are the three positions of three LED lights within a 2D space. Trilateration or multilateration helps to find the intersecting points which indicate two possible positions of the camera. The lighting server chooses one of these two positions, especially where the direct distance between camera and LED light is minimized.

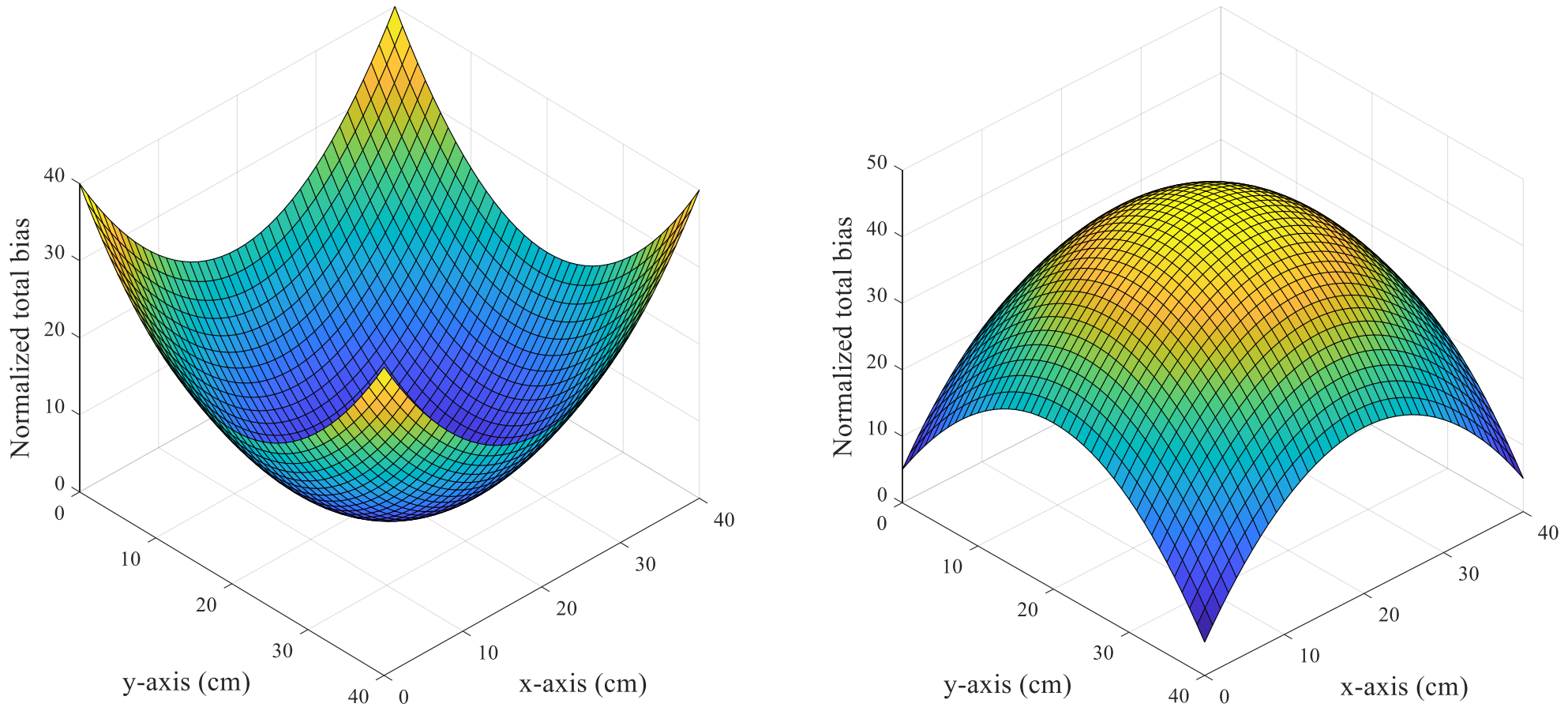

**Figure 15.** Systematic errors at (a) horizontal (b) vertical direction.

Figure 15 shows that systematic estimation error is generated when the lighting server estimates the position of the smartphone. The error is minimum for the horizontal bias (in Figure 15(a)) of the indoor environment and is much higher for the vertical bias (in Figure 15(b)). Therefore, system performance is much degraded when measuring vertical position.

The possible positions of the smartphone camera around the cluster of LED lights are revealed in Figure 16. Solid lines in the figure represent fixed distance between LED lights whereas the dotted lines represent the optical link between the LED light fixtures and the camera. Figure 16(b), (c), (d), (g) shows some similarities among them, e.g., the distance from camera to any two among three LED lights are equal. There are also other cases where there is no equality in the point of distance issue with others. Distances from LED light to camera are completely different as in Figure 16(e), (f), (h). However, all distances from the camera to the LED lights are equal in Figure 16(a).

Considering a scenario of the final stage of coordinate estimation for smartphone at the lighting server is shown in Figure 17. Three LED lights are simultaneously focused by the camera's FOV. In this case, each LED light keeps a constant 150 cm distance from other LED lights. Here, $P_1$ (200, 0), $P_2$ (200, 150), and $P_3$ (200, 300) are the coordinates of LED lights in the unit of centimeter whose are only focused by the camera at a certain time. Though all these LED lights are aligned in a line; therefore, the *x*-coordinates of all LED lights are equal.

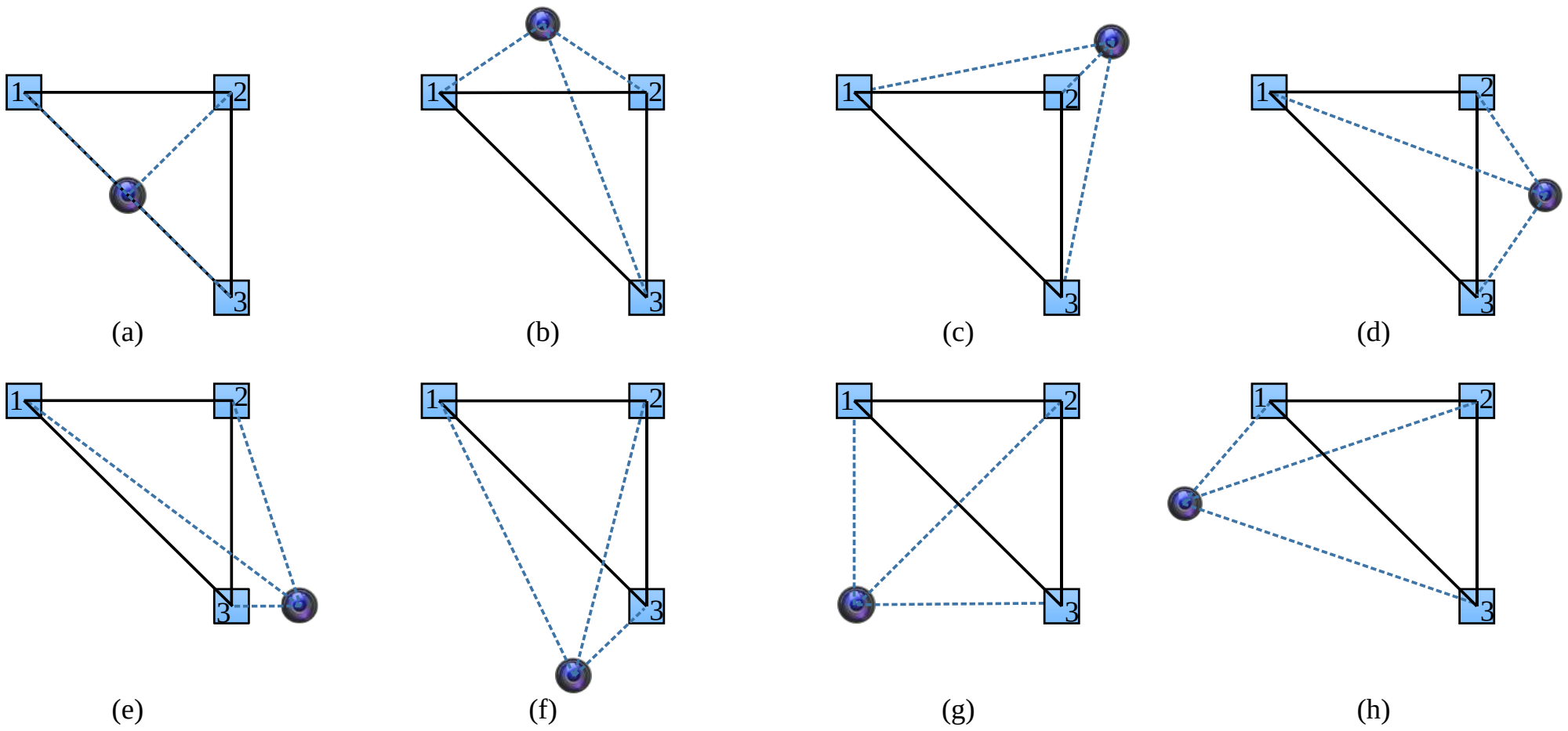

**Figure 16.** Smartphone's possible position around 3 LEDs.

Let consider that a smartphone is somewhere in-between the LED lights $P_1$ (200, 0) and $P_2$ (200, 150) and little far away from the LED light $P_3$ (200, 300) compare with other two. Using photogrammetry, measured direct distance from the camera to $P_1$ (200, 0), $P_2$ (200, 150), and $P_3$ (200, 300) are 320 cm, 336.05 cm, 410.37 cm, respectively. The relative distance from the camera to the smartphone helps the lighting server to estimate possible smartphone coordinates. Estimation shows smartphone is 40 cm away from the first LED light $P_1$ (200, 0) and is 110 cm away from $P_2$ (200, 150). Here, 40 cm is the smartphone's *y*-coordinate. The *z*-coordinate of the camera of the smartphone can be measured with the Pythagorean theorem, which will be 317.5 cm. Finally, estimated coordinate of the smartphone is $P_C$ (200, 40, 318).

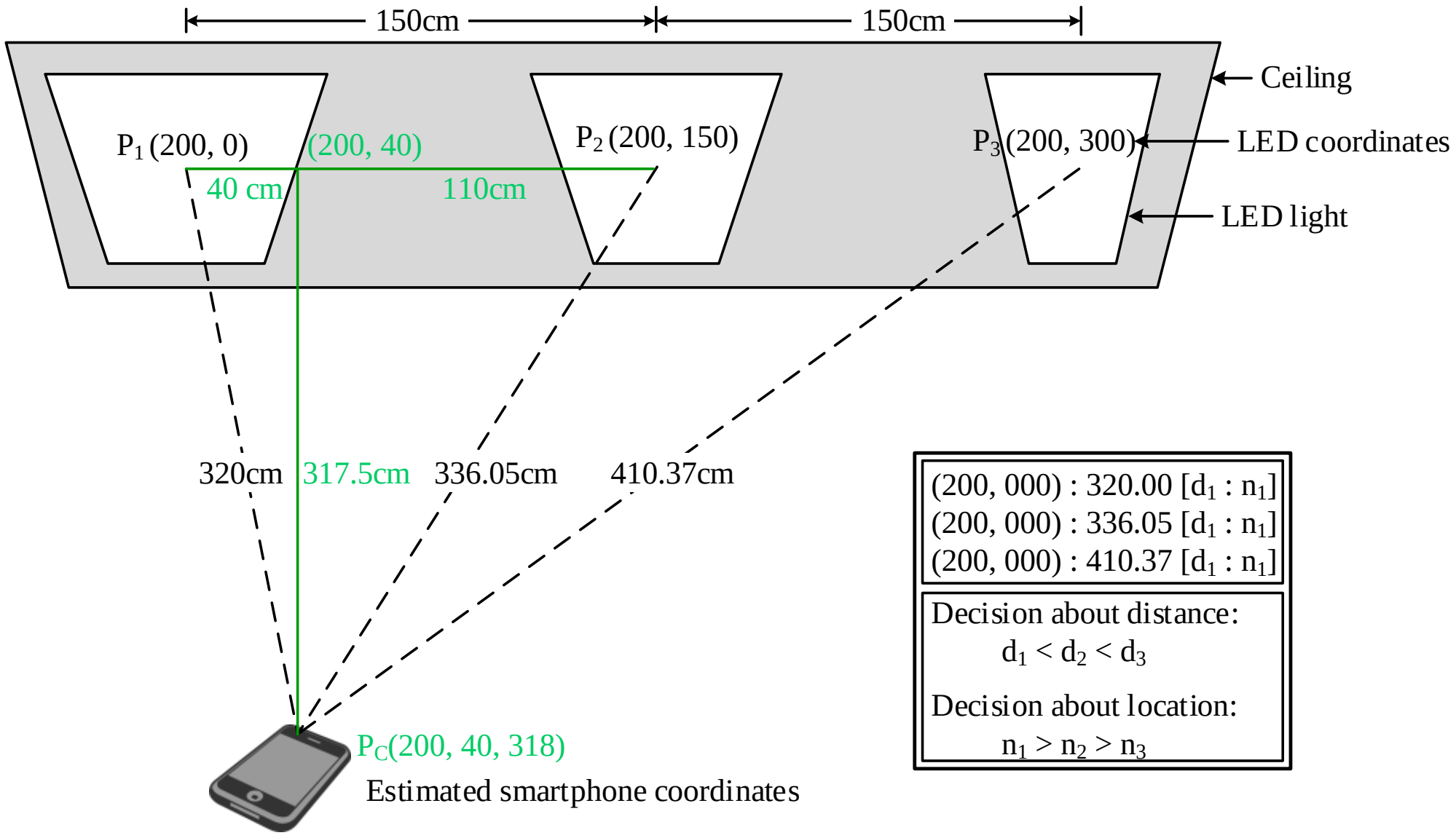

**Figure 17.** Estimate smartphone's position using distance information.

### 4.3.2 Smartphone possible position estimation using Kalman filter

After deciding smartphone coordinates, lighting server updates this information compare with previous data (if available), stores the information for future purpose, and more importantly sends it to the smartphone. Due to the dynamic nature of the smartphone, this server needs to run another algorithm in parallel, e.g., Kalman filter to estimate the next possible position along with the velocity, acceleration of the smartphone. This filter bank on the present input dimension instead of smartphone previous information (e.g., velocity, acceleration) [104].

The Kalman filter states as a linear filter and recursive estimator; most use case is error estimation in navigation claims under noise processes in a least squares sense by applying the minimum variance. There are three important blocks such as current approximation, Kalman filter gain, and new error in the approximation is shown in Figure 18. Previous approximation and the current measured value have an impact on the current approximation. Concurrently, Kalman filter gain puts importance on the error in the dimension and the error in the approximation. It also puts importance in-between previous approximation and present measured value. Moreover, new error in the approximation will pass on to future approximation error and feeds both current approximation and Kalman filter gain.

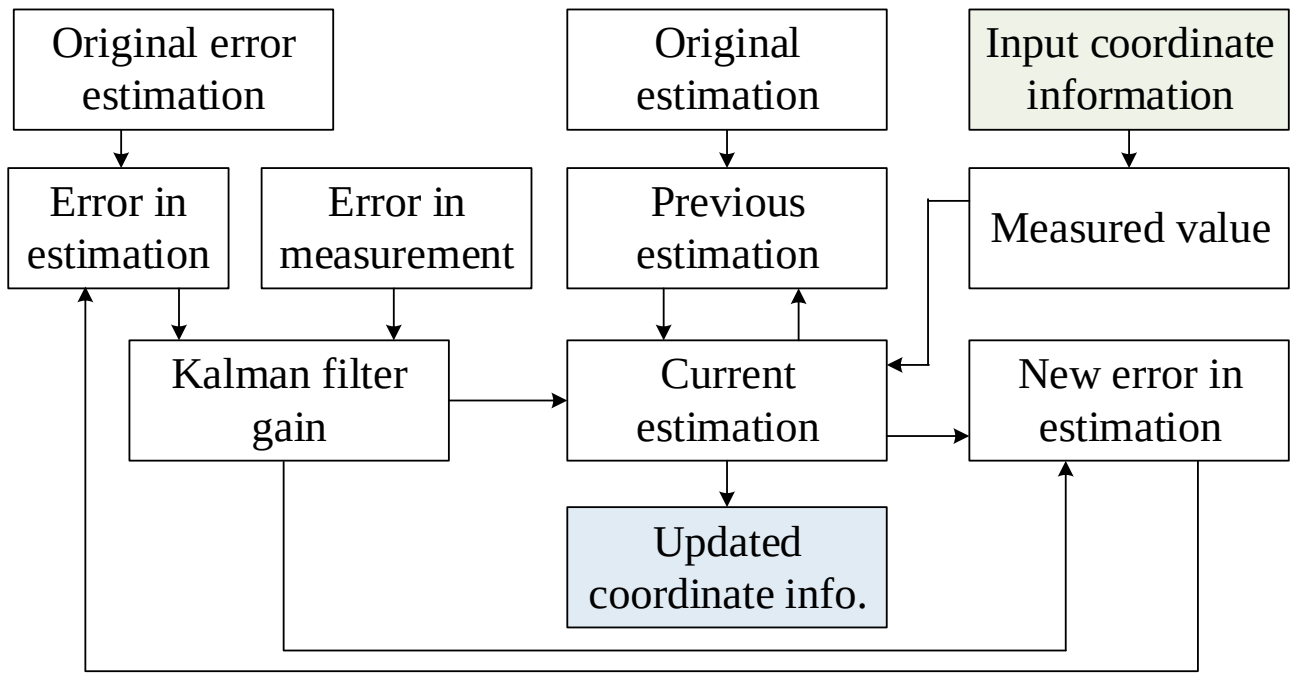

**Figure 18.** Estimating next possible position of smartphone using Kalman filter.

The smartphone preliminary projected location can define as follows:

$$X_{k_p} = JX_{k-1} + w_k \tag{35}$$

where $J$ is the adoption (or state) matrix, $X_{k-1}$ is smartphone's initial location, and $w_k$ is the additive noise from the initial location.

The measurements and state vectors are weighted by their respective processes' covariance matrices. The error in position approximation (or process covariance matrix) can be stated as follows:

$$P_{k_p} = JP_{k-1}J^T + Q_k \tag{36}$$

where $P_{k-1}$ is the initial process covariance matrix whereas $Q_k$ is additive noise.

With large variance, Kalman filter de-weights the measured value and low gain in comparison to the state estimate. This will lead the filter to rank the prediction instead of measurements. Furthermore, during small variance and high gain; the measured value is weighted more over the predicted value. The Kalman gain relies on the error in the approximation and error in the dimension. This gain, $K_g$ is explained as the ratio of error in estimate to overall error in both estimate and measurement as follows:

$$K_g = \frac{P_{k_p} T^T}{T P_{k_p} T^T + O_{error}} \tag{37}$$

where $T$ is a transformation matrix to convert a covariance matrix into Kalman filter gain matrix and $O_{error}$ is the error in measurement or observation.

The Kalman gain defines using the range of value between 1 to 0 and this gain always shows $0 \leq K_g \leq 1$. The value of $K_g$ being close to 1 implies that the error in the dimension is close to 0. This also concludes an unstable or larger estimation error and it is required to make the precise approximation again. Concurrently, the value of $K_g$ is close to 0 implies that there is an error in the dimension. The current estimation can be written as follows:

$$X_k = X_{k_p} + K_g \left[ Y_k - V X_{k_p} \right] \tag{38}$$

where $X_{k_p}$ is symbolized as previous approximation, $X_k$ is the present approximation, and $Y_k$ is the measured coordinates of the smartphone.

Correspondingly, the present error in the approximation will be minimized by the larger Kalman filter gain as follows:

$$P_k = [I - V K_g] P_{k_p} \tag{39}$$

## 4.4 Suspended signal propagation

It should be unnecessary to continue running the algorithm when the smartphone is not available. The lighting server stops to broadcast location information to the smartphone when it fails several times to get any signal from the smartphone.

# Chapter 5
# Vehicle Localization

## 5.1 Introduction

With the increase demand of autonomous vehicle, the concern for localizing vehicle also gets prioritized due. The proper localization scheme of the vehicle can ensure the prevention of an intended accident on the road. Vehicle localization can be either active or passive localization. Active features comprise setting region of interest (ROI) and computing the option for communicating with other vehicles and upholding safe distance from other vehicles to avoid unsolicited collisions by measuring temporal and spatial situations [105]. On the other hand, passive features comprise obtaining localization information from individual vehicles, which can be used by traffic control center and exploiting in an operative way to alleviate traffic congestion.

Forwarding vehicle (FV) is localized by the camera of the HV using the both OCC technology and photogrammetry. The taillight of the FV helps to transmits modulated data signal to the camera. FV broadcast this signal as FV-Identity (FV-ID) and data being extracted after demodulated and decoding it. Extracting data help to fix the ROI as well as measure the distance using photogrammetry. HV compares the position of the street light (SL) to measure its own position, because it is tough to measure the location of the FV when both HV and FV are moving. Later the determining coordinate of the FV turns as the foundation of the Cartesian coordinate structure to determine FV’s position. Figure 19 shows overall scenario of the proposed localization scheme [106], [107].

This chapter starts with the overview of vehicle localization scheme. The distance between HV and FV changes frequently due to their dynamic characteristic. Before describing localizing FV from HV, the process of calculating virtual coordinates of HV with respect to the SL describes, which solve coordinate calculation problem due to the dynamic characteristic.

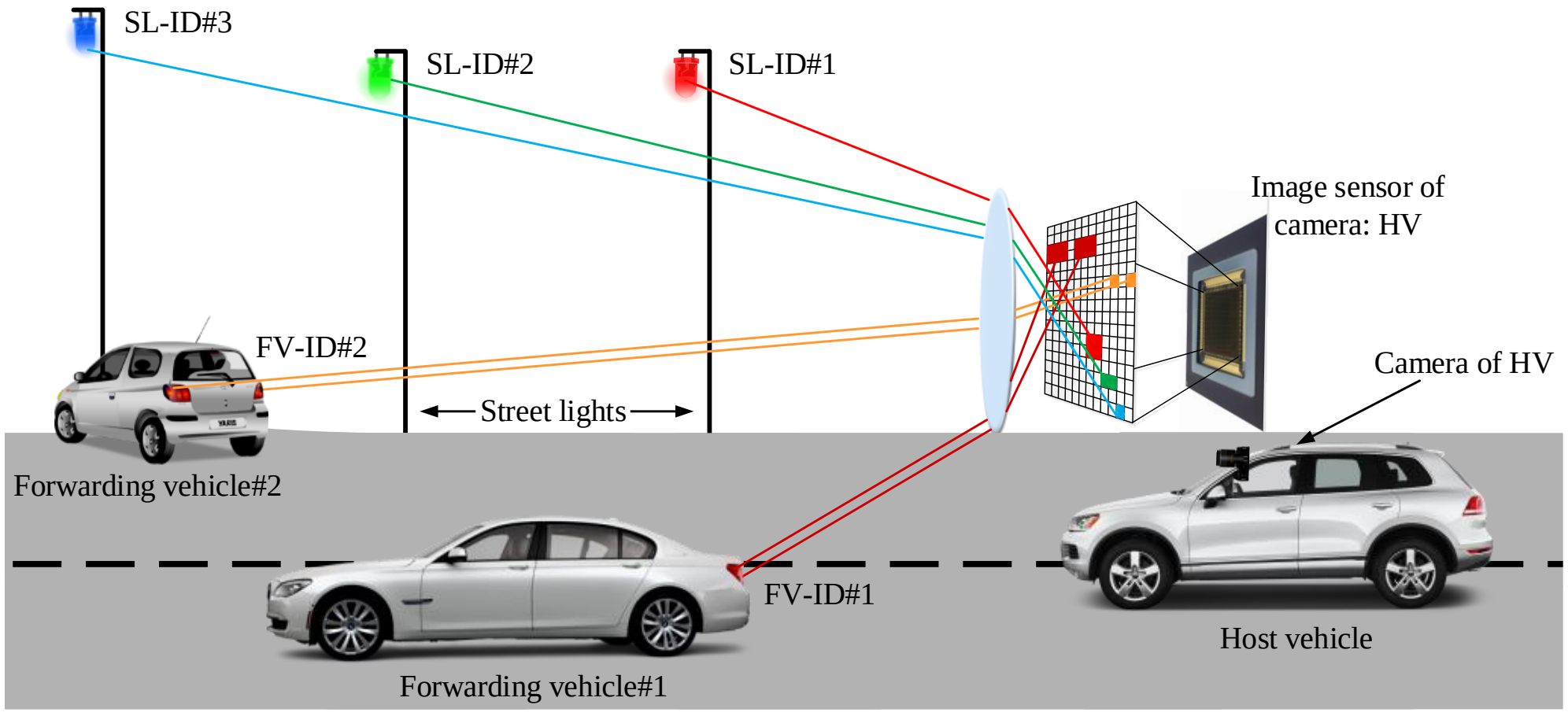

**Figure 19.** Integration of OCC and photogrammetry to localize vehicle.

## 5.2 Development of Vehicle Localization Scheme

In recent years' vehicle manufacturers have installed a camera (i.e., less than 30 frames/sec) inside the vehicles to support the drivers by providing a vision of their blind spot and monitor the outdoor circumstances for security issues. Applying OCC technology, a pair of tailing light of the FV turns as a transmitter and camera turns as receiver of HV [108], [109]. Using spatial-two-phase-shift-keying (S2-PSK) [110] modulation scheme, encrypted data transmits from the taillight as LED-ID, which is also known as FV-ID. The SLs transmit same way as the FV, where transmitting IDs denote as SL-Identity (SL-ID). Using OCC, simultaneously HV receives FV-IDs and SL-IDs from both FVs and SLs, respectively. Additionally, the distance between target FVs and SLs measures using photogrammetry. A single camera can perform both communication and distance measurement task simultaneously. In different phase, a pair of LED broadcast modulated light (i.e., S-PSK) beam to the camera of the HV. A flicker-free signal will generate from the transmitter with a constant clock rate (e.g., 125 or 200 Hz). The camera has a distinctive functionality called MIMO, which helps to characterize FV-IDs from SL-IDs. Unique ID for each and every transmitter ensure the ROI for vehicle localization. ROI specifies a certain region of an image for communication and supports to minimize the scope of false-position consequences. Additionally, in the point of distance measurement issue; same camera can use to measure the direct distance or side-to-side of the FV with respect to HV. The image of the projected FV changes on the IS according to the change of these vertical and horizontal positions shifting. Though, there is a relative motion among FVs and HV; therefore, measure the position of the FVs from the HV is quite impossible. However, from the known position information of the HV; it is quite easy to measure the position of FVs. Measuring the position of the HV is a challenging issue after measuring the position of the FVs. The position of HV can be measured from a fixed point on the road. SLs inhabit a fixed position on the road with respect to the vehicles.

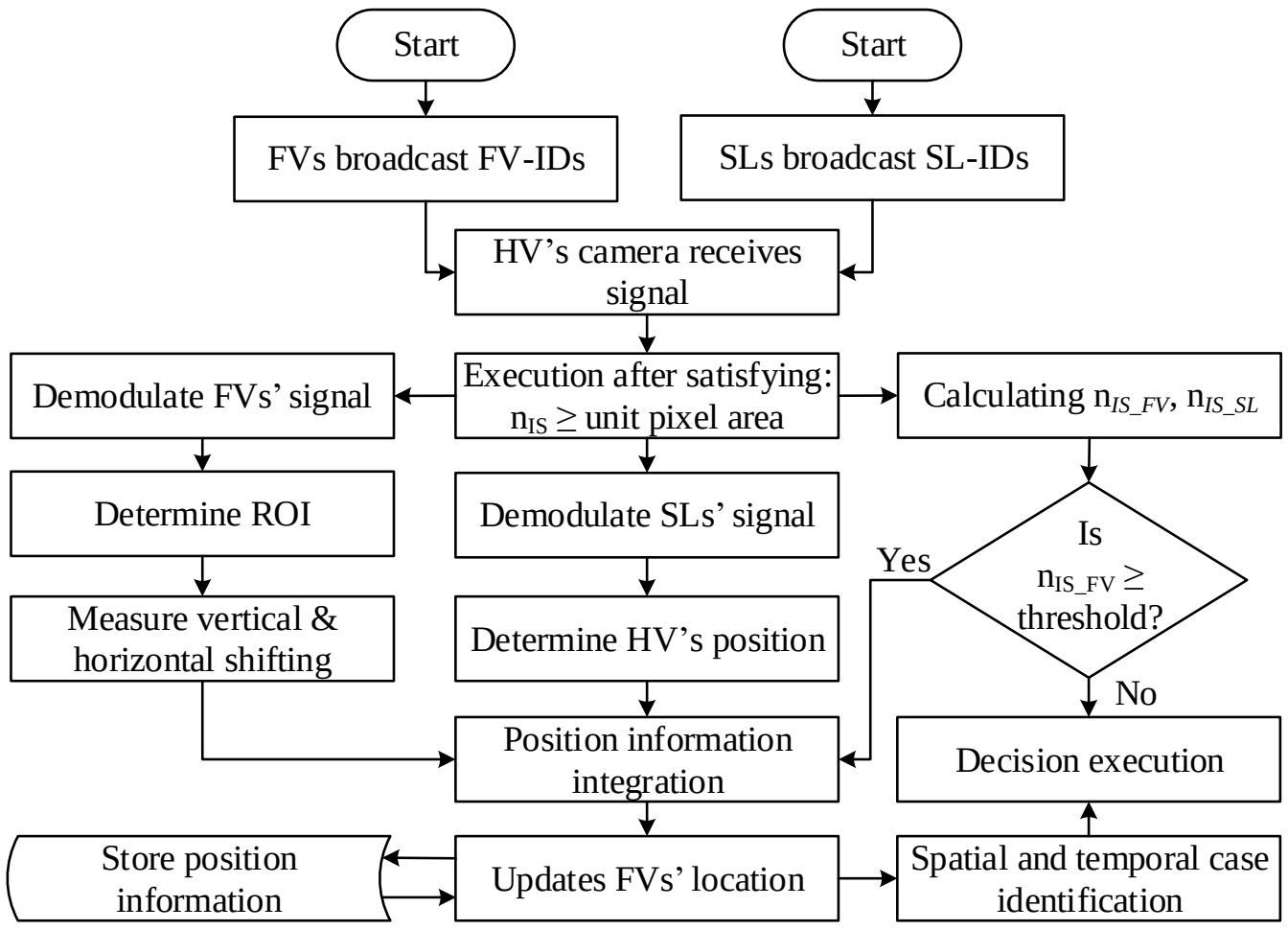

**Figure 20.** Vehicle localization with comparative location information.

Therefore, SLs can become the origin in the coordinate systems to measure the position of the HV. The image area of the IS for the detecting FVs and SLs is important. If this size is too small, then the data from the ID can be excluded as in Figure 20. Moreover, when the measure image areas for the projecting FVs are equal or greater to a threshold value, than HV will take a quick decision to avoid intended accident.

### 5.2.1 LED-ID from SL and FV

At the back of the FV, a pair of LED transmits modulated waveform (i.e., S2-PSK) to the camera of the HV as FV-ID [110] as in Figure 21. The SL follows the same way by separating a single LED array into two pairs of LED to transmit signal to the same camera. The transmitting signal phases from the pairs of LED array varies with the sequence of the input bits. Symmetric Manchester symbol helps to map each and every symbol. Transmitting from the pairs of LED in different phase for bit 1 and same phase for bit 0, using an approach of spatial under-sampling. The bit interval $s_1(t)$ for one of the pairs of LED is as follows:

$$s_1(t) = \sum_{k=0}^{\mathbb{N}} s_1(t_k + kT) \ \text{ where } \begin{cases} 0 \le t < T_{bit} \\ s_1(t_k) = \begin{cases} 1, & 0 \le t_k < T/2 \\ 0, & T/2 \le t_k < 0 \end{cases} \end{cases} \quad (40)$$

where for $\mathbb{N}$ bit-interval phases $k$ is an unsigned integer, $T$ is the signal cyclical interlude and $T_{bit}$ is a bit interval.

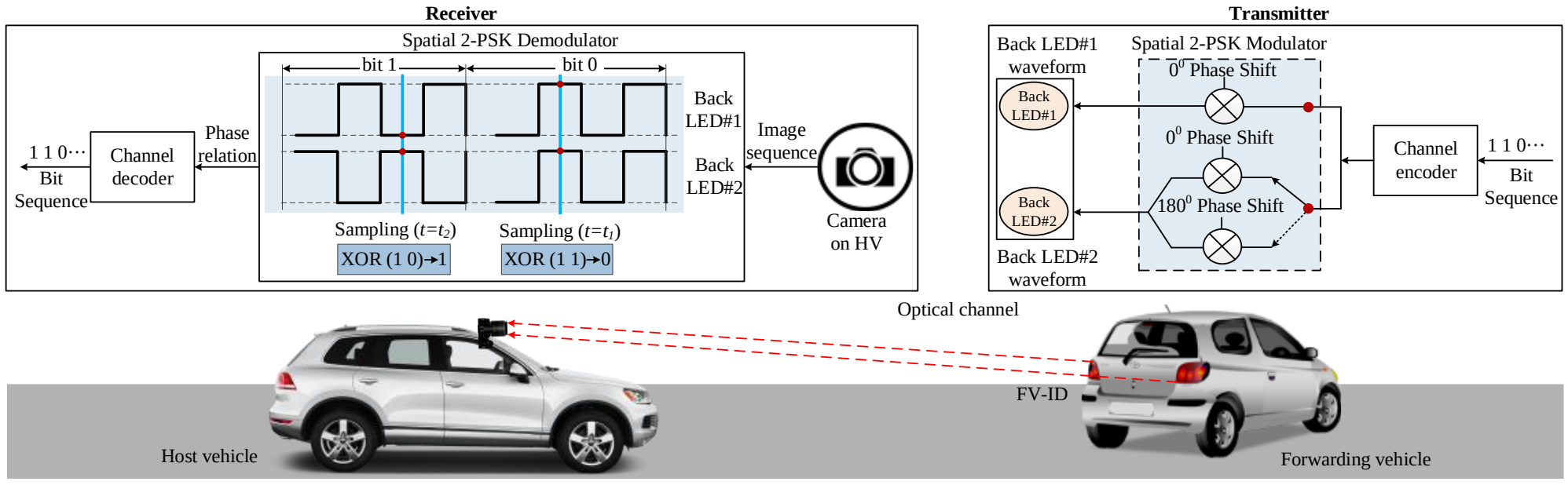

**Figure 21.** Encoding and decoding technique for getting LED-ID using OCC.

For another taillight, bit interval $s_2(t)$ states as follows:

$$s_2(t) = \sum_{k=0}^{N} s_2(t_k + kT) \ \text{ where } \begin{cases} 0 \le t < T_{bit} \\ s_2(t_k) = \begin{cases} 1, & \overline{s_1}(t_k) \\ 0, & s_1(t_k) \end{cases} \end{cases} \quad (41)$$

After demodulation and decoding, a bit reconstructs from different states of the pair of LED. At $t_s$ sampling time, 0 bits reconstruct from the same phase of the pairs of LED and vise-versa for reconstructing bit 1. In the same image, the value of bit captured obtains of XOR operations as follows:

$$bit = s_1(t_s) \oplus s_2(t_s) \quad (42)$$

where at sampling time $t_s$; $s_1(t_s)$ and $s_2(t_s)$ are states of the pairs LED.

The BER obtains within an image in the modulation scheme like S2-PSK is comparatively lower than the available modulation pattern such as undersampled phase shift on-off keying [20]. The residual BER is removed with this modulation scheme by a nonlinear XOR classifier [21] as follows:

$$P_{e,S2-PSK} = 2\delta p_e(1-\delta p_e) \tag{43}$$

where $p_e$ is bit error probability of the LED state and $\delta$ is enhancement of error rate.

## 5.3 Localizing the Forwarding Vehicle

### 5.3.1 Calculating Virtual Coordinates of HV position

In polar and Cartesian coordinate systems, position of each and every object measures on the basis of an origin within either 2D or 3D space of the coordinate system. Localizing the FV from the HV is nearly impossible when both are in a relative motion and FV changes its position frequently over the period of localization. Therefore, there is no origin to measure the position of the FV. The only solution can be applied when the position of the HV is known to determine the positions of the FVs all over the period.

The positions of SLs are always fixed compare with other moving transports on the road. The position of the HV measures by comparing with the position of SLs. Comparative location information of HV recognize as the virtual coordinate which acts as the origin to determine the position of FVs. The on-board diagnostic II (OBD II) system merges with the virtual coordinate of HV to meet the accuracy in localization measurement. Herein, the IDs, which transmits from the SLs should be unique from the IDs transmits from the FVs. Moreover, one ID from a single SL should be distinguished from other SL-IDs. Additionally, several other information related to the SL such as on the same side of the road the height of a single SL, distance between SLs includes with the SL-IDs. Adjacent SL-IDs helps to catch the understanding the right or wrong side of the road.

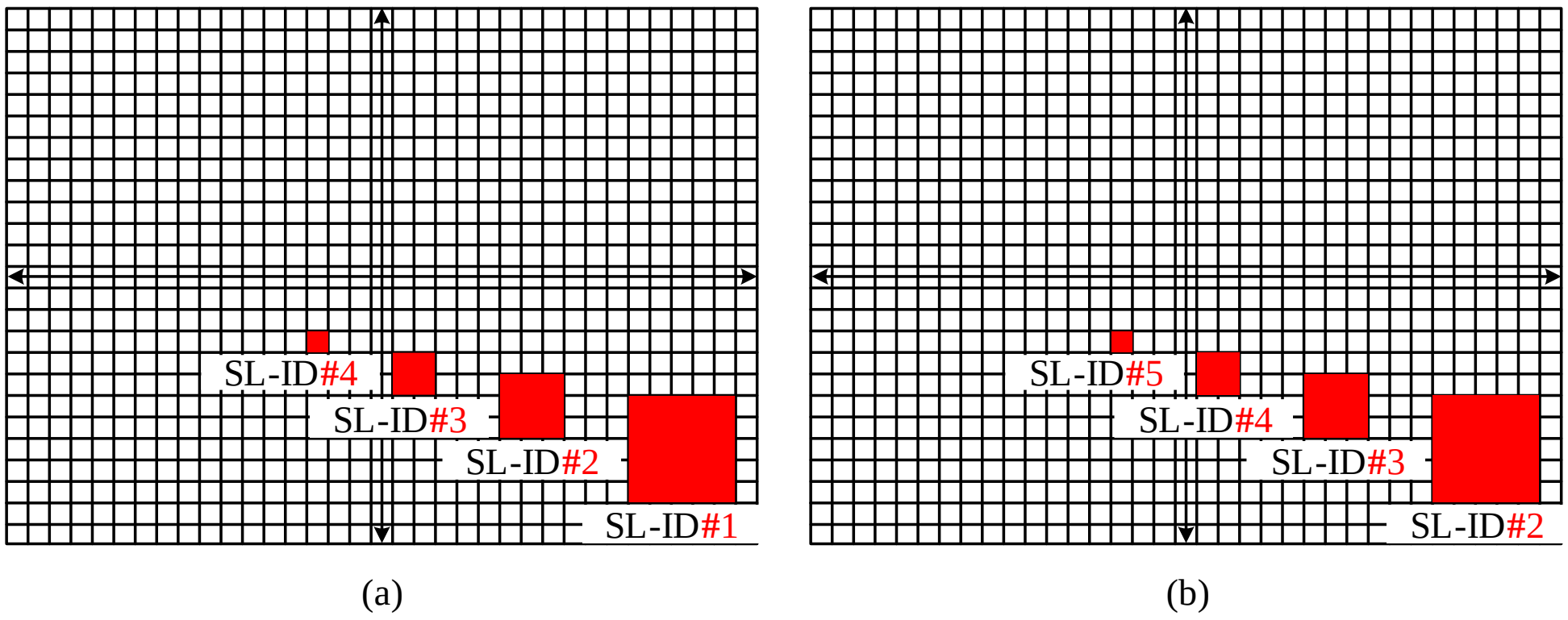

**Figure 22.** Observing change of SL-IDs within time lapse from (a) t to (b) t+1.

After picking the ROI, photogrammetry helps to calculate distance from camera to SLs' LEDs. In Figure 22, the receive SL-IDs change with the change of the position of the HV as well as camera FOV. At time t, receive IDs from the SLs observe between SL-ID#1 and SL-ID#4 as in Figure 22(a). Furthermore, at times (t+1) these receive IDs varies between SL-ID#2 ~ SL-ID#5 as in Figure 22(b). The projectile image area size subjugates large area of the IS for the nearest SL compare with the other SLs.

The camera inside the HV can decode the data from the SL-IDs using OCC. Figure 23(a) demonstrations, constant distance $d_n$ between two SLs and SL's height $(SL_h_n)$ where $n^{th} (n = 1, 2, ..., \mathbb{N})$ associated to the SLs' number. The direct distance $D_{SL_j - HV}$ from LED of the SLs to the camera in the HV measures using photogrammetry where $j^{th} (j = 1, 2, ..., \mathbb{N})$ shapes the iteration number sequence over a period. The flat direct space from LED of SL to camera is $a_j$. Applying the Pythagorean theorem to calculate this flat distance on a right triangle. Here, the height of the SL and $D_{SL_j - HV}$ are the remaining two edges of that right triangle.

After decoding this distance, a few triangles can be formed from the Figure 23(a). Applying the Pythagorean theorem again on the triangles $CSL_1H_t$ and $CSL_1SL_2$ at a specific time; it can be observed as follows:

$$a_1^2 = c^2 + h^2 \quad (44)$$

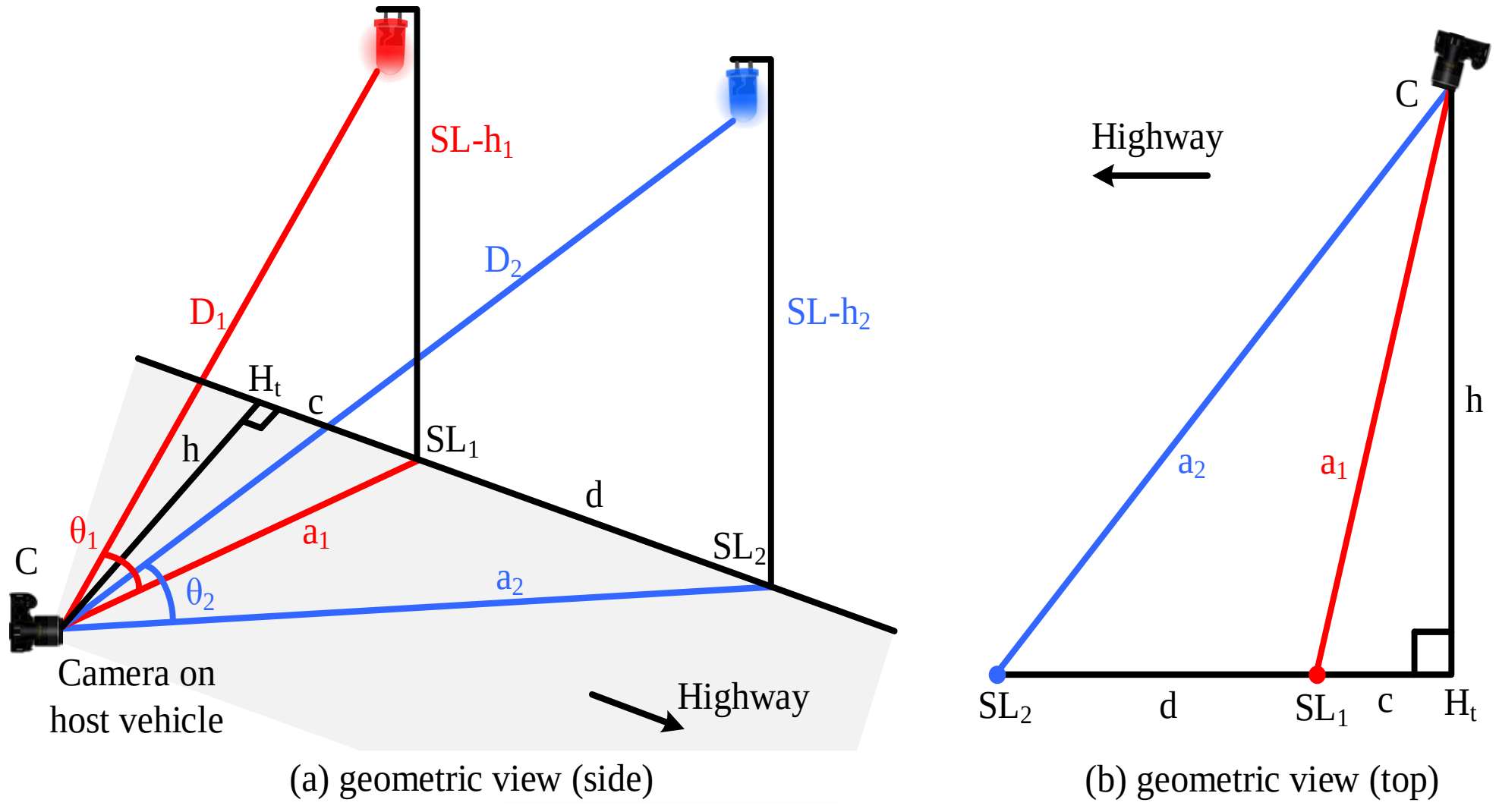

**Figure 23.** Generating virtual coordinates for Host Vehicle.

$$a_2^2 = (d_{SL-SL} + c)^2 + h^2 \tag{45}$$

where $h$ is the flat distance between the pavement and camera, $c$ implies the space between the shortest distance from the SL to the cross section and a cross section of the flat line, $a_1$ and $a_2$ are the flat distance for $SL_1$ and $SL_2$, respectively. In Figure 23(b), the condition $c < d_{SL-SL}$ will always satisfy. Equation (44) and (45) can write as follows:

$$c = \frac{(a_2^2 - a_1^2) - d_{SL-SL}^2}{2d_{SL-SL}} \tag{46}$$

Linking Equations (44) and (46) can help to determine the flat distance between the pavement and the camera as follows:

$$h = \sqrt{a_1^2 - \left\{\frac{(a_2^2 - a_1^2) - d_{SL-SL}^2}{2d_{SL-SL}}\right\}^2} \tag{47}$$

In all cases, the flat position $h_j$ is the positive quantity that has an influence on the position of the HV whereas the angular position of the SL $\theta_{SL_j-HV}$, image area of SL $n_{IS_SL}$, HV's velocity $V_{HV}$ have individual impact on the positioning of the HV. The factors associated with the HV's position will change with the movement of HV. Total position shifts of HV between the time lapse of $t$ and $\Delta t$ is stated as follows:

$$P_{HV}(t+\Delta t):\left\{h_j \pm \Delta h;\ \Delta\theta_{SL_j-HV};\ n_{IS_SL} \geq \rho^2;\ V_{HV}\left(\Delta[c_j + d_{SL-SL_n}]/\Delta t\right)\right\} \quad (48)$$

The flat distance from the pavement to the HV depending on the varying quantity of the flat direct distance $a_j$, the distance from one SL to another, i.e., $d_{SL-SL_n}$, and distance between the cross point of the flat line and the shortest distance from cross point to SL, i.e., $c_j$, where all these values change with change of the angular position $\theta_{SL_j-HV}$.

$$h(a_j,\ c_j,\ d_{SL-SL_n}):\left\{\Delta\theta_{SL_j-HV}\right\} \quad (49)$$

The direct distance from the SL to HV depends on the angular position $\theta_{SL_j-HV}$ and the area of the SL's LED $n_{IS_SL}$ on the IS. For those SLs whose are far away from the HV and projected image area $n_{IS_SL}$ is less than the unit pixel area, i.e., $\rho^2$ on the IS; that value can be discarded from the overall calculation. Additionally, at the edge of the road terns or with the bending of the road, the angular position $\theta_{SL_j-HV}$ as well as $D_{SL_j-HV}$ changes. Therefore, the direct distance from LED of the SLs to the HV states as follows:

$$D_{SL_j-HV}(n_{IS_SL},\ \Delta\theta_{SL_j-HV}):\left\{n_{IS_SL} \geq \rho^2;\ \Delta\theta_{SL_j-HV}\right\} \quad (50)$$

The changes of angular position, measure from the comparison successive angular position within the time interval between $t$ to $\Delta t$ as follows:

$$\Delta\theta_{SL-HV}(t+\Delta t):\left\{\theta_{SL_1-HV} \sim \theta_{SL_2-HV}\right\} \quad (51)$$

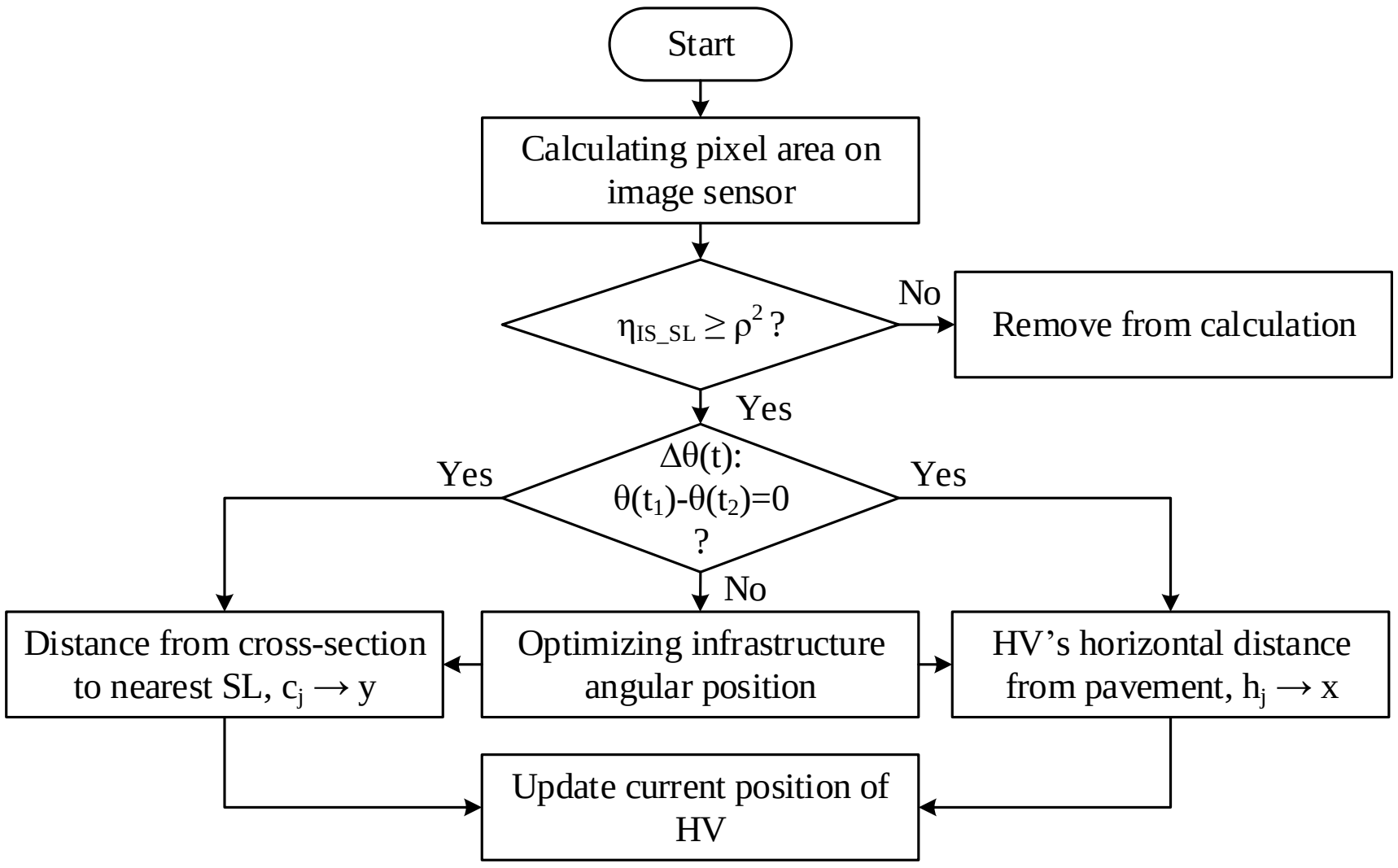

**Figure 24.** Optimizing virtual position with angular position.

The x-coordinate and y-coordinate of HV with respect to the nearest SL are analogous to the flat distance $h_j$ and $c_j$, respectively. Coordinate information of HV updates with the change of its own position. Figure 24 presents the algorithm which uses to calculate and correct process of flat position information of HV. From Equation 51, the condition of the road, i.e., curvature or edge can be determined by comparing two successive angular position, i.e., $\theta_{SL_1-HV}$ and $\theta_{SL_2-HV}$. The position of the SL should very with the condition of the road conditions. Therefore, it is required to optimize the impact of road and infrastructures at road-side. For smooth and flat road scenarios, this impact can be omitted from the calculation.

### 5.3.2 Determine the position of FVs from the HV

Each FV transmits modulated signal from a pair of taillight LED to the receiver, e.g., camera of the following vehicle (i.e., HV) using OCC technology. The transmitted signal, which is denoted as FV-ID; conveys some basic emergency information about traffic conditions along with the other e.g., the physical area of the taillight. Each FV-ID should be unique compare with the other FV-IDs.

Before localizing FVs, it is mandatory to ensure a perfect communication link between FV and HV within the optical channel. For the S2-PSK modulation scheme, camera of HV should clearly visualize both taillights of FV. There are various worst scenarios will arise, such as one of the taillight of one FV blocked by the other FV during communication link set-up. Due to the interruption of data reception, cause signal loss or increase BER. Moreover, one of the important factor is that when signal from one vehicle (e.g., first vehicle) interrupts by the other vehicle (e.g., second vehicle) means the first vehicle is far from the HV. Therefore, it is not required to concentrate on the first vehicle rather than focus on the second vehicle. In our algorithm, we set the priorities of decision execution on the basis of the projected taillight on the IS. Therefore, the HV may take the initiative when the first vehicle will appear at the site of the HV. Concurrently, FV-IDs from different FVs may provide solution to distinguish different FVs when they are moving parallel with each other. The introductory condition for efficacious communication and localization is LED-ID based vehicle identification, which brings the benefits to confirm the ROI. Every SL broadcasts its SL-ID along with the FV as FV-ID as Figure 25. In Figure 25(a), camera of the HV makes the background of the captures image as black by controlling the exposure time to identify the signal from both FVs and SLs. All IDs accumulate from the received image after demodulation and decoding processing using OCC technology as Figure 25(b).

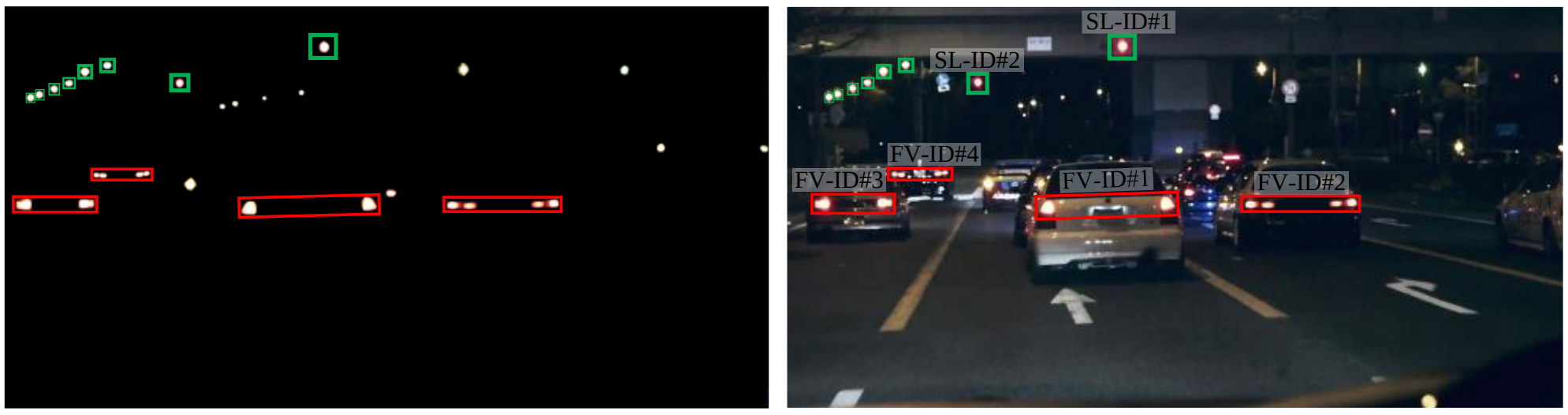

(a) (b)

**Figure 25.** Using OCC, (a) selecting ROI and (b) receiving IDs from FVs and SLs.

With the relative distance variation between the FV and HV; the projected taillight area of the FV varies on the IS. There are two positions shifts can cause projected image area variation on the IS such as vertical shifting and horizontal shifting. When the FV moves side-by-side cause the horizontal shifting whereas vertical shifting will cause by the change of direct distance between FV and HV. Figure 26 represents a

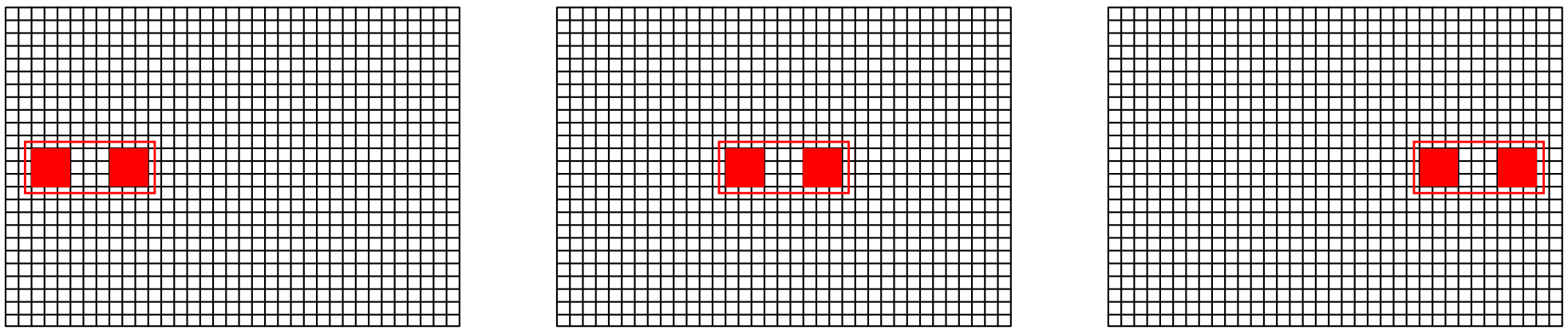

(i) FV at the right side of the HV (ii) FV stays straight of the HV (iii) FV at the left side of the HV

**Figure 26.** Visualizing horizontal shifting of FV.

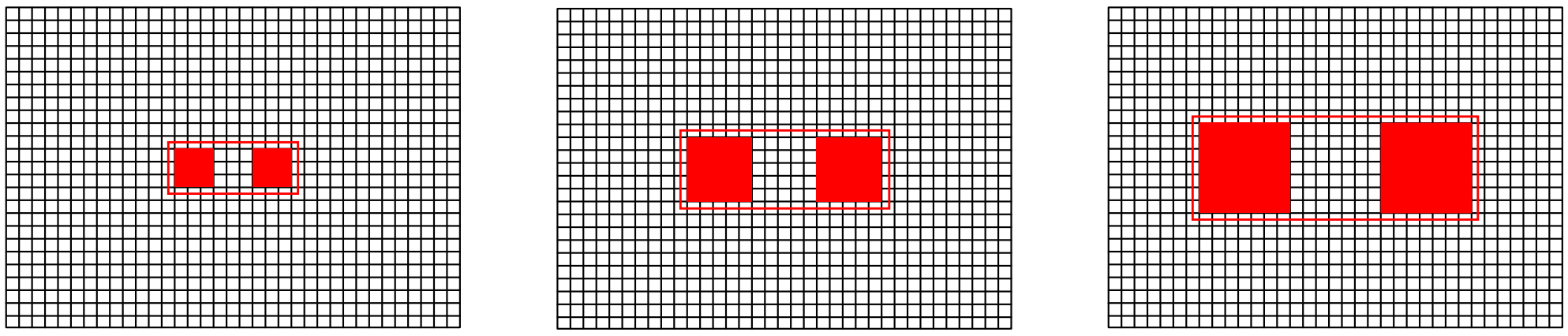

(i) Safe distance from FV (ii) Min. distance from FV (iii) Critical distance from FV

**Figure 27.** Visualizing vertical shifting of FV.

scenario of FV's movement where two red squares resemble the a pair of taillight from a FV on the surface of the IS. In Figure 26(a), the source of the projected pairs of taillight locates at the right side of the HV though light is always changing direction after being refracted by the lens of the camera. Sequentially, Figure 26(b) and (c) shows that FV moves from the right side to the middle and later moves to the left with respect to the HV. Furthermore, Figure 27 brings the example for vertical shifting of the FV from the HV. The projected image in the Figure 27(a) is much smaller than the Figure 27(b) and 26(c) simplify that the direct distance between FV and HV decrease gradually.

Commonly, an IS builds with 2D photodetector pixel array and transistors, horizontal and vertical access circuitry and readout circuitry. This horizontal and vertical access circuitry along with readout circuitry reads signal value from each and every pixel. The measurement of angular position helps to mitigate the error in the image area size measurement based localization during the dense traffic scenarios where the FVs move at a high speed. In Figure 28, a center plane, considers at middle section of the IS, which separate the IS into two vertical parts. Angular displacement of FV with respect to the HV $\theta_{FV_k}$ can be measured by calculating flat displacement of the projected image of FV on the IS $Hd_{IS_FV_k}$ as in Figure 28. Here, the number of receiving FV-ID on the camera of the HV is $k^{th}(k=1,2,...,\mathbb{N})$. Multiple colors for different projected image area on the IS states the distinguish among different FV-IDs from various FVs. If the image projected at the center of the IS, both angular displacement $\theta_{FV_k}$ and flat displacement $Hd_{IS_FV_k}$ become zero. There are several advantages of calculating angular displacement instead of helping localization calculation such as extracting lane change information, error alleviation from position approximation for right-left side of the road, depth approximation. The flat displacement $Hd_{IS_FV_k}$ for corresponding FV is a function of the image area of the FV's taillight $n_{IS_FV}$ and its angular displacement $\theta_{FV_k}$ as follows

$$Hd_{IS_FV_k}(n_{IS_FV} \geq \rho^2):\left\{\theta_{FV_k}\right\} \tag{52}$$

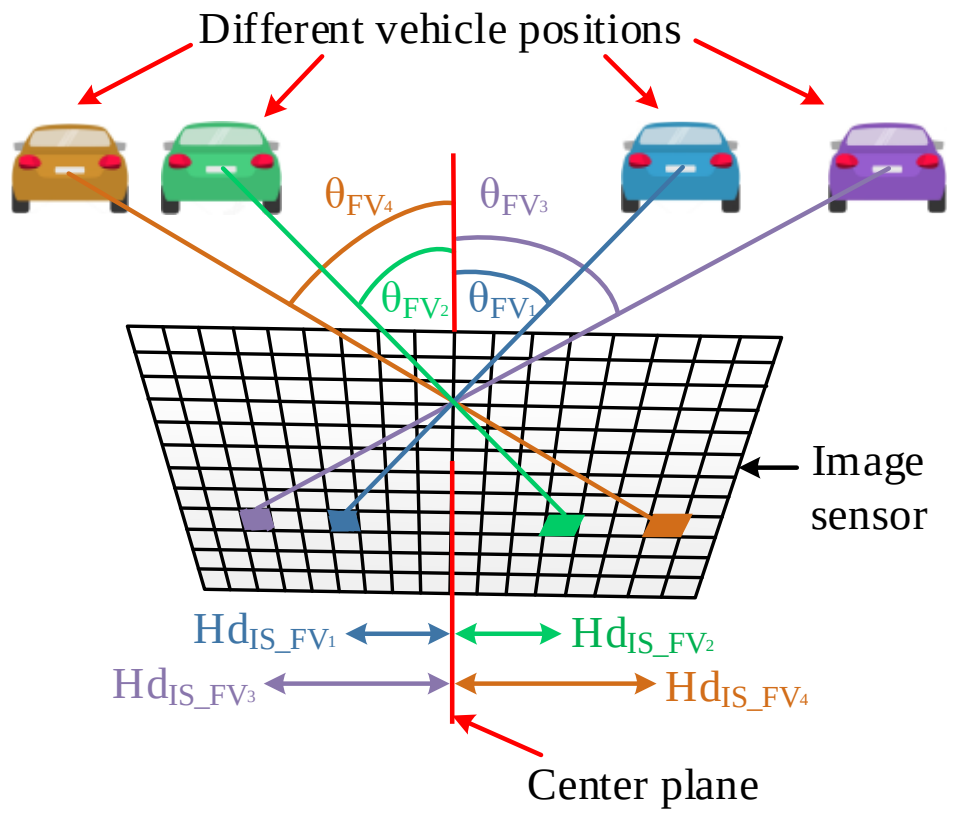

**Figure 28.** Measuring angular position from flat displacement on IS.

Finally, to calculate position of FV taking time difference $\Delta t$ from initial time $t$ and these vary with the variation of the virtual position $P_{HV}$ of the HV, projected image area of the taillight of the FV (i.e., $n_{IS_FV}$), flat displacement for corresponding FV (i.e., $Hd_{IS_FV_k}$), and difference of speed of FV (i.e., $\Delta V_{FV}$) as follows:

$$P_{FV}(t+\Delta t):\left\{P_{HV}(t+\Delta t);\ n_{IS_FV} \geq \rho^2;\ Hd_{IS_FV_k};\ \Delta V_{FV}\right\} \tag{53}$$

# Chapter 6
# Result and Discussion

The considering parameter for both indoor and vehicle localization are enlisted in table 2. The simulation will vary with the variation of the parameter set for the localization.

## 6.1 Simulation Results for Indoor Localization

The simulation result for the indoor localization consider a specific indoor environment of 1600 sqft.

**Table 2.** Localization parameter for transmitter and receiver.

| **Localization parameter** | **Performance value** |
|---|---|
| Parameters for light fixtures (transmitter) | |
| LED panel size | 10x10 $cm^2$ |
| Modulation scheme | OOK (indoor), S2-PSK (outdoor) |
| Encoding technique | Manchester coding |
| Data rate | 20 kbps |
| Parameters for camera (receiver) | |
| Detection range | 30-150 m |
| FOV | $90^{\circ}$-$120^{\circ}$ |
| Resolution of image | 2-10 MP |
| Physical size of sensor | 36 × 24 $mm^2$ |
| Frame rate | 30 fps |
| Focal length | 16-25 mm |
| Pixel size | 2.5-4 μm |
| Lens aperture | 4 |
| Exposure time | 1/2000 to 1/15 sec |
| Height of street light | 7 m |
| Inter-distance between street light | 25 m |
| Lane width | 10 m |
| Vehicle speed | 0-80 km/h |

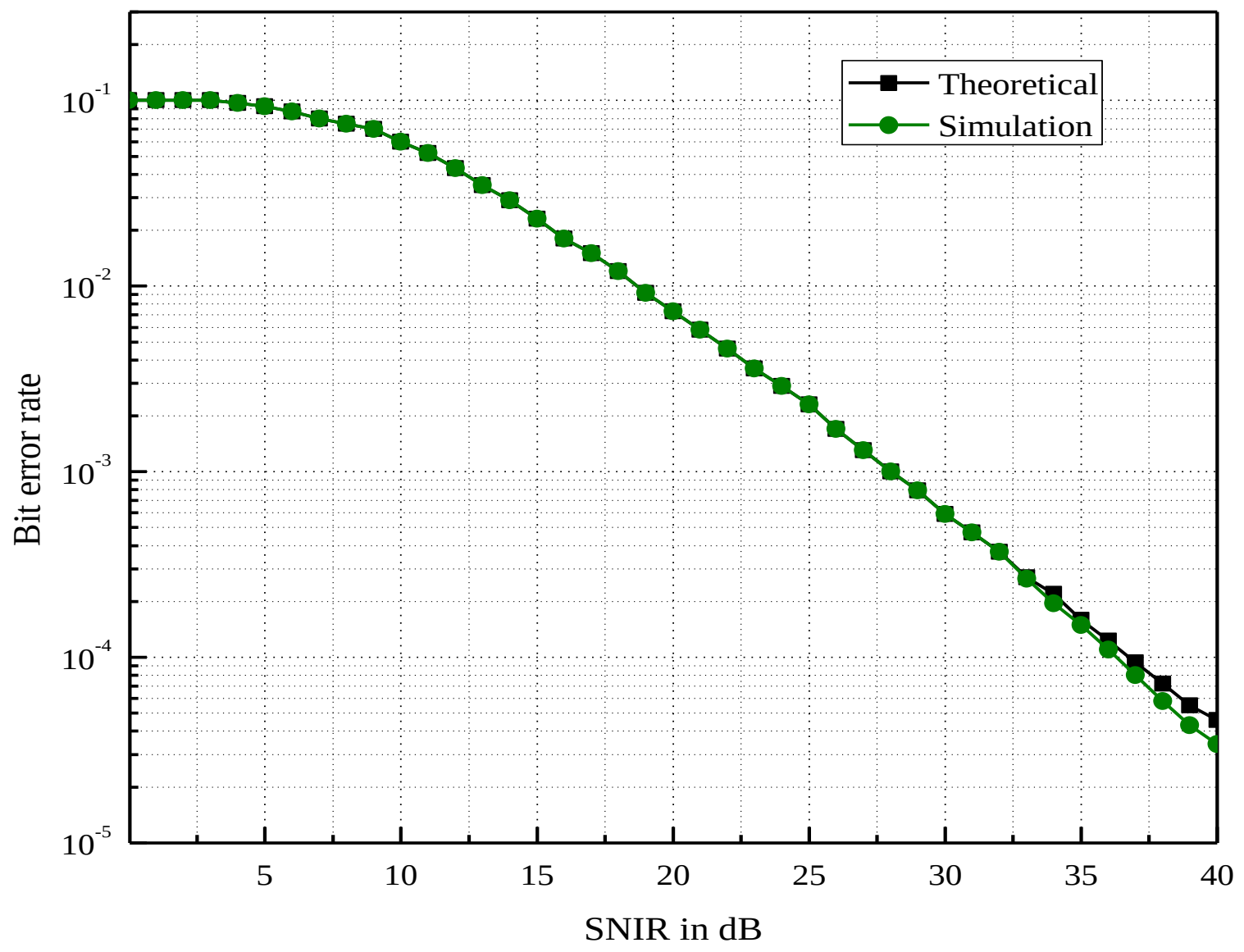

**Figure 29.** SNIR vs BER for indoor scenarios.

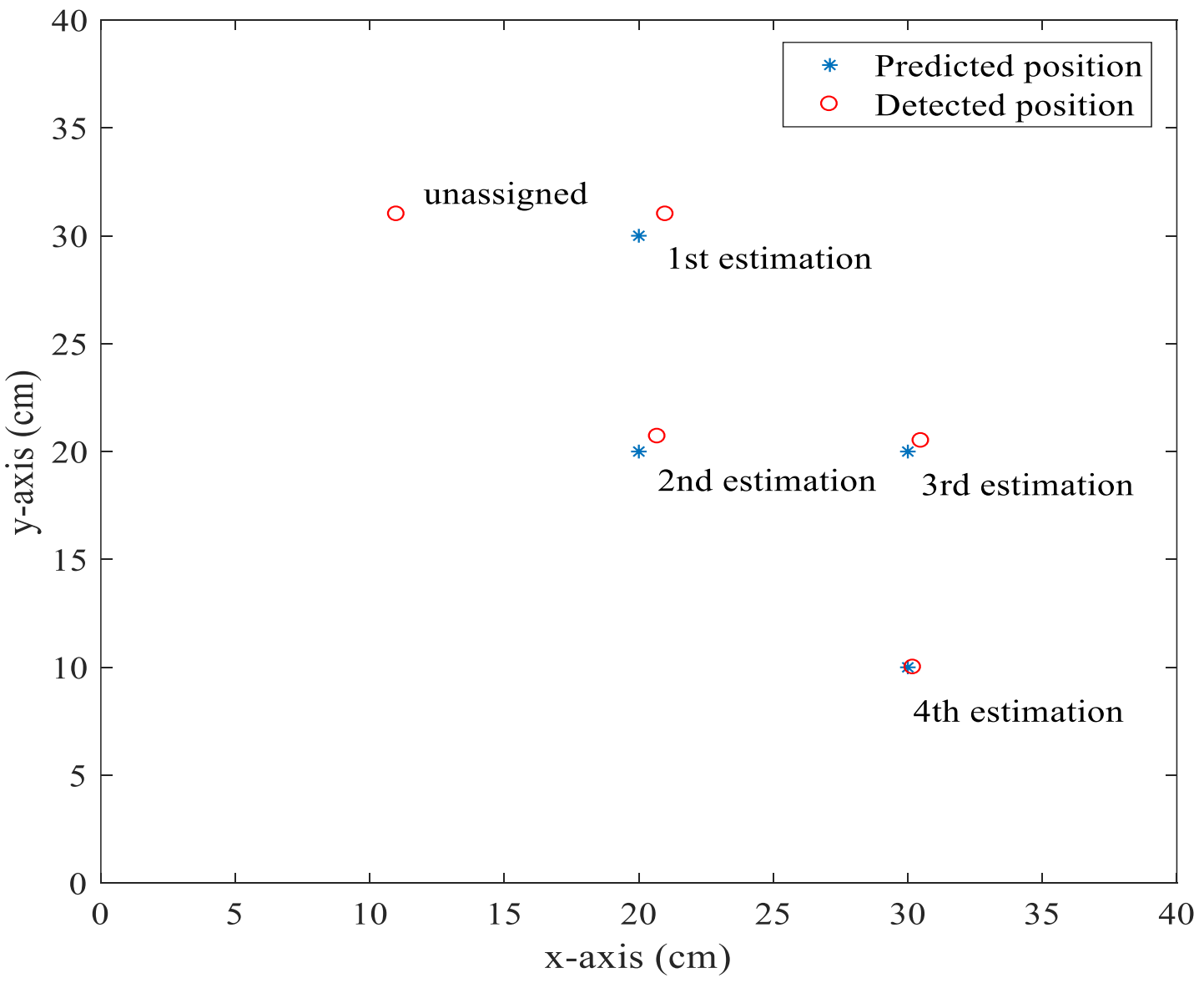

**Figure 30.** Performance evaluation of estimations with detection smartphone's position.

A comparison data are plotted in Figure 29 based on the theoretical result and simulation result for visualizing performance between BER and SNIR. The simulation result is based on a perfect environment where noise from ambient light is ignored. Therefore, both curves are well fitted and there is limited variation in between them. It states that with the increase of SNIR, the BER decrease accordingly.

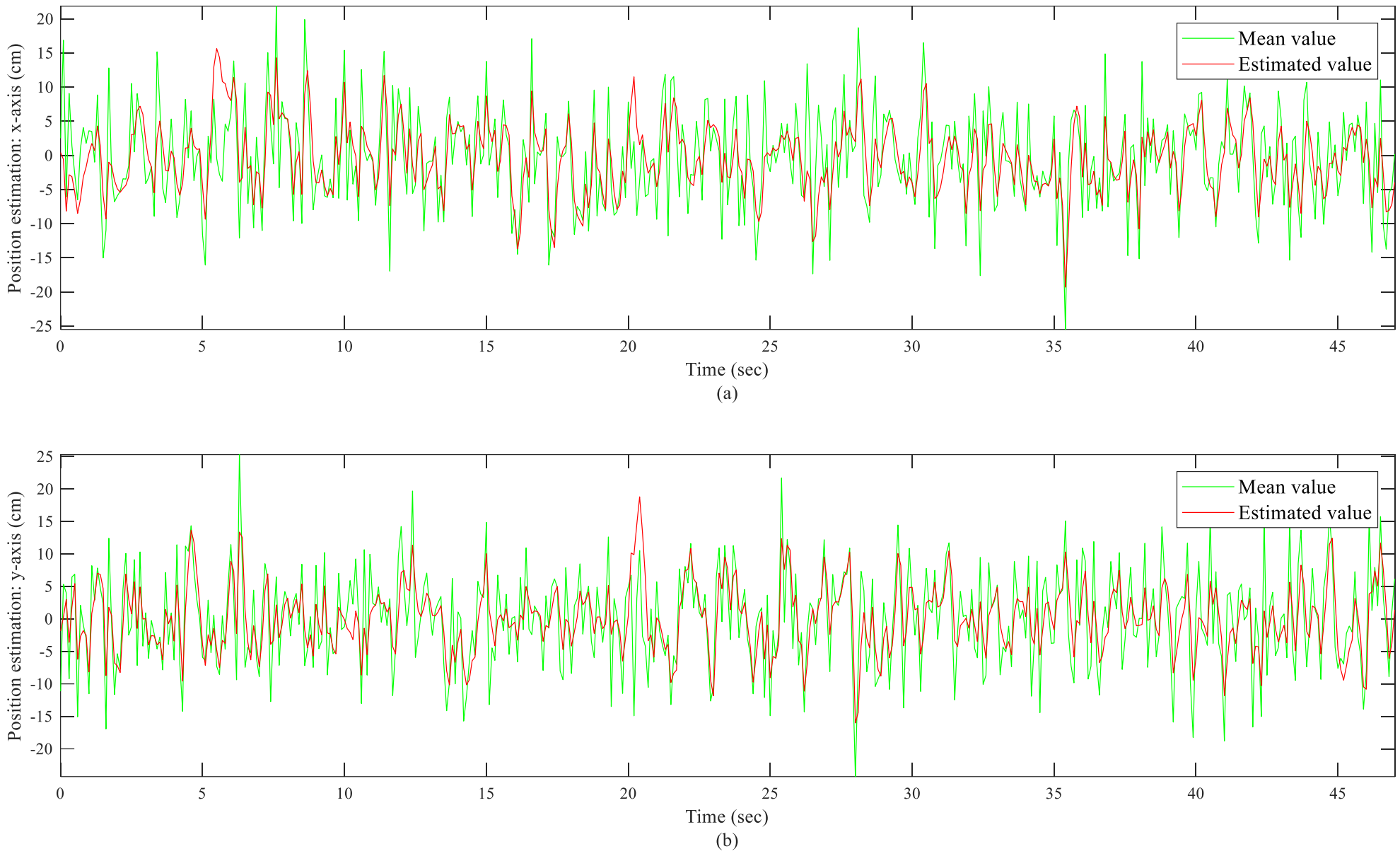

**Figure 31.** Tracking position mean value of smartphone using Kalman filter.

At the very beginning of the simulation, the position of the smartphone is not identified. Therefore, it shows unidentified location information of the smartphone in Figure 30. Applying Kalman filter in successive interval, the estimated position of the smartphone will perfectly identify at the fourth stage. It can be encapsulated from the Figure 30 that within 9 to 10 cm, the position of the smartphone is estimated using the tracking algorithm.

To remove the complexity from the simulation result, the variation of smartphone coordinates in the z-direction is ignored. The variation of x- and y-coordinate are recorded along with recording its performance of the tracking algorithm in Figure 31. Green tracks explain the mean value of smartphone movement, whereas the red line means the estimation with the Kalman filter. The shift of smartphone position is recorded both at x direction and y direction, which illustrations in Figure 31(a) and Figure 31(b), respectively. Sampling frequency for iteration is 1 Hz, whereas the total run time is 50 sec. Total 50 samples were taken for this simulation. It can be summarized from the Figure 31 that the deviation of the estimation is existed in all samples from the mean value of the position.

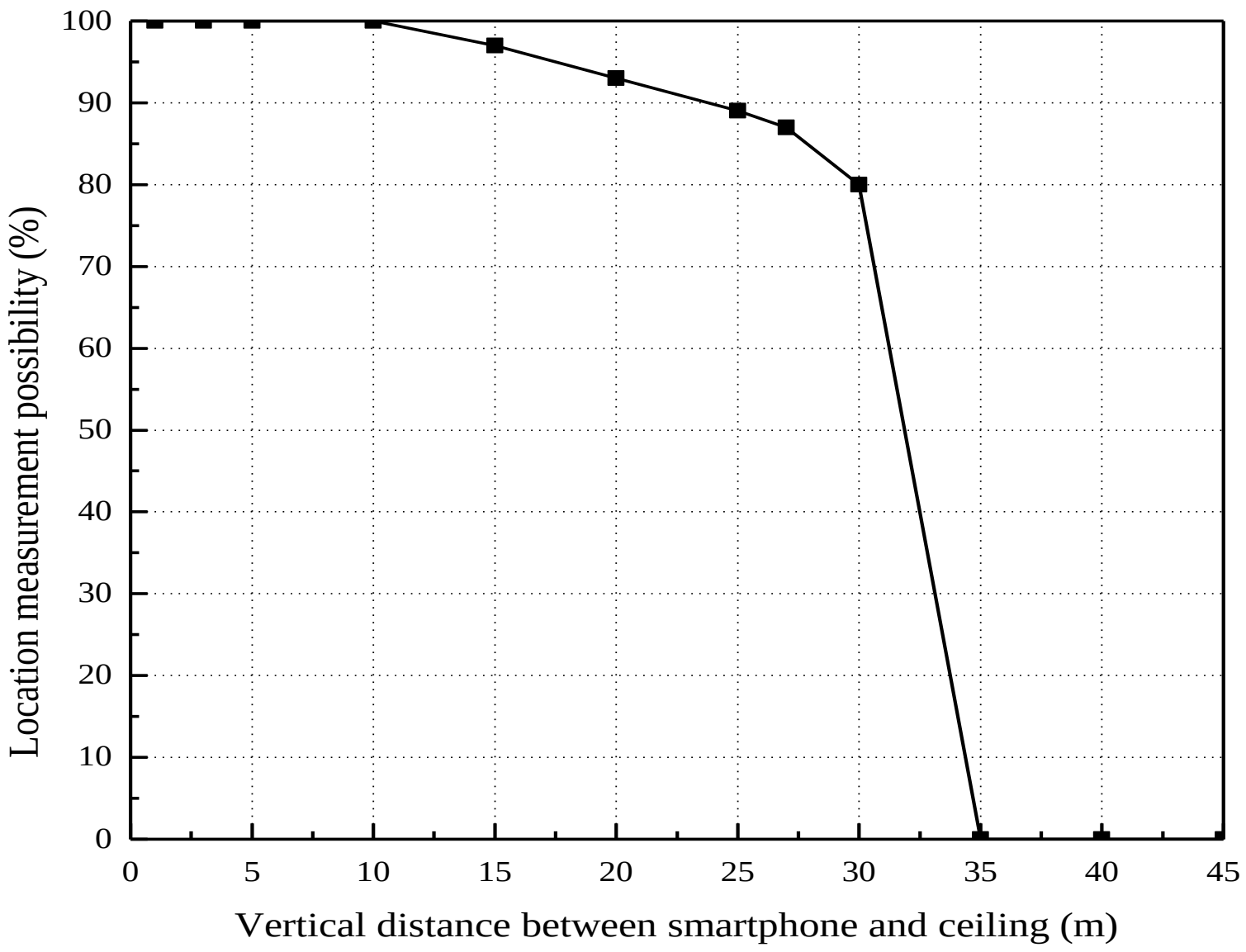

**Figure 32.** Comparing localization possibility with indoors height.

Using photogrammetry, the variation of distance measurement measures with the variation of the variation of the projected image on the IS. With the increase of distance of smartphone from the LED light, the occupied projected image on the IS decrease. Theoretically the possibility of distance measurement stays until the projected image on the IS occupies greater or equal to the unit image area. When the ceiling is much higher, the chance of the projected image will be less than one-unit area of the IS. Simulation result plotted in Figure 32 shows that the projected image area is 4-pixel area when the vertical distance from camera to ceiling is 10 m and this point possibility of localization is much higher than other points. The pixel area decreases, i.e., $4 > \eta_i \geq 1$ with the increase of the distance from 10 to 35 m, the possibility of localization also decreases accordingly. Nevertheless, it is difficult to confirm that the projected image merges with pixel edge-by-edge. After 35 m distance in the simulation case and specific transmitter case, the occupying pixel area of the IS becomes $\eta_i < 1$ and makes the localization possibility dramatically zero.

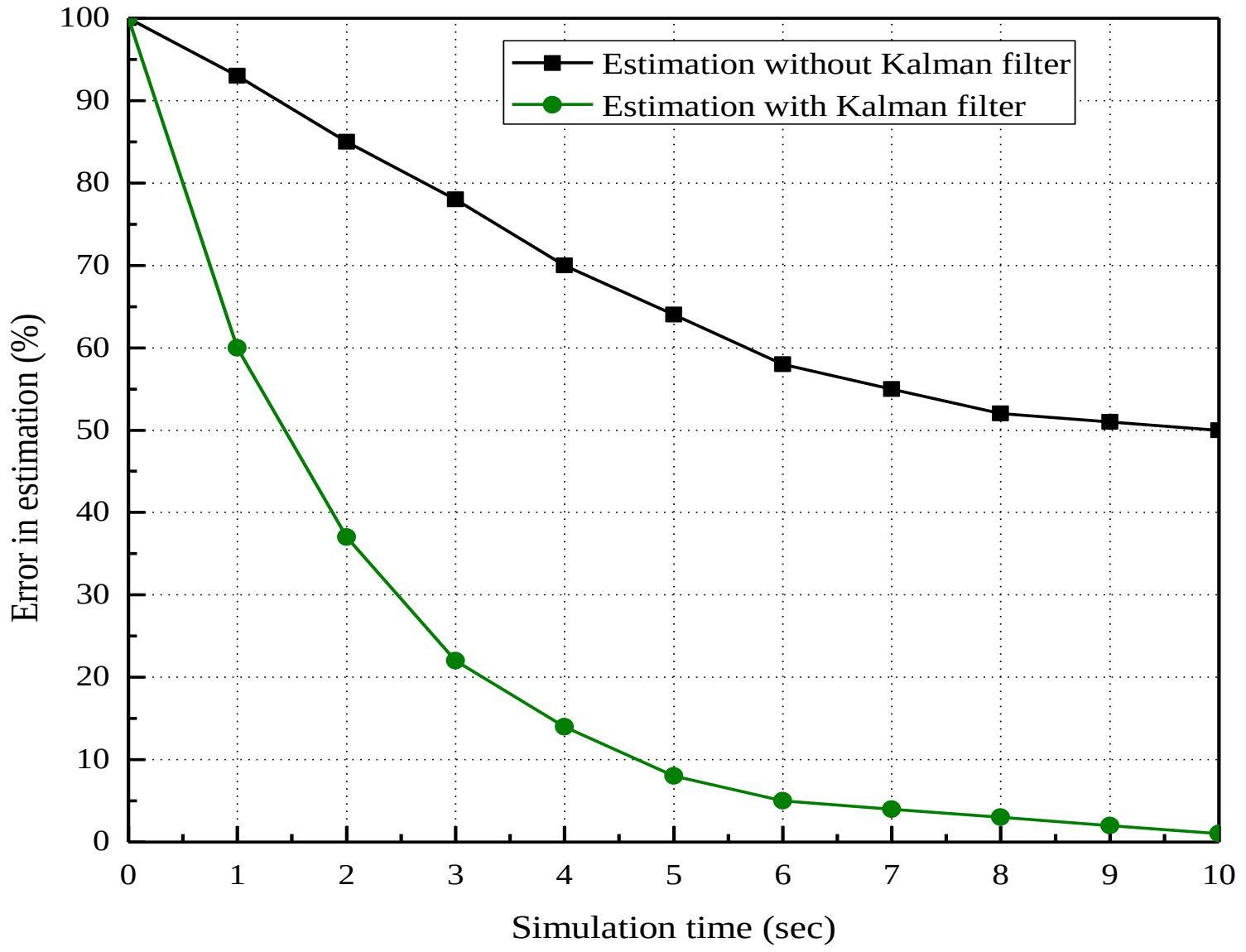

**Figure 33.** Impact of Kalman filter on error estimation for smartphone position.

With the change of the transmitter size, it is possible to increase the possibility of localization accordingly.

In our lab test environment, it is required 80 ms to receive signal from the LED and process the signal for extract data from the signal. Finally, to localize the smartphone, less than 500 ms is required. During the time of localization calculation, the smartphone may change its position and calculate the coordinates may not provide an impact as well. Therefore, it is required to deploy a tracking algorithm to estimate the next possible smartphone location utilizing its previous position information along with the velocity, acceleration, etc. The simulation records the position estimation result in both cases, e.g., with the presence of the Kalman filter and without the presence of any tracking algorithm in Figure 33. The recorded data are plotted and the findings show a significant performance deviation. At the very beginning, error in position estimation was higher in both cases. Over time, both cases able to decrease the percentage of error. Moreover, the error mitigation in location estimation reduces more when the algorithm uses a Kalman filter to estimate the next possible smartphone location. At 10 sec periods, approximation error is near about zero for Kalman filter based approximation. Meanwhile, at same time, another approximation

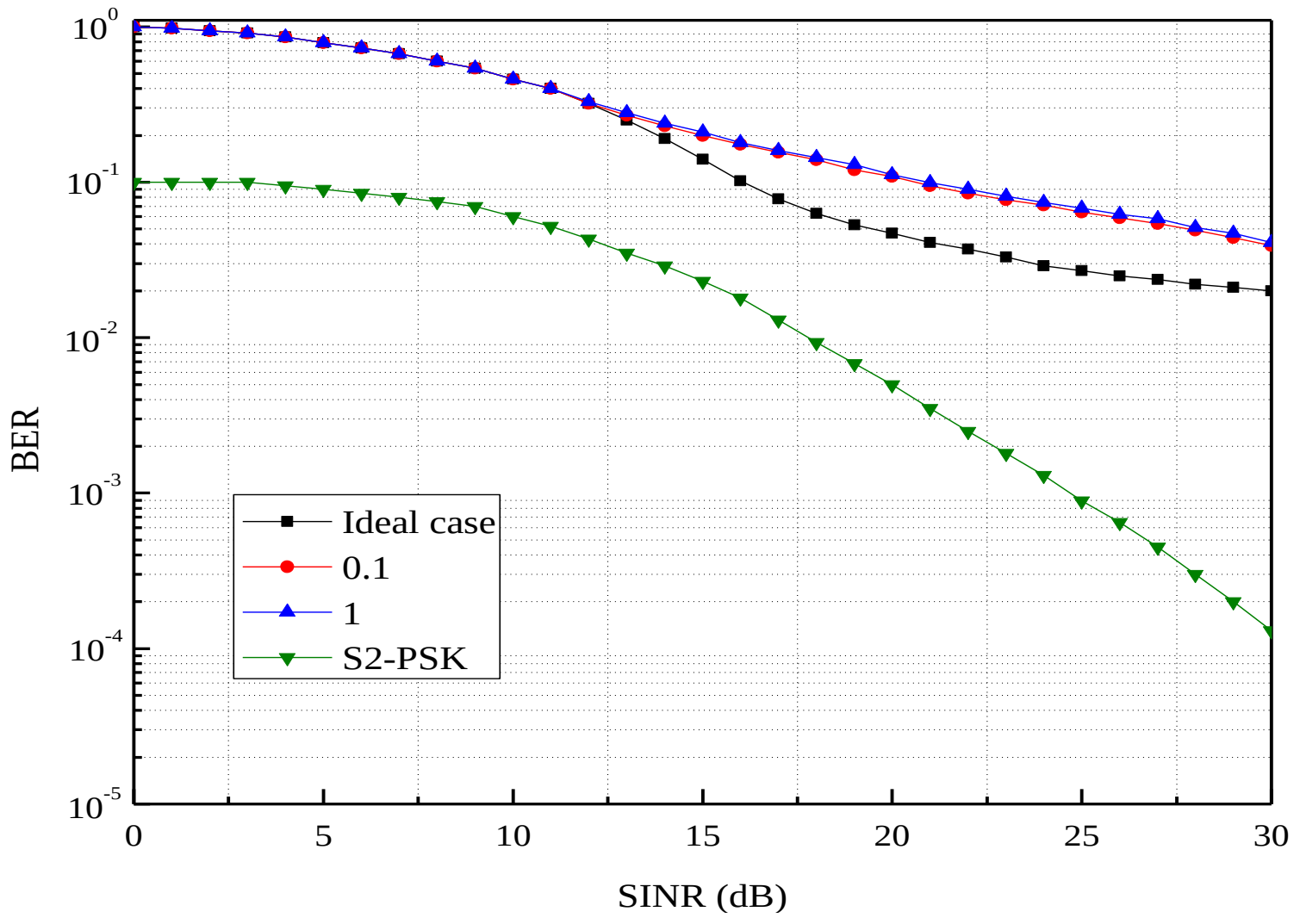

**Figure 34.** SNIR vs BER for vehicle localization.

(i.e., without Kalman filter) shows 50 % error in overall approximation.

## 6.2 Simulation Results for Vehicle Localization

It has to be consider several facts and environmental influences to achieve accuracy in localization. In the simulation result, turbulence in road scenarios as well as the worst weather conditions, e.g., fog, snow, and rain; have to be ignored to judge the proposed localization scheme in a vehicular environment. It keeps constant other parameters to evaluate performance of a single parameter on the localization correctness.

The BER performance with respect to the SNIR estimated with a low pass filter, i.e., Gaussian filter as a blurring filter for image processing. In ideal state, the variance $\sigma_c^2$ (=0.5) was considering zero for channel filtering in Figure 34. Curves for estimating valiances (=0.1, 1.0) for the same low pass filter were also plotted in the same figure to evaluate the impact of channel filter estimation errors. Simulation results conclude with the improve BER performance over SNIR.

The frame rate of the camera has an impact on the communication data rate. Theoretically, each frame rate is responsible for detecting one-bit data. Such as, for the 30 frames per second (fps) camera it is possible to detect 30-bit data from the LED source. The Manchester coding technique uses in S2-PSK modulation scheme for data encoding. Therefore, one-bit data will recover from two sequential frame and the data rate is 15 bits per second for 30 fps camera. In Figure 35, at varying data speed the performance of BER is plotted. For lower BER, LED power is required to increase for higher data bit transmission.

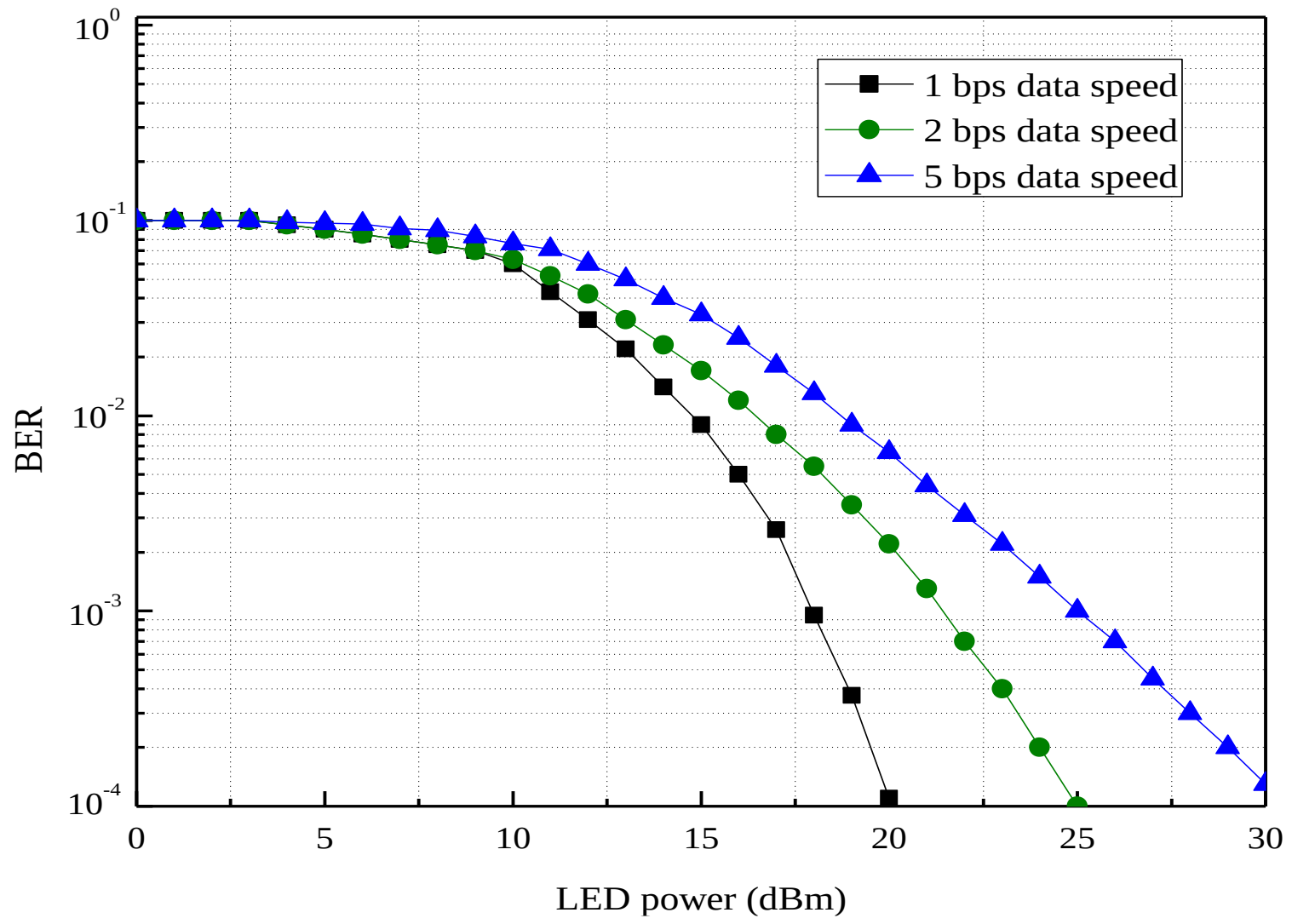

**Figure 35.** LED power vs BER for vehicle localization.

The distance error happens when there is an inconsistency between tangible and measure distance. Systematic error triggered by environmental particulars, observation approaches, and tool leads to this false measurement in such vehicular environment and necessity to be minimized to accomplish improved positioning precision. Average error takings from series of repetitive measurements, whereas maximum error produces from a single measurement. In the point of average and maximum error, the association of a distance error with different camera parameters supports to advance the performance of distance measurement tactic. Among other camera parameter IS resolution is on one of them. The resolution is defined by the number of pixels which has an impact on distance calculation. With higher camera resolution, the possibility to calculate the projected image area increase as well as more precise localization is possible. Better resolution provides more detail to detect projected image area of the LEDs on the IS. On the contrary, error increase during distance measurement due to the low resolution camera. From the Figure 36, at 1 megapixel, both average and maximum distance measurement error is 17.5 cm and 13.3 cm, respectively, which is higher than other cases. In the camera of 5 to 10 megapixel, a linear variation found for maximum distance measurement error whereas average error remains constant.

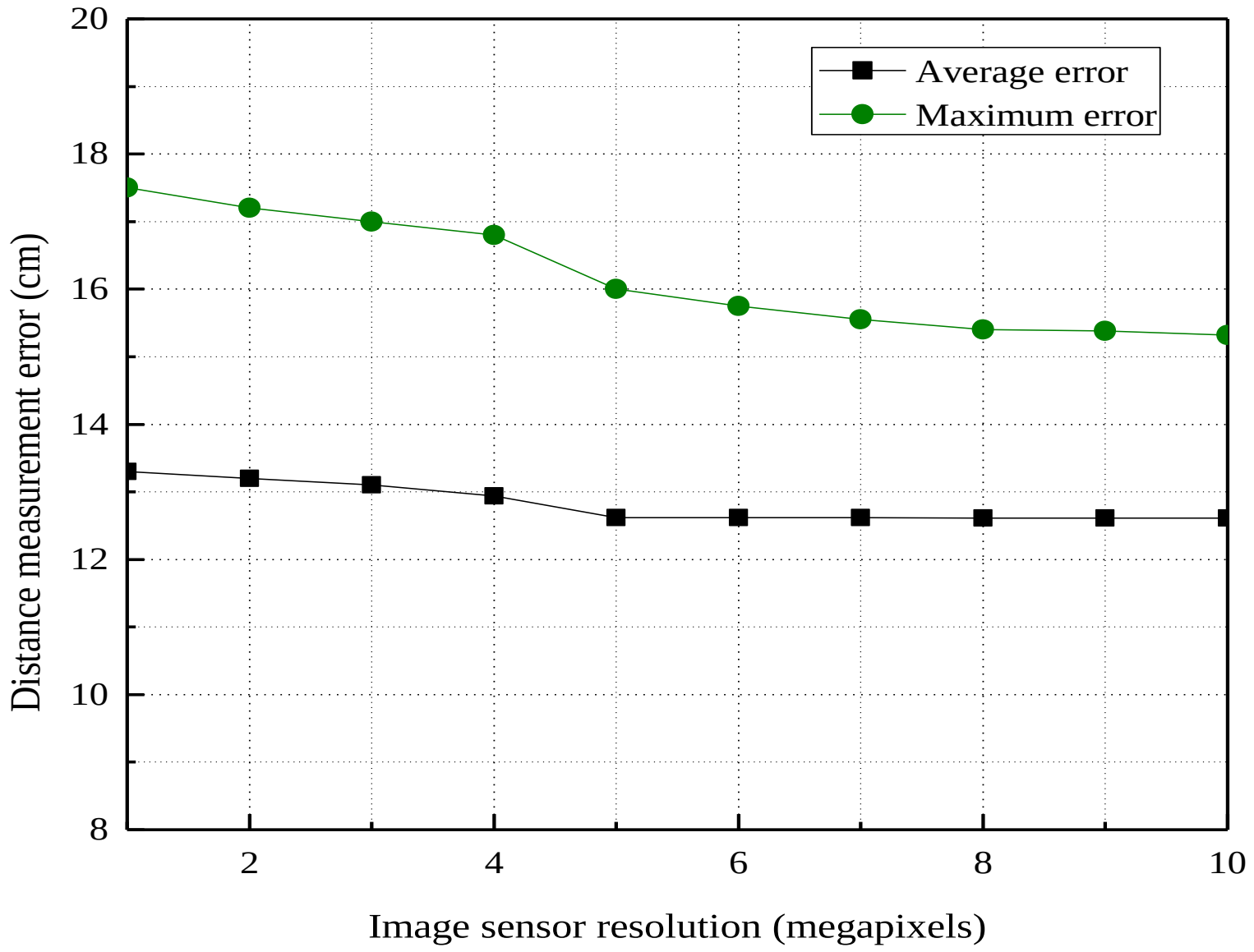

**Figure 36.** Distance measurement error vs image sensor resolution of HV's camera.

The vehicle localization scheme should perform perfectly in all sorts of scenarios as well as with the variable speed of FV. Camera of the HV receives signal from the taillight of the FVs even when they are moving at top speed. The IS should entirely expose under the illumination with each aspect of the targeted LEDs i.e., forwarding vehicle's taillight, streetlights; during this dynamic circumstances. The camera shutter speed (also called exposer time) will ensure a period when the amount of light will be exposed to the IS. In the high speed vehicular environment, longer camera exposure time causes a blurred image and shorter exposure time helps to capture detail sparks of light from a target object. The exposure time has an influence on assessing the performance of distance calculation due to the dependence on the receive image quality at the IS. Error in distance measurement increased in both cases of maximum and average with the increased value of camera shutter speed as in Figure 37. At lower shutter speed, e.g., 1/2000; higher localization accuracy obtained at the point of low measurement error, whereas maximum error obtained at the point of higher shutter speeds, e.g., 1/15.

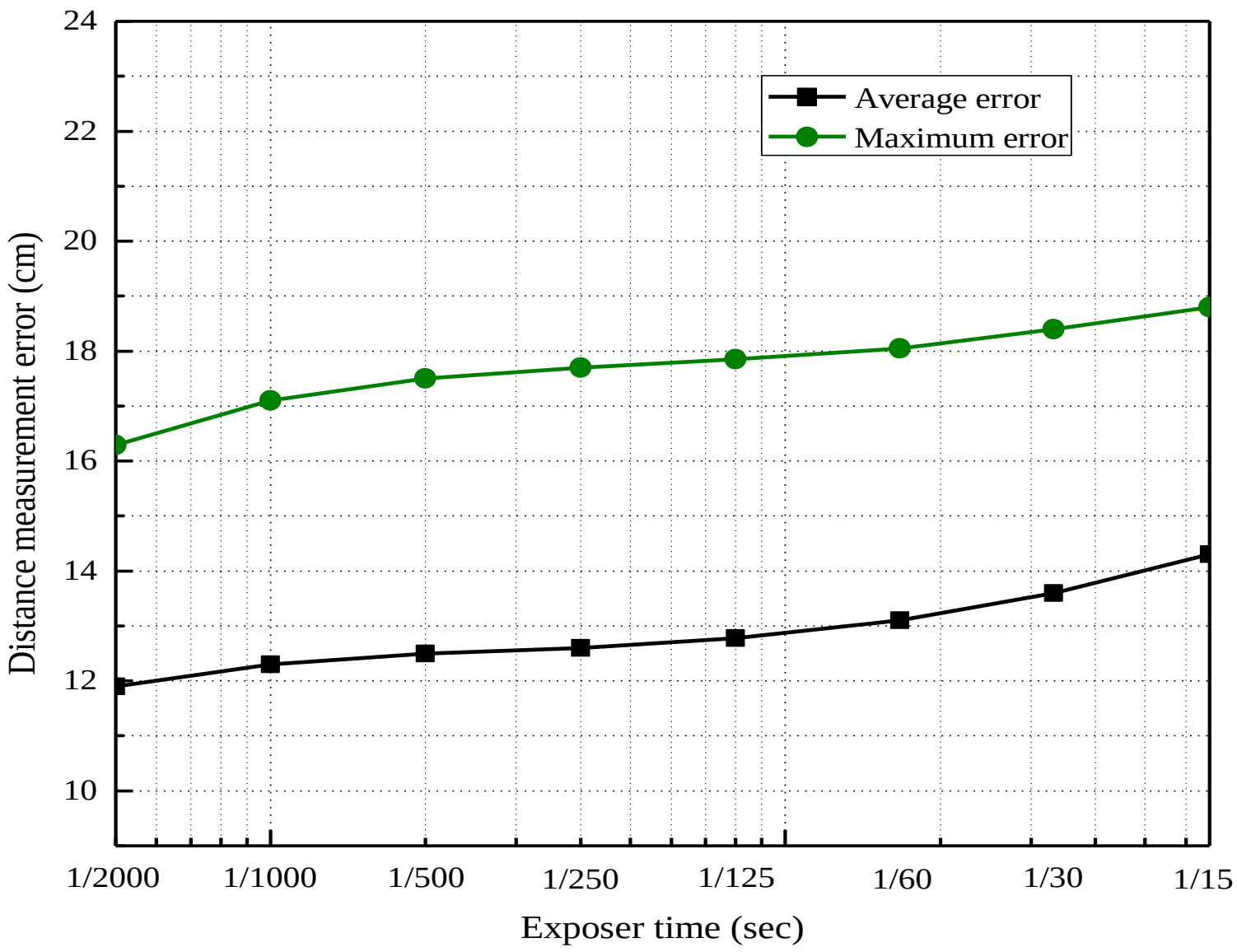

**Figure 37.** Distance measurement error vs the exposure time on HV's camera.

There are two important facts should be considered during the distance measure between the two vehicles, i.e., shift position and vehicle speed. To evaluate impact of vehicle speed on the vehicle positioning accuracy, mute the influence of position shifting of FV and plotted the simulation result in Figure 38. Within 200 m distance, FV's speed varies from 0 to 110 km/h and HV's speed remains constant (i.e., 30 km/h). Due to the relative distance between FV and HV, the error was found at the start of the simulation. Up-to 110 km/h speed of FV, both maximum and average distance measurement error were increased steadily. It was happening due to the requirement of execution time for location estimation of FV.

At a constant speed of the HV i.e., 50 km/h; simulating position measurement accuracy wherein distance between SLs varied from 10 to 150 m. At the beginning, in the Figure 39, lower accuracy for distance measurement obtained due to the data extraction from the SL-IDs as well as the speed of HV. When the distance between two adjacent SLs was 10 to 40 m, the localization accuracy raised-up sharply. This was happening due to the number of cross SLs was higher within a very small period of time. Additionally, the number of SLs have an impact on the performance of localization accuracy. It confirms the calculation of virtual coordinates accurately.

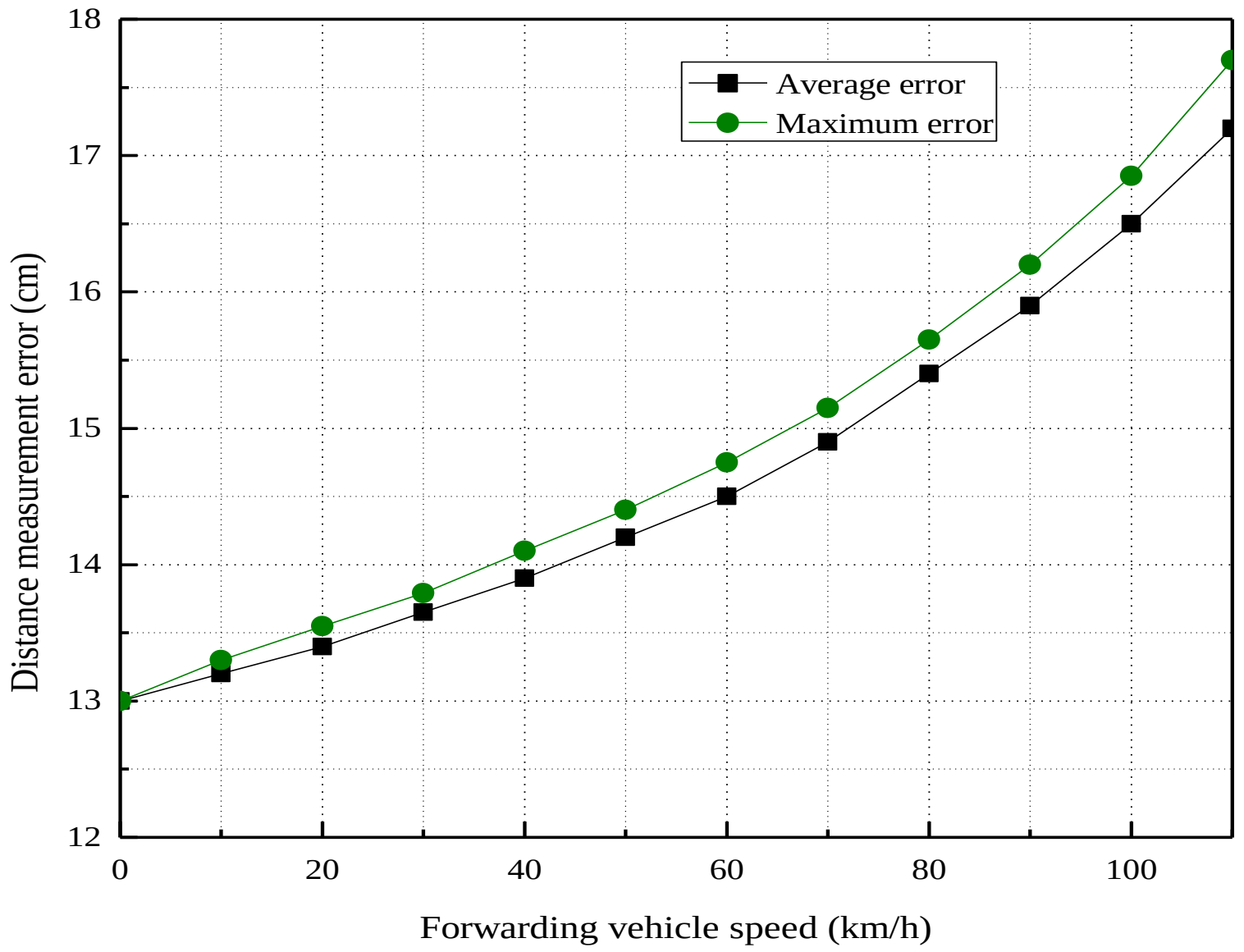

**Figure 38.** Distance measurement error vs speed of Forwarding Vehicle.

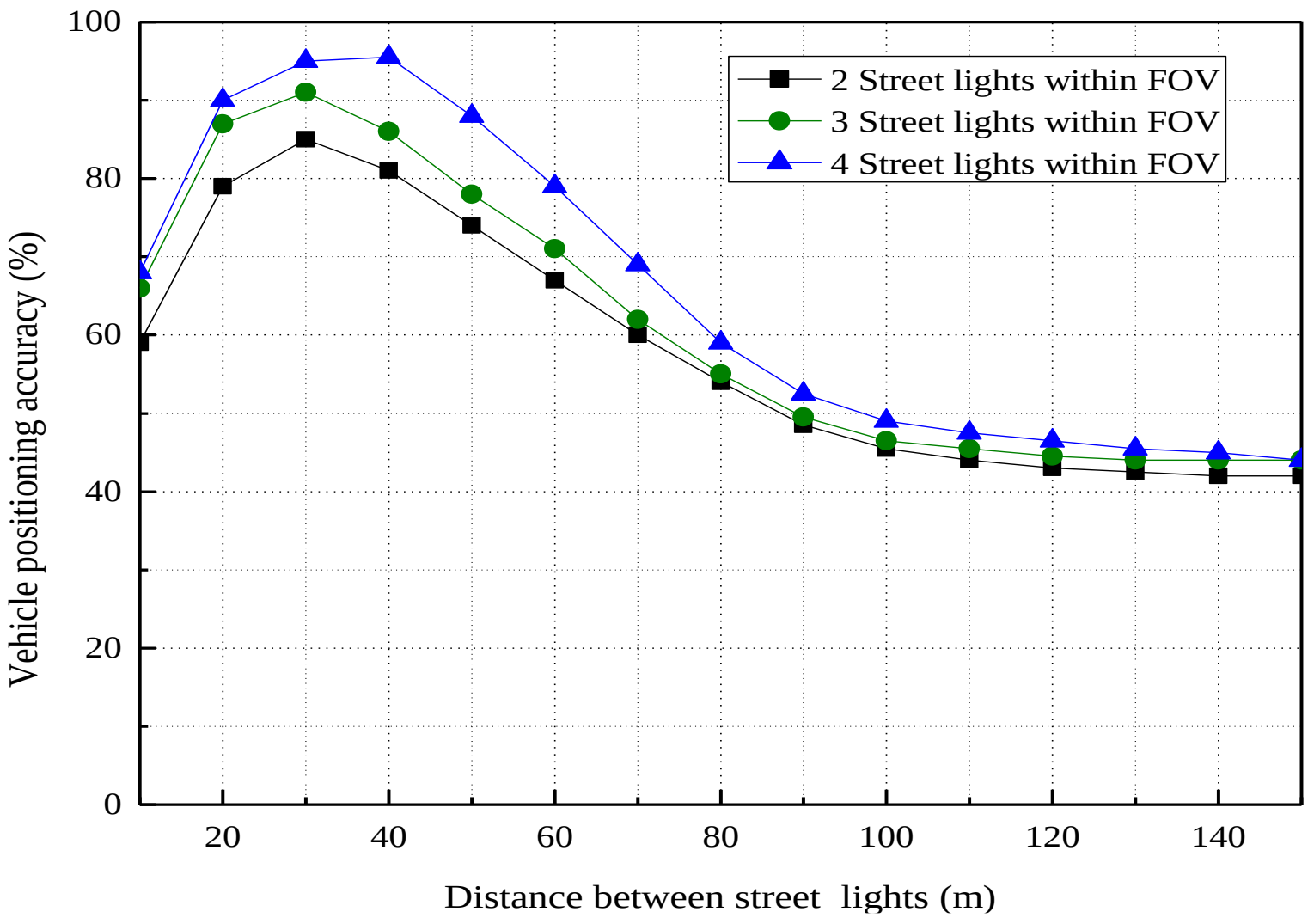

**Figure 39.** Vehicle positioning accuracy vs distance between two street lights.

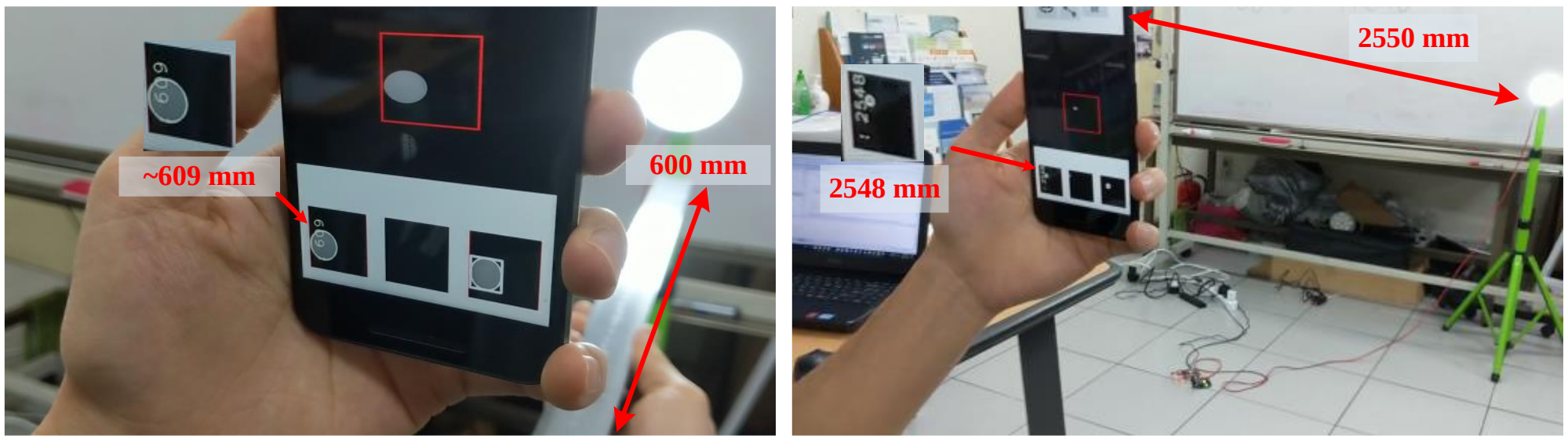

(a) (b)

**Figure 40.** Experimental arrangement for distance measurement.

At the maximum pitch of the graph shows that the distance measurement accuracy approximately 90% when the distance between two SLs was 40 m and the camera receives SL-ID from four adjacent SLs simultaneously. Moreover, with the increase of distance between SLs minimize the chance of accurate HV positioning. Therefore, the slope of overall location measurement accuracy moves down.

## 6.3 Experimental Set-up for Distance Measurement

The key concept of this research work to measure distance from the camera to the LED light fixtures by measuring its contour on the IS. Figure 40 shows the experimental set-up and performance of distance measurement procedure using existing lab facilities under ambient light condition (e.g., within the Lab environment). The contour of a circular LED measures with the camera from various distances. The measuring distance was 609 mm when the camera located approximately 600 mm away from the camera in Figure 40(a). Concurrently, this distance is 2550 mm when the measured distance shows some error, i.e., 2548 mm in Figure 40. With this experimental result a data plot generates as in Figure 41. The plotted result demonstrations the measurement error percentage with respect to the direct distance between camera and LED. For most of the cases the measurement error resolution remains within 1%. The experiment on distance measurement endorsed the possibility of the proposed localization scheme.

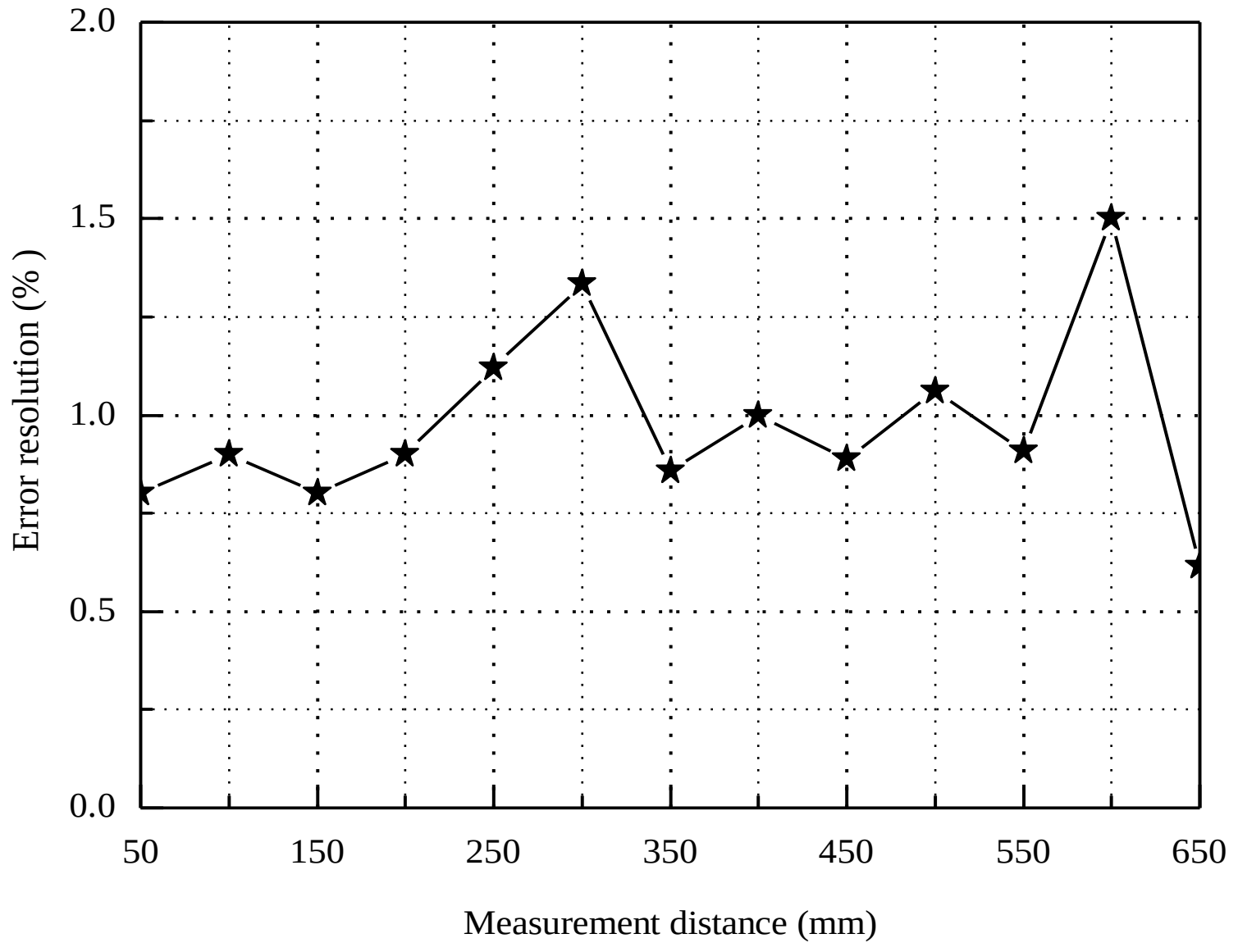

**Figure 41.** Experimental error resolution vs measurement distance.

## 6.4 Challenges in Localization

An Optical band based localization scheme still not mature enough and facing several challenges in the long run of commercialization. Distortion of captured image camera and the effect of ambient light sources of the signal data are the most considered challenges.

## 6.5 Error Mitigation in Localization

The preparation for error mitigation is not enough for OCC and photogrammetry based indoor [111] and vehicle localization. It is far more unavoidable the impact of lens distortion from localization. Therefore, the method of back propagation algorithm can be applied to face this challenge. Concurrently, calculating light intensity on the higher dynamic range helps to minimize the influence of ambient light during image projection.

### 6.5.1. Normalizing the effect of distortion

The overall distortion can be classified as a lens or optical distortion and perspective distortion. Both kinds of distortion are responsible for image deformation. Curved projected image generates corresponding to the straight line in real world coordinates due to the lens distortion. This lens distortion can be classified as barrel distortion and pincushion distortion. On the other hand, perspective distortion caused due to the relative position between the object and the camera.

Applying back propagation algorithm, the error in localization calculates first and later optimize the error by comparing with the real-time measured value.

### 6.5.2. Cancelation the effect of ambient light

Importantly, a recent development of high-dynamic range imaging technique reduces noise and enhances the image quality under daylight. Therefore, it is expected that ambient light no longer poses a problem for OCC, even when the transmitter possesses a near-infrared (NIR)-optical band.

# Chapter 7

# Application of Localization in Wireless Networking

## 7.1 Introduction

Application of localization in location aware wireless network is increasing. In this chapter, the application of OCC and photogrammetry based localization in several scenarios, such as IoT, digital signage, autonomous vehicle, indoor cases; are described for networking.

## 7.2 Application Scenarios

### 7.2.1 Internet of things

IoT ensures the connectivity several physical devices with a network. Each of those devices has a unique identity to identify remotely. It will create opportunities of interaction pathways among devices and improve the overall accuracy, efficiency, and financial benefits for the human being. These devices accumulating essential data to share with other devices. Figure 42 shows the connectivity of various devices with the cloud.  Localization is a key feature for IoT.

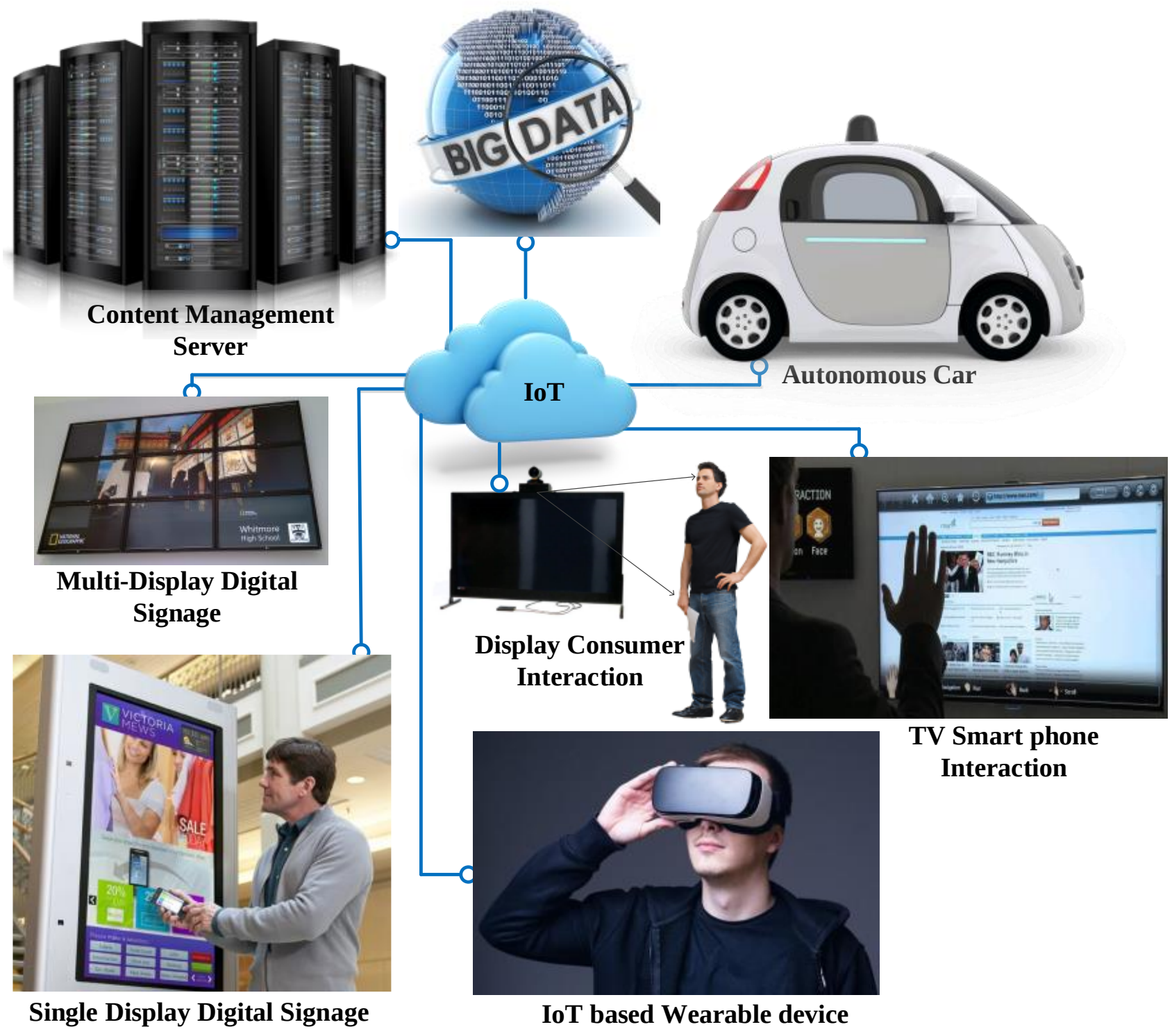

**Figure 42.** Localization is the key features for internet-of-things.

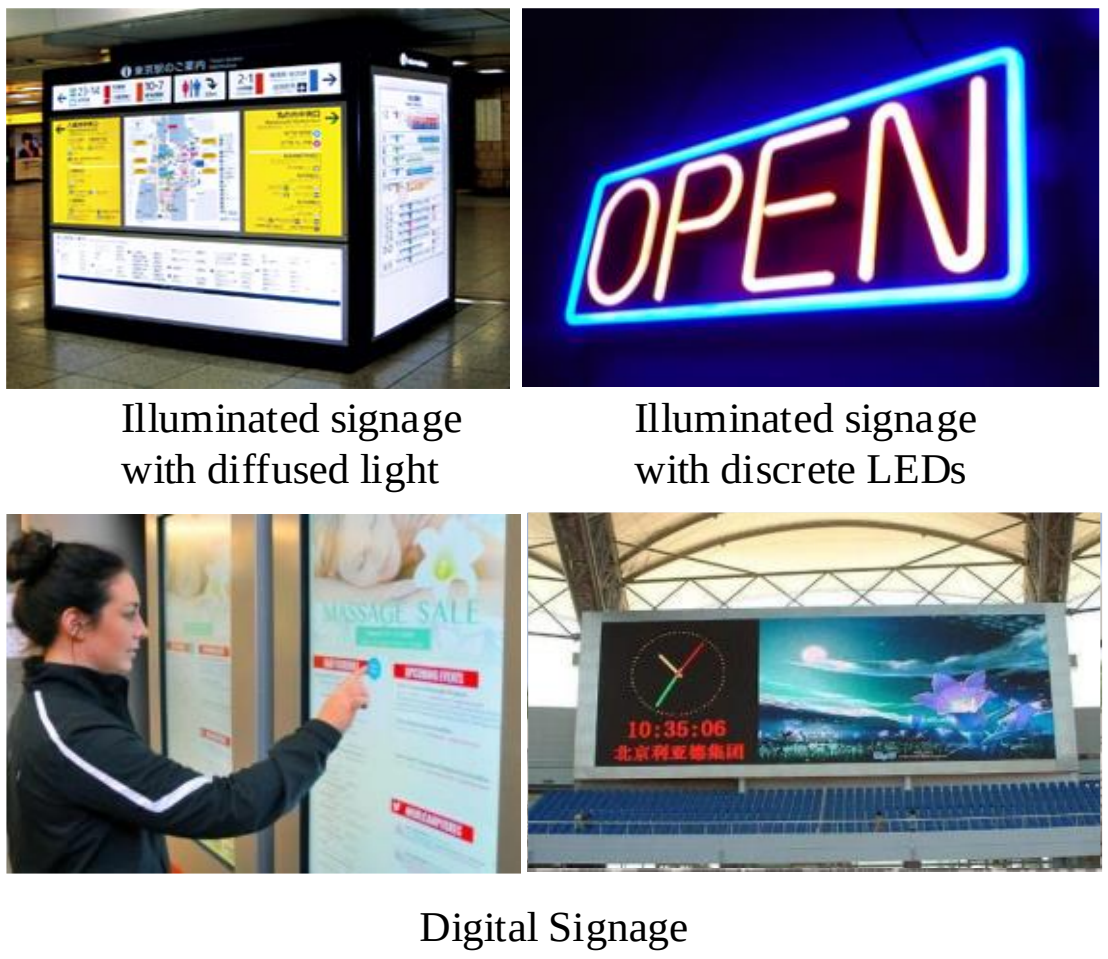

**Figure 43.** Application scenarios of location based services.

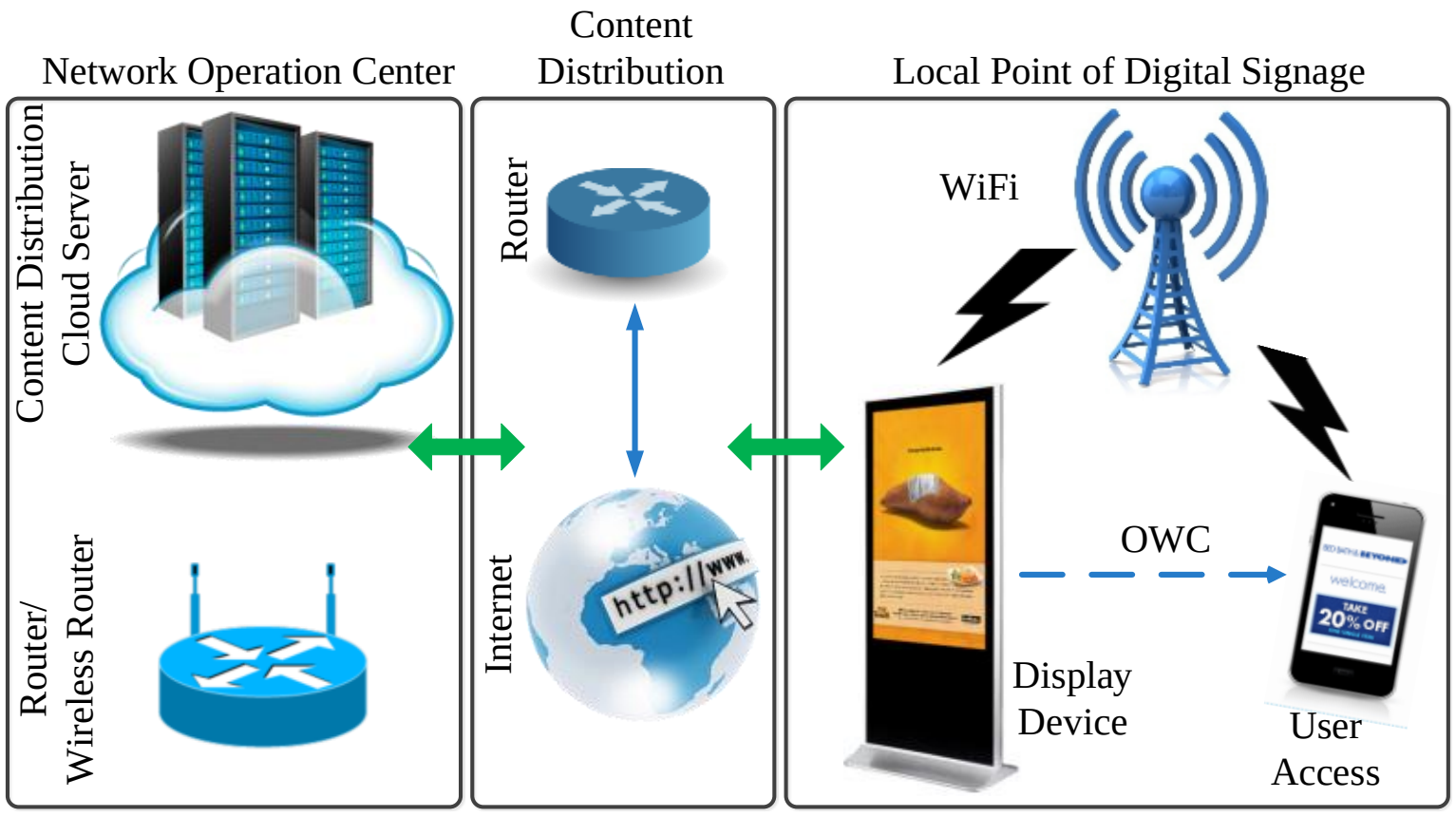

**Figure 44.** Back-hole network for location based services.

### 7.2.2 Location based services

LBS is a software-based services promote to the users or consumers in the context of location information. The application of LBS is basically found in health care, indoor sensor search, information about promotional data, and so on. It ensures a connectivity between service provider and the consumers. Figure 43 shows several application scenarios for LBS. Meanwhile, Figure 44 represents the networks behind the LBS scenarios.

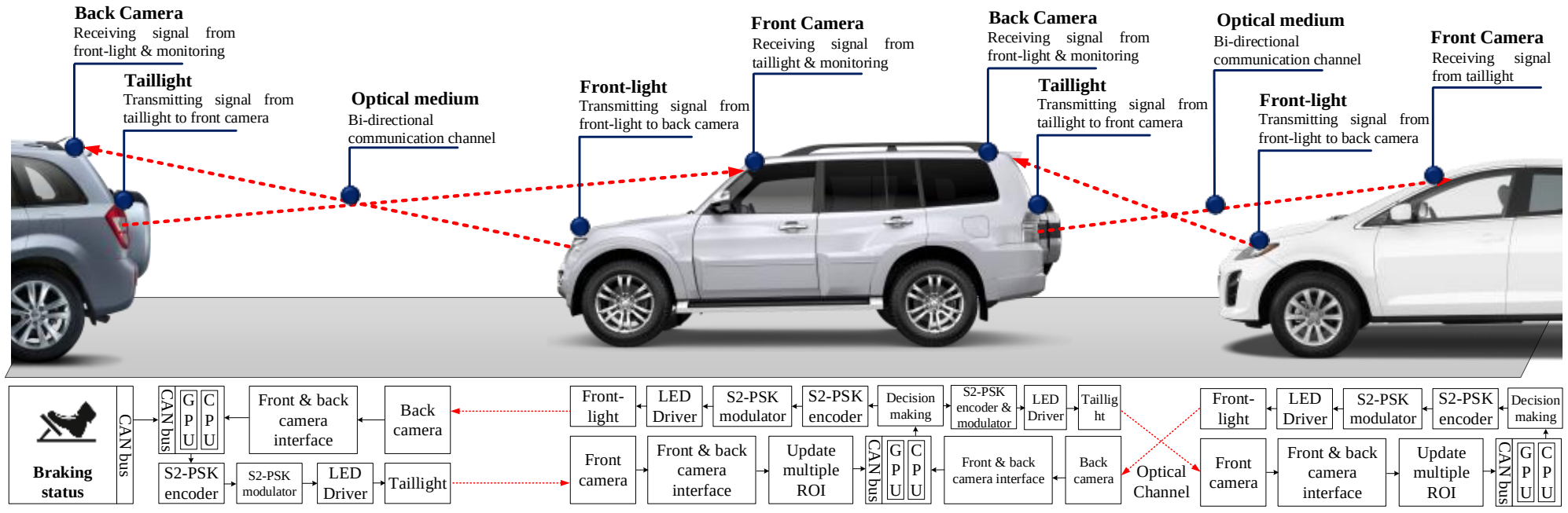

**Figure 45.** Relaying Emergency Information.

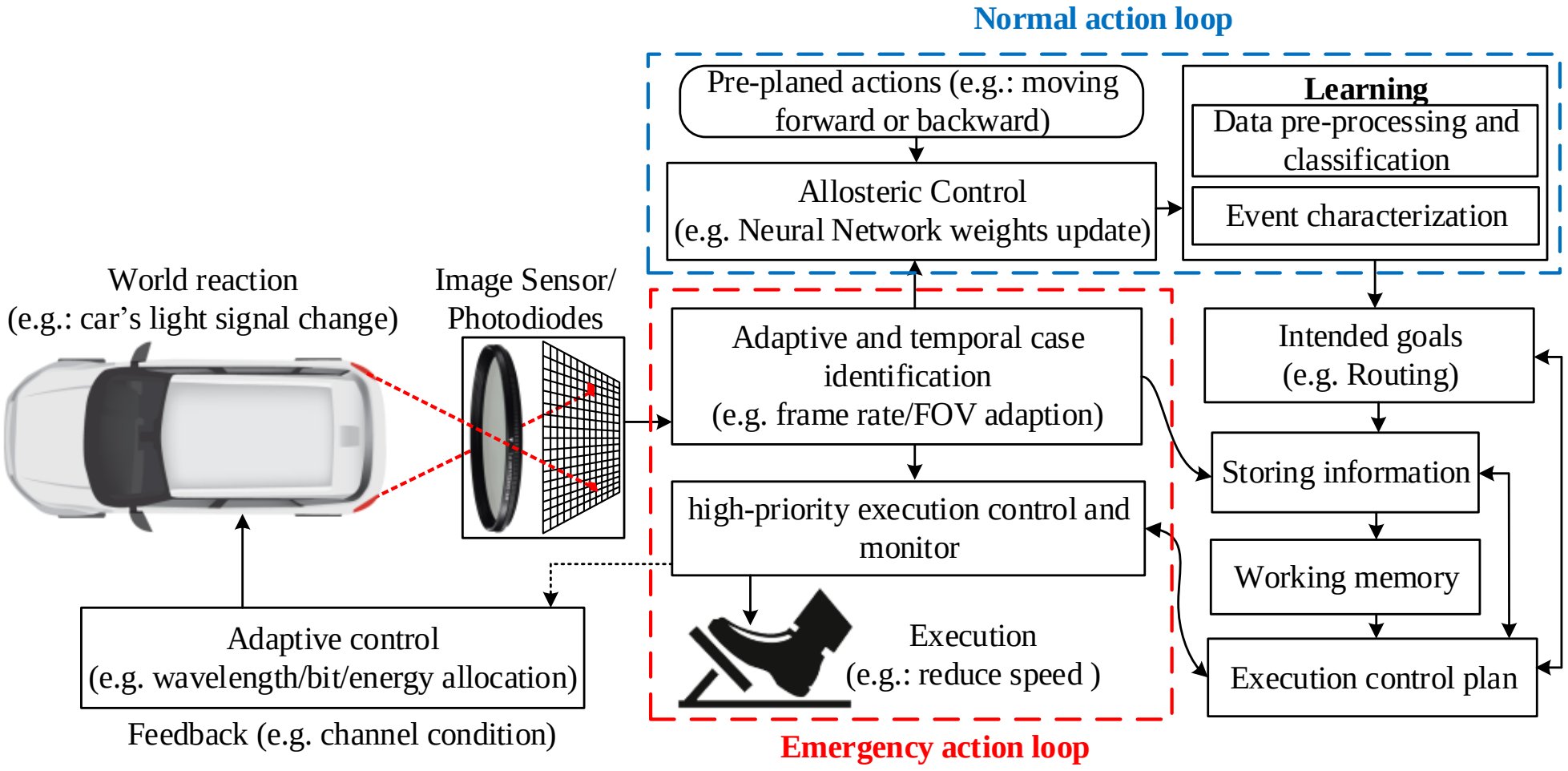

**Figure 46.** Emergency decision execution (e.g., braking, slowing down speed).

### 7.2.3 Emergency decision execution in vehicular environment

Vehicle localization data helps to keep monitoring other vehicles from a vehicle for internet of vehicle [112]. It identifies the adaptive and temporal cases of the FVs from the HV. The following vehicle always checking the status of the FV and relay this information to its following vehicles as in Figure 45. All these cases can be stored in the storage system and can be used for executing emergency decision e.g., reduce speed or pressing the breaks as in Figure 46.

# Chapter 8
# Conclusion and Future Directions

## 8.1 Summary

Throughout this study, an OCC and photogrammetry based indoor localization scheme has proposed. Deploying these technologies in the mainstream for localization is a novel idea. The simulation results show the 10 cm level localization resolution for indoor environment. The lighting server deals with the processing task to process the smartphone positioning calculation. Both the transmitter-receiver presents within the same indoor environment and the chance of interruption by other system is nearly zero, which ensure an extremely secure localization scheme for smartphone. Moreover, use of the tracking algorithm ensures mitigation of error from the overall scheme. Kalman filter tracks the next possible position of the smartphone. Previously the lighting fixtures used for illumination, later it uses for the communication and now it's possible to use the same infrastructure for localization.

A vehicle localization scheme introduced by employing the technology of OCC and photogrammetry. The distance measurement part is covered by the photogrammetry whereas OCC helps to make a communication link between two vehicles. Therefore, overall performance also depends on the involvement of the OCC as well. Measuring the variation of the projected image area on the IS, the variation of the distance between HV and FV can be measured. Meanwhile, using OCC technology, ROI is specified, which helps to fix the target to calculate required distance. The choice of localizing HV with respect to the HV is required to minimize the error of FV localization. The communication with the SLs would be the same way as the HV communicate with the FVs. The proposed vehicle localization scheme is tested by several simulation results.

## 8.2 Future Approaches

In the future, different environmental scenarios, e.g., staircase, escalator; that is, different heights of indoor environment will consider to evaluate performance of indoor localization scheme. Moreover, there are several facts has to be considered for example: different size of LEDs, localizing under different color of LEDs, rotation of the smartphone; is the most important than all. Additionally, it can predict that the development of the current proposed scheme will bring the stand-alone technology to localize smartphone without using Kalman filter.

On the other hand, for vehicle localization, there are still several challenges existing, for example, different size and shape of taillight to be detected. It is required to bring the neural network for detecting those different shapes. Moreover, there should deploy deep-learning based algorithm to eliminate the result of bad weather such as rain, fog, snow; in front side of the camera.

# List of Publications

## Journal

## Conference

## Patent

## Magazine

## IEEE Standard Contribution